\documentclass[aps,pra,twocolumn,superscriptaddress,longbibliography]{revtex4-1}
\usepackage{amsmath, verbatim, amsfonts, color}
\usepackage{graphicx}
\usepackage[T1]{fontenc}

\begin{document}

\title{Tight-binding theory of spin-spin interactions, Curie temperatures, and quantum Hall effects
in topological (Hg,Cr)Te in comparison to non-topological (Zn,Cr)Te, and (Ga,Mn)N}

\author{Cezary \'Sliwa}
\email{sliwa@magtop.ifpan.edu.pl}
\affiliation{International Research Centre MagTop, Institute of Physics, Polish Academy of Sciences,
al.\ Lotnik\'ow 32/46, PL-02668 Warsaw, Poland}
\affiliation{Institute of Physics, Polish Academy of Sciences,
al.\ Lotnik\'ow 32/46, PL-02668 Warsaw, Poland}

\author{Tomasz Dietl}
\affiliation{International Research Centre MagTop, Institute of Physics, Polish Academy of Sciences,
al.\ Lotnik\'ow 32/46, PL-02668 Warsaw, Poland}

\date{2024-06-05}

\begin{abstract}

Earlier theoretical results on $p$-$d$ and $d$-$d$ exchange interactions for zinc-blende semiconductors with Cr$^{2{+}}$ and Mn$^{3{+}}$  ions
are revisited and extended by including contributions beyond the dominating ferromagnetic (FM) superexchange term [i.e., the interband
Bloembergen-Rowland-Van Vleck contribution and antiferromagnetic (AFM) two-electron term], and applied to
topological Cr-doped HgTe and non-topological (Zn,Cr)Te and (Ga,Mn)N in zinc-blende and wurtzite crystallographic structures.
From the obtained values of the $d$-$d$ exchange integrals $J_{ij}$, and by combining the Monte-Carlo simulations with the percolation theory for randomly
distributed magnetic ions, we determine magnitudes of
Curie temperatures $T_{\text{C}}(x)$ for $\mathrm{Zn}_{1-x}\mathrm{Cr}_x\mathrm{Te}$ and $\mathrm{Ga}_{1-x}\mathrm{Mn}_x\mathrm{N}$
and compare to available experimental data. Furthermore, we find that
competition between FM and AFM $d$-$d$ interactions can lead to a spin-glass phase in the case of
Hg$_{1-x}$Cr$_x$Te. This competition, along with a relatively
large magnitude of the AF $p$-$d$ exchange energy  $N_0\beta$ can stabilize the quantum spin Hall effect, but may require the application of tilted magnetic field
to observe the quantum anomalous Hall effect in HgTe quantum wells doped with Cr, as confirmed by
the Chern number determination.

\end{abstract}

\maketitle

\section{Introduction}
Extensive studies of (Ga,Mn)As and other dilute ferromagnetic semiconductors (DFSs), in which
band holes mediate exchange interactions and magnetic anisotropies, have allowed
demonstrating several functionalities, such as the effects of light, electric fields and currents
on the magnetization magnitude and direction \cite{Dietl:2014_RMP,Jungwirth:2014_RMP}. Similarly striking phenomena
have been discovered in DFSs without band carriers, the prominent examples being the piezo-electro-magnetic
effect in wurtzite (Ga,Mn)N \cite{Sztenkiel:2016_NC} or the parity anomaly \cite{Mogi:2022_NP} and the quantum anomalous Hall
effect \cite{Yu:2010_S,Chang:2013_S} in topological (Bi,Sb,Cr)Te and
related systems \cite{Ke:2018_ARCMP,Tokura:2019_NRP,Bernevig:2022_N,Chang:2023_RMP}, the latter opening a prospect
for developing a functional low magnetic field resistance standard \cite{Goetz:2018_APL,Fox:2018_PRB,Okazaki:2022_NP,Rodenbach:2023_arXiv}.
In agreement with indications coming from synchrotron studies of magnetic configurations of V and Cr
impurities in (Bi,Sb)$_2$Te$_3$ \cite{Peixoto:2020_QM},
a theory developed by us for dilute magnetic semiconductors (DMS) such as
topological (Hg,Mn)Te and topologically trivial (Cd,Mn)Te \cite{Sliwa_2021} shows that,
if a spurious self-interaction term is disregarded, the Anderson-Goodenough-Kanamori superexchange
dominates over the interband Bloembergen-Rowland-Van Vleck mechanism \cite{Yu:2010_S,Ke:2018_ARCMP,Tokura:2019_NRP,Bernevig:2022_N,Chang:2023_RMP}
on the both sides of the topological phase transition in insulator magnetic systems \cite{Sliwa_2021}.
Ferromagnetic insulators also serve to generate giant Zeeman splitting of bands in adjacent
layers: the quantum anomalous Hall effect in (Bi,Sb)Te/(Zn,Cr)Te quantum wells (QWs) \cite{Watanabe:2019_APL}
and topological superconductivity in InAs nanowires proximitized by an EuS
layer \cite{Escribano:2022_QM} constitute just two recent examples.


Exchange interactions between the spins of electrons residing in partially occupied $d$-shells, mediated by
hybridization $\hat{V}_{\text{hyb}}$ between $d$ shells of magnetic ions and $p$ orbitals of neighboring anions, can be either
ferromagnetic (FM) or antiferromagnetic (AFM), orbital dependent, and sensitive to Jahn-Teller distortions \cite{Goodenough_1963}.
The progress in understanding the underlying mechanism of these interactions allowed to formulate
Anderson-Goodenough-Kanamori rules \cite{Goodenough_1963}, which predict the character of the interactions in a given system
provided that the nature of the relevant chemical bonds is known. More recently, Blinowski, Kacman,
and Majewski \cite{Blinowski_1996} --- by a complex calculation involving the fourth-order perturbation
theory in $\hat{V}_{\text{hyb}}$ --- found that in tetrahedrally coordinated II-VI Cr-based dilute magnetic semiconductors
the superexchange component of the spin pair interaction is FM.
This FM coupling results from the partial filling of $t_2$ orbitals occurring for the high spin $d^4$ electronic
configuration of the magnetic shell in the case of the substitutional $\mathrm{Cr}^{2{+}}$
ions in the tetrahedral environment. On the experimental side, FM ordering
was indeed found in (Zn,Cr)Te but a meaningful determination of Curie temperature $T_{\text{C}}$ as a
function of the Cr concentration $x$ is hampered by aggregation of Cr cations \cite{Kuroda:2007_NM}. Accordingly,
the values of $T_{\text{C}}(x)$ determined recently for high quality $\mathrm{Zn}_{1-x}\mathrm{Cr}_x\mathrm{Te}$
epilayers \cite{Watanabe:2019_APL} provide an upper limit of $T_{\text{C}}$ expected for a random distribution of magnetic
ions. The generality of the superexchange scenario was then demonstrated in the case of wurtzite (wz) (Ga,Mn)N containing
randomly distributed $\mathrm{Mn}^{3{+}}$ ions, for which the experimental dependence $T_{\text{C}}(x)$ \cite{Sarigiannidou:2006_PRB,Sawicki:2012_PRB,Stefanowicz:2013_PRB} was reasonably
well explained theoretically employing Blinowski's {\em et al.} theory for the zinc-blende structure. Monte-Carlo
simulations served to obtain $T_{\text{C}}(x)$ magnitudes from the computed values of the exchange
integrals $J_{ij}$ for Mn spin pairs at the distances $R_{ij}$ up to the 16th coordination
sphere \cite{Sawicki:2012_PRB,Stefanowicz:2013_PRB,Simserides_2014}.

Here, the above-mentioned theoretical results for the $d^4$ configuration in DMS without band carriers
are revisited and extended by including additional contributions beyond
the superexchange, which results from the fourth order perturbation theory in $\hat{V}_{\text{hyb}}$, such as the interband
Bloembergen-Rowland term
\cite{Bloembergen:1955_PR,Lewiner:1980_JPCol,Larson_1988,Dietl:2001_PRB,Kacman:2001_SST,Sliwa:2018_PRB,Sliwa_2021},
known in the topological literature as the Van Vleck mechanism \cite{Yu:2010_S,Ke:2018_ARCMP,Tokura:2019_NRP,Bernevig:2022_N,Chang:2023_RMP}.
Furthermore, our theory is developed for both zinc-blende and wurtzite semiconductors, allowing us to assess the crystal structure's role in magnetic interactions. We also demonstrate quantitative agreement in the
values of $T_{\text{C}}$ obtained by the Monte-Carlo simulations and from the percolation theory for random
ferromagnets \cite{Korenblit_1973}. This agreement makes us possible to
determine $T_{\text{C}}(x)$ from $J(R_{ij})$ values for the whole $x$ range, $0 < x < 1$, without Monte-Carlo simulations
that are computationally expensive for disordered magnetic systems. The present developments are possible by making
use of contemporary software tools for formula derivations, which allowed us to correct several inaccuracies
in the original formulae \cite{Blinowski_1996} for superexchange between pairs of spins residing on $d^4$ shells.
We also correct some previous inconsistencies in values of Mn parameters \cite{Simserides_2014} and discuss
our results in comparison to experimental data on $T_{\text{C}}(x)$ for wz-$\mathrm{Ga}_{1-x}\mathrm{Mn}_x\mathrm{N}$
\cite{Sarigiannidou:2006_PRB,Sawicki:2012_PRB,Stefanowicz:2013_PRB} and zb-(Zn,Cr)Te \cite{Watanabe:2019_APL}.



Comparing the present results for Cr$^{2{+}}$ and Mn$^{3{+}}$ to the
previously investigated Mn$^{2{+}}$ case [realized in (Hg,Mn)Te and
(Cd,Mn)Te] \cite{Sliwa_2021},  the superexchange term (denoted as $hh$)
has the opposite sign for those two configurations. In contrast, the
interband $he$ and $ee$ contributions behave similarly, i.e., $he$ is FM
at small distances between magnetic ions and AFM for distant pairs. Our
results demonstrate, in agreement with experimental observations, that FM
interactions prevail in the case of (Zn,Cr)Te and wz-(Ga,Mn)N. At the same
time, our $T_{\text{C}}$ values are much smaller compared to {\em ab
initio} results \cite{Sato:2010_RMP}, as local functional approximations tend
to underestimate localized character of $3d$ orbitals in semiconductors.

According to our theory, the ferromagnetic $hh$ term entirely dominates in
(Ga,Mn)N. We show that the theoretical $T_{\text{C}}(x)$ magnitudes
for wz-(Ga,Mn)Te are only slightly larger than experimental values, which
may point out to an influence of the Jahn-Teller distortion neglected in
the present approach.

However, the situation is more involved in the case of (Hg,Cr)Te
and (Zn,Cr)Te. In those systems, due to the importance of the competition
between FM and AFM couplings, the theoretically
expected  critical temperatures are rather sensitive to the employed
tight-binding model of the host band structure. Actually, in the case of
topological (Hg,Cr)Te, although the sign of Curie-Weiss temperatures points
to a prevailing role of the FM interactions, a competition between FM and
AFM contributions may result in the spin-glass freezing at low
temperatures.

The striking properties of semiconductors with magnetic ions mentioned above result from strong $sp$-$d$ exchange interactions
between effective mass carriers and electrons residing in open $d$ shells of magnetic ions \cite{Bonanni:2021_HB}. While the signs and magnitudes
of $s$-$d$ and $p$-$d$ exchange integrals have been extensively studied theoretically \cite{Kacman:2001_SST,Dietl:2008_PRB} and experimentally \cite{Mac:1996_PRB,Suffczynski:2011_PRB} for wide band gap Cr-doped II-VI compounds and wz-(Ga,Mn)N, we present here theory of $sp$-$d$ coupling for Cr-doped topological HgTe. We show that the presence of Cr ions
should improve the quantization accuracy of the quantum spin Hall effect in the paramagnetic phase. We also enquire on whether the quantum anomalous Hall effect, examined so far
theoretically for Mn-doped HgTe QWs \cite{Liu:2008_PRL}, could be observed in the Cr-doping case.


\section{Impurity and band structure parameter values}
\label{sec: bands}
\subsection{Cr and Mn impurity levels}
We place zero electron energy at the valence band top, $E_v = 0$. We are interested in three energy levels introduced by cation-substitutional
$\mathrm{Cr}$ impurities in II-VI compounds (HgTe and ZnTe) and $\mathrm{Mn}$ impurities in GaN:
(i) the donor level $E_d$ corresponding to $\mathrm{Cr}^{2{+}/3{+}}$ and $\mathrm{Mn}^{3{+}/4{+}}$ states with the spin $S = 2$;
(ii) two acceptor states of $\mathrm{Cr}^{2{+}/1{+}}$ and $\mathrm{Mn}^{3{+}/2{+}}$  located at energies $E_d +U$ and $E_d + U +J$ corresponding to five $d$ electrons
but different spin value, $S = 5/2$ and $S = 3/2$, respectively. The energies $(E_d, U, J)$ are
explicitly given in terms of Parmenter's $(E_0, U, J)$ in Sec.\ \ref{sec: theory} (any reference to Parmenter's notation is indicated explicitly).
Spectroscopic free-ion data imply the exchange energy $J = 3$\,eV \cite{Blinowski_1996}. For GaN:Mn we take $J = 2$\,eV \cite{Simserides_2014}.
Experimental studies of ZnTe:Cr point to $E_d = 0.2$\,eV and $U = 1.1$\,eV \cite{Kuroda_2007}.
Assuming the valence band offset between HgTe and ZnTe $W = 0.5$\,eV, we anticipate $E_d = -0.3\pm0.1$\,eV in HgTe:Cr.
Similarly, experimental results for wz-GaN:Mn lead to $E_d = 1.1$\,eV and $U = 0.7$\,eV \cite{Graf_2003,Han_2005,Hwang_2005}.
(In Ref.\,\onlinecite{Simserides_2014}  $E_d = 1.8$\,eV was incorrectly used).
Because of the lack of reliable data for the internal strain parameter $u$,
we assume that Cr and Mn impurities are located at the center of the tetrahedron of the nearest-neighbor anions;
the distortion due to Jahn-Teller effect is neglected.

\begin{table*}
\caption{Summary of the electronic configurations involved in the leading order of perturbation
theory for a $d^4$ ground state and spherical symmetry.}
\begin{tabular}{c|ccccc}
\vspace{0.2cm}
Configuration& Number of electrons& Spin ($S$)& Orbital angular momentum ($L$)& Energy (Parmenter)& Total degeneracy\\
\hline
$d^3$& 3& $3/2$& 1, 3& $3 E_0 - 3 J + 3 U$& 40\\
$d^4$& 4& 2& 2& $4 E_0 - 6 J + 6 U$& 25\\
$d^5$& 5& $5/2$& 0& $5 E_0 - 10 J + 10 U$& 6\\
$d^{5\downarrow}$& 5& $3/2$& 1, 2, 3, 4& $5 E_0 - 5 J + 10 U$& 96\\
\hline
\end{tabular}
\label{tab: parmenter}
\end{table*}

\subsection{Tight-binding  parameters}
One of standard approaches to determine electronic energy bands in solids is the tight-binding approximation (TBA).
Within that model, empirical or \emph{ab initio} data provide on-site and overlap energies
which serve to construct the Bloch Hamiltonian. We reuse the $sp^3$ band structure parameters given previously for HgTe \cite{Sliwa_2021}.
together with
$V_{pd\sigma} = -0.64 \,\mathrm{eV}$, $V_{pd\pi} = -0.45 V_{pd\sigma}$, and $V_{sd\sigma} = 1.08 V_{ps\sigma}$ for
Cr ions \cite{Blinowski_1996}. The tight-binding parameters of Ref.\ \onlinecite{Bertho_1991} are employed for ZnTe.
For zb-GaN we reuse the parameter set of Ref.\ \onlinecite{Simserides_2014}, whereas a set of parameters published by
Yang \emph{et al.} have been employed for wz-GaN, although we include all the second-neighbor overlaps (only half of them
have been originally included). We neglect possible inconsistencies in the parameter set that may arise as a
result.

In order to verify whether the present results are sensitive to details of the band-structure, additional models have been considered for (Hg,Cr)Te and (Zn,Cr)Te.
These models include explicitly the $d$ orbitals of both the cation and anion, thus allowing to include also $d$-$d$ hybridization.
In the band-structure models which include the $d$ orbitals of the anion we assume for the hybridization matrix elements,
following the universal ratios of tight-binding overlaps \cite{Shi:2004_PRB}:
$V_{dd\sigma} = -3.47 \,\mathrm{eV}$, $V_{dd\pi} =  1.87 \,\mathrm{eV}$, and $V_{dd\delta} = -0.51 \,\mathrm{eV}$.

\subsection{Splitting and shift of bands resulting from $p$-$d$ hybridization}

We begin our discussion of exchange interactions in our systems by introducing the second-order effective Hamiltonian for a band carrier coupled to a $d^4$ magnetic shell, including the effect of the $p$-$d$ hybridization. The matrix elements between the initial and final configurations
of a set of band states (denoted $\{ k \}$) and the $d^4$ shell, in the occupation-number representation as $\{ n_k \}$, are given by
\begin{widetext}
\begin{eqnarray}
  \lefteqn{ \left< d^{4}_{fin}; \{ n_{k;fin} \} \middle| H^{(2)}_{eff} \middle| d^{4}_{ini}; \{ n_{k;ini} \} \right>} \nonumber \\
    & = & \sum_{d^{5}_{int}; k, k'} A_{+1}(E_k, E_{k'}) \left< d^{4}_{fin} \middle| a_d(k') \middle| d^{5}_{int} \right> \left< d^{5}_{int} \middle| a_d^{\dagger}(k) \middle| d^{4}_{ini} \right>
      \left< \{ n_{k;fin} \} \middle| a_{k'}^{\dagger} a_k \middle| \{ n_{k;ini} \} \right> + {} \nonumber \\
    & & {} + \sum_{d^{5\downarrow}_{int}; k, k'} A_{+1\downarrow}(E_k, E_{k'}) \left< d^{4}_{fin} \middle| a_d(k') \middle| d^{5\downarrow}_{int} \right> \left< d^{5\downarrow}_{int} \middle| a_d^{\dagger}(k) \middle| d^{4}_{ini} \right>
      \left< \{ n_{k;fin} \} \middle| a_{k'}^{\dagger} a_k \middle| \{ n_{k;ini} \} \right> + {} \nonumber \\
    & & {} + \sum_{d^{3}_{int}; k, k'} A_{-1}(E_k, E_{k'}) \left< d^{4}_{fin} \middle| a_d^{\dagger}(k) \middle| d^{3}_{int} \right> \left< d^{3}_{int} \middle| a_d(k') \middle| d^{4}_{ini} \right>
      \left< \{ n_{k;fin} \} \middle|  a_k a_{k'}^{\dagger} \middle| \{ n_{k;ini} \} \right>,
\end{eqnarray}
\end{widetext}
where $A_{+1}$, $A_{+1\downarrow}$, and $A_{-1}$ are temperature-dependent numerical coefficients ($A_{+1\downarrow}$ is obtained from $A_{+1}$ by the substitution $U \to U + J$):
\begin{widetext}
\begin{eqnarray}
  \lefteqn{A_{+1}(E_k, E_{k'})} \nonumber \\
  & = & \frac{1}{(E_d + U - E_k) (E_d + U - E_{k'})
    \left[ \exp\left(\frac{E_k - E_d - U}{k_B T}\right) - \exp\left(\frac{E_{k'} - E_d - U}{k_B T}\right) \right]} \times {} \nonumber \\
  & & {} \times \biggl[ (E_d + U - E_{k'}) \exp\left(\frac{E_{k'} - E_d - U}{k_B T}\right) -
    (E_d + U - E_k) \exp\left(\frac{E_k - E_d - U}{k_B T}\right) + {} \biggr. \nonumber \\
  & & \qquad \biggl. {} + (E_{k'} - E_k) \exp\left(\frac{E_k - E_d - U}{k_B T}\right) \exp\left(\frac{E_{k'} - E_d - U}{k_B T}\right) \biggr], \\
  \lefteqn{A_{-1}(E_k, E_{k'})} \nonumber \\
  & = & \frac{1}{(E_d - E_k) (E_d - E_{k'})
    \left[ \exp\left(\frac{E_k - E_d}{k_B T}\right) - \exp\left(\frac{E_{k'} - E_d}{k_B T}\right) \right]} \times {} \nonumber \\
  & & {} \times \left[ (E_d - E_{k'}) \exp\left(\frac{E_{k} - E_d}{k_B T}\right) -
    (E_d - E_k) \exp\left(\frac{E_{k'} - E_d}{k_B T}\right) + (E_{k'} - E_k) \right],
\end{eqnarray}
\end{widetext}
which simplifies in the limit $T \to 0$, $\mu \to E_F$ to
\begin{eqnarray}
  A_{+1}(E_k, E_{k'}) & = & \frac{1}{\min(E_k, E_{k'}) - E_d - U}, \\
  A_{-1}(E_k, E_{k'}) & = & \frac{1}{E_d - \max(E_k, E_{k'})}
\end{eqnarray}
under the assumptions $E_d < E_F, E_k, E_{k'} < E_d + U$.

\begin{table}
\begin{tabular}{|c|cc|}
\hline
subspace& $C$& $B$\\
\hline
$d^5$& $\frac{1}{5}$& $\frac{2}{15}$\\
$d^{5\downarrow}$& $\frac{7}{15}$& $-\frac{7}{15}$\\
$d^{3}$& $\frac{1}{3}$& $\frac{1}{3}$\\
\hline
\end{tabular}
\caption{The numerical coefficients (matrix elements of the creation/annihilation operators) for $p$-$d$ interaction of $d$ orbitals of $t_2$ symmetry,
averaged over the three Jahn-Teller configurations in order to restore the cubic symmetry.
Notation: $C$ is a spin-independent one-electron energy shift; $B$ multiplies Heisenberg term $\mathbf{S} \cdot \mathbf{s}$.}
\label{tab: CB}
\end{table}

The summation over an orthonormal set of intermediate states in each of the subspaces $(d^{5}_{int}, d^{5\downarrow}_{int}, d^{4}_{int})$ yields a one-particle hamiltonian
for the band carrier. It is sensitive to the Jahn-Teller distortion of the magnetic impurity. At the $\Gamma$ point, the carriers hybridize to $t_2$ orbitals only,
and with the same strength $V_{pd}$ to each of them. Now, averaging over the three Jahn-Teller configurations restores the cubic symmetry. In such circumstances the interaction
assumes the form $C + B \, \mathbf{S} \cdot \mathbf{s}$, which is a constant
one-particle energy (contributing to the band offset) plus the Heisenberg $p$-$d$ exchange term. The numerical coefficients are summarized in Table\ \ref{tab: CB}.

In the case of a conduction-band carrier, the symmetry allows also for an $s$-$d$ term, originating from the potential exchange. Since the Bloch function for the conduction band
is a mixture of both cation and anion orbitals, the corresponding exchange energy is reduced with respect to the free-ion.

\section{Theory of spin-spin interactions}
\label{sec: theory}

We follow the established approach \cite{Larson_1988,Sliwa:2018_PRB,Sliwa_2021} based on the leading-order perturbation theory (4th-order in $p$-$d$ hybridization).
We use here the convention for the Hamiltonian representing spin pair interactions, $H_{ij} = -2 J_{ij} \mathbf{S}_i \cdot \mathbf{S}_j$, where $J_{ij}$ results from angular averaging to be discussed later, and --- in general --- contains three contributions, $hh$ corresponding to superexchange, $he$ describing the interband Bloembergen-Rowland coupling mechanism, and $ee$ giving the two-electron contribution. We adapt here our previous theory \cite{Sliwa_2021} for the $d^5$  magnetic shells for the $d^4$ case.

The components of the full the Hamiltonian,
\begin{equation}
  \hat{H} = (\hat{H}_k + \hat{H}_d) + \hat{V}_{\text{hyb}},
\end{equation}
correspond to a set of bands for non-interacting electrons ($\hat{H}_k = \sum_k E_k a_k^{\dagger} a_k$), the $d$-shells with the on-site Coulomb
repulsion ($\hat{H}_d = \sum_i \hat{H}_{d,i}$, e.g., Parmenter's Hamiltonians \cite{Parmenter_1973}), and hybridization between the band states and the $d$-shells ($\hat{V}_{\text{hyb}}$).
Application of the perturbation theory involves summing over processes in which the electron jumps from one $d$-shell (state $d_1$)
to a band state ($k$), then to the $d$-shell of another magnetic ion (state $d_2$), then to another band state ($k'$), and then back
to state $d_1$ --- forming a closed path. Terms consisting of two independent loops (performed by two electrons) cancel in the total result.
One has also to take into account, by means of the quasi-degenerate perturbation theory \cite{Winkler_2003}, that we are interested
in the Hamiltonian just for the ground state of the two magnetic ions, $d^4 \otimes d^4$. Each term is a product of a phase due to the anticommutation rule
for the fermionic creation and annihilation operators, an energetic denominator, and four matrix elements of the hybridization operator
(the latter independent of the occupations of the band states or the order in which the transitions happen). We write the matrix elements of the effective
Hamiltonian (between the initial and final states) as a sum over an orthonormal set of intermediate states $(d_{int,1}, d_{int,2})$:
\begin{eqnarray}
  \lefteqn{\left< d_{fin,1}, d_{fin,2} \middle| H_{eff}^{(4)} \middle| d_{ini,1}, d_{ini,2} \right>} \nonumber \\
  & = & \sum_{k,k'} \sum_{d_{int,1}, d_{int,2}} A_{x_1,x_2}(E_k, E_{k'}, \mu) \nonumber \\  & & {} \times
          \left< d_{fin,1} \middle| a_{d,1,x_1}(k, k') \middle| d_{int,1} \right> \nonumber \\ & & {} \times
          \left< d_{int,1} \middle| a_{d,1,x_1}(k', k)^{\dagger} \middle| d_{ini,1} \right> \nonumber \\ & & {} \times
          \left< d_{fin,2} \middle| a_{d,2,x_2}(k', k) \middle| d_{int,2} \right> \nonumber \\ & & {} \times
          \left< d_{int,2} \middle| a_{d,2,x_2}(k, k')^{\dagger} \middle| d_{ini,2} \right>
\end{eqnarray}
Here, $x_i \in \{ {+1}, {-1} \}$ is ${+1}$ for acceptor excitations (intermediate state $d_{int,i} \in d^5 \oplus d^{5\downarrow}$,
see below) or ${-1}$ for donor excitations ($d_{int,i} \in d^3$), and $a_{d,i,x_i}(k, k')$ is an annihilation [$a_{d,i,{+1}}(k, k') = a_{d,i}(k)$]
or creation [$a_{d,i,{-1}}(k, k') = a_{d,i}(k')^{\dagger}$] operator for the $d$-shell $i$, respectively. By definition, the hybridization
operator can be written as the following sum over band states $k$:
\begin{eqnarray}
  \hat{V}_{\text{hyb}} & = & \sum_{i,k} [a_k^{\dagger} a_{d,i}(k) + a_{d,i}(k)^{\dagger} a_k], \\
  a_{d,i}(k) & = & \sum_{n,s,m} \left< k \middle| n, s \right> \left< n \middle| V \middle| d, i, m \right>_{\kappa(k)} \nonumber \\
    & & \qquad {} \times e^{-i \kappa(k) \cdot \mathbf{R}_i} a_{d,i,m,s},
\end{eqnarray}
where $n$ is a band orbital index, $s \in \{ \frac{1}{2}, -\frac{1}{2} \}$ is a spin index, $m$ is a $d$-shell orbital index,
and $\left< n \middle| V \middle| d, i, m \right>_{\kappa(k)}$
is the tight-binding overlap [at the wave vector $\kappa(k)$ corresponding to band state $k$] between band orbital $n$ and $d$-shell
orbital $m$, the latter located at position $\mathbf{r}_i$ inside the unit cell,
and at absolute position $\mathbf{R}_i = n_1 \mathbf{a}_1 + n_2 \mathbf{a}_2 + n_3 \mathbf{a}_3 + \mathbf{r}_i$ [$(n_{\alpha})_{\alpha=1,2, 3}$ number lattice cells].
The phase factor $e^{-i \kappa(k) \cdot \mathbf{R}_i}$ compensates the fact that we are working with Bloch functions
(rather than with wave functions).

The relevant subspace of the Hilbert space for the $d$-shells is a direct sum of the subspaces corresponding to the electronic
configurations being involved in the leading order: $d^4$ (ground state), $d^3$, $d^5$ (high-spin) and $d^{5\downarrow}$ (spin-flipped).
Since even in a tetrahedral symmetry this amounts to a 32-parameter Hamiltonian, we are virtually forced to reuse the
three-parameter Hamiltonian of Parmenter \cite{Parmenter_1973}. The relevant energies are given in terms of Parmenter's $E_0$, $U$, $J$ in
Table\ \ref{tab: parmenter}. We will denote as $E_d - E_v$, $E_d + U - E_v$, and $E_d + U + J - E_v$ the energies of the
transitions $d^3 \to d^4 + h$, $d^4 \to d^5 + h$, $d^4 \to d^{5\downarrow} + h$ (respectively; $h$ stands for a hole
at the valence-band top, and $E_v$ for the valence-band-top energy). The parameters $E_d$, $U$, and $J$ adopted in the present Article are related to the original Parmenter's parameters $E_0$, $U$, and $J$ according to: $E_d \rightarrow E_0 + 3 (U - J)$,
$U \rightarrow (U - J)$, $J \rightarrow 5 J$, where the r.h.s. is expressed in terms of Parmenter's $(E_0, U, J)$.

Explicit expressions for $A$'s (including the phase factors) can be written for $d^3$ and $d^5$ excitations in an insulator
($E_k < E_F$ or $E_k > E_F$) in the low-temperature limit, $T \to 0^{+}$, $\mu \to E_F$, $f(E_k) \to \Theta(E_F-E_k)$,
under the assumption that the ground state of the unperturbed Hamiltonian ($H_k + H_d$) corresponds to $d^4$ electronic configuration, i.e.,
$E_d <E_F$, $E_d + U > E_F$, and $E_d + U + J > E_F$ (the expression for $d^{5\downarrow}$ excitations are obtained
by replacing $U$ with $U + J$ for the corresponding shell $d_i$; zero denominators, e.g. $E_{d,1} - E_{d,2}$, must be eliminated by taking appropriate limits:
$E_{d,1}, E_{d,2} \to E_d$, and/or formal algebraic transformations).

\begin{table}
\caption{A summary of the reduced matrix elements for the $d$-shells annihilation operators; cf.\ Eqs.~(\ref{eq: ared}--\ref{eq: adagred}).}
\begin{tabular}{ccccc|c}
\vspace{0.2cm}
$n$& $S$& $L$& $S'$& $L'$& $\left| \left<\left< a \right>\right>_{L, S; L', S'} \right|^2$\\
\hline
3& $3/2$& 1& 2& 2& $6/5$\\
3& $3/2$& 3& 2& 2& $14/5$\\
4& 2& 2& $5/2$& 0& 5\\
4& 2& 2& $3/2$& 1& $5/4$\\
4& 2& 2& $3/2$& 2& $5/4$\\
4& 2& 2& $3/2$& 3& $5/4$\\
4& 2& 2& $3/2$& 4& $5/4$\\
\hline
\end{tabular}
\label{tab: redmatel}
\end{table}

\begin{widetext}

\begin{eqnarray}
  A_{+1,+1} & = & \frac{1}{2} \frac{f(E_k) - f(E_{k'})}{E_k - E_{k'}} [w_{1u}(E_k) w_{2u}(E_k) + w_{1u}(E_{k'}) w_{2u}(E_{k'})] + \nonumber \\
    & & {} - \frac{1}{2} [f(E_k) + f(E_{k'})] \frac{w_{1u}(E_k) w_{1u}(E_{k'}) - w_{2u}(E_k) w_{2u}(E_{k'})}{E_{d1} + U_1 - E_{d2} - U_2}
  \label{eq: Afirst}
\end{eqnarray}

\begin{eqnarray}
  A_{+1,-1} & = & \frac{1}{2} \frac{f(E_k) - f(E_{k'})}{E_k - E_{k'}} [w_{1u}(E_k) w_{2}(E_k) + w_{1u}(E_{k'}) w_{2}(E_{k'})] + \nonumber \\
    & & {} - \frac{1}{2} [f(E_k) + f(E_{k'})] \frac{w_{1u}(E_k) w_{1u}(E_{k'}) - w_{2}(E_k) w_{2}(E_{k'})}{E_{d1} + U_1 - E_{d2}} + {} \nonumber \\
    & & {} - \frac{w_{2}(E_k) w_{2}(E_{k'})}{E_{d1} + U_1 - E_{d2}}
\end{eqnarray}

\begin{eqnarray}
  A_{-1,+1} & = & \frac{1}{2} \frac{f(E_k) - f(E_{k'})}{E_k - E_{k'}} [w_{1}(E_k) w_{2u}(E_k) + w_{1}(E_{k'}) w_{2u}(E_{k'})] + \nonumber \\
    & & {} - \frac{1}{2} [f(E_k) + f(E_{k'})] \frac{w_{1}(E_k) w_{1}(E_{k'}) - w_{2u}(E_k) w_{2u}(E_{k'})}{E_{d1} - E_{d2} - U_2} + {} \nonumber \\
    & & {} + \frac{w_{1}(E_k) w_{1}(E_{k'})}{E_{d1} - E_{d2} - U_2}
\end{eqnarray}

\begin{eqnarray}
  A_{-1,-1} & = & \frac{1}{2} \frac{f(E_k) - f(E_{k'})}{E_k - E_{k'}} [w_{1}(E_k) w_{2}(E_k) + w_{1}(E_{k'}) w_{2}(E_{k'})] + \nonumber \\
    & & {} - \frac{1}{2} [f(E_k) + f(E_{k'})] \frac{w_{1}(E_k) w_{1}(E_{k'}) - w_{2}(E_k) w_{2}(E_{k'})}{E_{d1} - E_{d2}} + {} \nonumber \\
    & & {} + \frac{w_{1}(E_k) w_{1}(E_{k'}) - w_{2}(E_k) w_{2}(E_{k'})}{E_{d1} - E_{d2}}
  \label{eq: Alast}
\end{eqnarray}

\begin{eqnarray}
  w_{i}(E) & = & \frac{1}{E_{di} - E} \\
  w_{iu}(E) & = & \frac{1}{E_{di} + U_i - E}
\end{eqnarray}

\end{widetext}

Matrix elements of the annihilation operators from $d^{n+1}$ to $d^n$ can be written in terms of Clebsch-Gordan coefficients as:
\begin{widetext}
\begin{eqnarray}
  \lefteqn{\left<d^n, \alpha, L, S, L_{z,f}, S_{z,f} \middle| a_{L_{z,a}, S_{z,a}} \middle| d^{n+1}, \beta, L', S', L_{z,i}, S_{z,i} \right> = {}} \nonumber \\
    & & \left<\left< a \right>\right>_{\alpha, L, S; \beta, L', S'} 
        \left<L', L_{z,i} \middle| L, L_{z,f}; 2, L_{z,a}\right> \left<S', S_{z,i} \middle| S, S_{z,f}; \frac{1}{2}, S_{z,a}\right>,
\label{eq: ared}
\end{eqnarray}
\end{widetext}
where $(L_{z,a}, S_{z,a})$ are the quantum numbers of the annihilated electron.
Analogously, we have for matrix elements of the creation operators from $d^n$ to $d^{n+1}$:
\begin{widetext}
\begin{eqnarray}
  \lefteqn{\left<d^{n+1}, \alpha, L, S, L_{z,f}, S_{z,f} \middle| a^{\dagger}_{L_{z,c}, S_{z,c}} \middle| d^n, \beta, L', S', L_{z,i}, S_{z,i} \right> = {}} \nonumber \\
    & & \left<\left< a^{\dagger} \right>\right>_{\alpha, L, S; \beta, L', S'} 
        \left<L, L_{z,f} \middle| L', L_{z,i}; 2, L_{z,c}\right> \left<S, S_{z,f} \middle| S', S_{z,i}; \frac{1}{2}, S_{z,c}\right>,
\label{eq: adagred}
\end{eqnarray}
\end{widetext}
with $ \left<\left< a^{\dagger} \right>\right>_{\alpha, L, S; \beta, L', S'} = \left<\left< a \right>\right>_{\beta, L', S'; \alpha, L, S}^{*}$.
The squared absolute values of the reduced matrix elements are summarized in Table\ \ref{tab: redmatel}.

\begin{table}
\caption{Values of $j_{m,m_i}(d_{sub,i})$ in equations (\ref{eq: dint1}) and (\ref{eq: dint2}) for an excitation with an orbital quantum number $m$ from a $d^4$
ground state with an orbital configuration $m_i$ (rows correspond to sectors of the Hilbert space) under spherical symmetry.}
\begin{tabular}{c|cc}
\vspace{0.2cm}
$d_{sub,i}$& $m = m_i$& $m \ne m_i$\\
\hline
$d^3$& 0& $1/4$\\
$d^5$& $1/5$& 0\\
$d^{5\downarrow}$& $-1/5$& $-1/4$\\
\hline
\end{tabular}
\label{tab: j}
\end{table}

We further restrict our attention to the case when the orbital configurations $m_i$ of the $d$-shells remain unchanged.
In such a situation an excitation (electron or hole) which enters the $d$-shell with an orbital quantum number $m$,
leaves it with $m$ unchanged. Then, the spin-dependent parts of the subexpressions
\begin{eqnarray}
  \lefteqn{\sum_{d_{int,i} \in d_{sub,i}}
    \left< d_{fin,i} \middle| a_{d,i,m,s} \middle| d_{int,i} \right>
    \left< d_{int,i} \middle| a_{d,i,m,s'}^{\dagger} \middle| d_{ini,i} \right>} \nonumber \\
  & \cong & j_{m,m_i}(d_{sub,i})  \sum_{\alpha} \left< d_{fin,i} \middle| S_{i,\alpha} \middle| d_{ini,i} \right>
          \left< s \middle| \sigma_{\alpha} \middle| s' \right>, \qquad \label{eq: dint1} \\
  \lefteqn{\sum_{d_{int,i} \in d_{sub,i}}
    \left< d_{fin,i} \middle| a_{d,i,m,s}^{\dagger} \middle| d_{int,i} \right>
    \left< d_{int,i} \middle| a_{d,i,m,s'} \middle| d_{ini,i} \right>} \nonumber \\
  & \cong & j_{m,m_i}(d_{sub,i}) \sum_{\alpha} \left< d_{fin,i} \middle| S_{i,\alpha} \middle| d_{ini,i} \right>
          \left< s' \middle| \sigma_{\alpha} \middle| s \right>, \qquad
  \label{eq: dint2}
\end{eqnarray}
and $j_{m,m_i}(d_{sub,i})$ are constant numbers which depend on the Hilbert subspace, $d_{sub,i}$, and the relation between
the orbital quantum number $m$ and the orbital state of the $d$-shell, $m_i$, as summarized in Table\ \ref{tab: j}.

Equipped with the results (\ref{eq: Afirst}--\ref{eq: Alast}) and Table\ \ref{tab: j}, we rederive Eqs.~1--9 of Ref.\ \onlinecite{Blinowski_1996}.
In Ref.\ \onlinecite{Blinowski_1996}, the contribution with $m \ne m_i$ at both ions ($i = 1, 2$) is denoted $F$;
that with $m = m_i$ at both ions is denoted $G$; the remaining contribution is denoted $H$. One therefore has \cite[Eq.\ 2]{Blinowski_1996}
\begin{equation}
  J^{\gamma\delta}_{\alpha\beta}(\mathbf{R}_{12}) = F^{\gamma\delta}_{\alpha\beta}(\mathbf{R}_{12}) + H^{\gamma\delta}_{\alpha\beta}(\mathbf{R}_{12}) + G^{\gamma\delta}_{\alpha\beta}(\mathbf{R}_{12}).
\end{equation}
We let $f(E_k), f(E_{k'}) \to 1$ in our (\ref{eq: Afirst}--\ref{eq: Alast}) and find differences w.r.t. Ref.\ \onlinecite{Blinowski_1996}:
an incorrect normalization (the prefactor 2 in Eq.\ 1 of Ref.\ \onlinecite{Blinowski_1996} is extraneous),
wrong signs of $\mathcal{F}$ and $\mathcal{H}$, a missing 2 in the numerator of $1 / (e_2 + e_1)$ in $\mathcal{F}$, and a lacking symmetrization of $\mathcal{H}$
with respect to $\nu \leftrightarrow \nu'$ (the last issue does not affect numerical values). Still, the numbers (e.g.\ for ZnTe:Cr) in
Table I of Ref.\ \onlinecite{Blinowski_1996} are $-1 / 2$ the values that can be obtained from the incorrect formulas. However, as
shown in the next section, those corrections to the $hh$ contribution, together with $he$ and $ee$ terms taken into account here but neglect previously, while important quantitatively,
do not alter the main conclusion
of the previous works \cite{Blinowski_1996,Simserides_2014}:
spin-spin interactions are predominately FM for magnetic ions with high spin $d^4$ configuration in zinc-blende DMSs.

Even though the energy shifts due to Jahn-Teller effect are not taken into account in the present theory, the exchange integrals are sensitive to
orbital configurations. We report as $J$ its average over the three $T_2$ states of each of the two $d$-shells in question (with an appropriate
transformation of the quantization axis in the wurtzite case, as described in Appendix\ \ref{app: rot_d}).

For practical and efficient numerical evaluation of the double Brillouin zone integral, it is desirable to perform one more algebraic transformation
at this stage. We take $\mathcal{F}$ with $f(E_k), f(E_{k'}) \to 1$ as an example, and write
\begin{widetext}
\begin{eqnarray}
  \lefteqn{\frac{1}{E_d + U + J - E_{k}} \frac{1}{E_d + U + J - E_{k'}}
    \left( \frac{2}{U + J} + \frac{1}{E_d + U + J - E_{k}} + \frac{1}{E_d + U + J - E_{k'}} \right)} \nonumber \\
  & = & \left[ \begin{array}{cc} \left( 2 + \frac{U + J}{E_d + U + J - E_{k'}} \right) \frac{U + J}{E_d + U + J - E_{k'}} \\ \left( \frac{U + J}{E_d + U + J - E_{k'}} \right)^2 \end{array} \right]^{T}
    \left[ \begin{array}{cc} \frac{1}{2 (U + J)^3}& 0\\ 0& -\frac{1}{2 (U + J)^3} \end{array} \right] \left[ \begin{array}{c} \left( 2 + \frac{U + J}{E_d + U + J - E_k} \right) \frac{U + J}{E_d + U + J - E_k} \\ \left( \frac{U + J}{E_d + U + J - E_k} \right)^2 \end{array} \right],
\end{eqnarray}
\end{widetext}
which resembles $LDU$ decomposition of some matrix. In this form, the dependence on $E_k$ has been separated from that on $E_{k'}$.
Therefore, one can group this energy dependence in the product with the hybridization matrix elements; then the multiplication by the
remaining plane wave phase factor and summation over momentum is just a Fourier transform (and it can be performed by means of the FFT algorithm).
In this way we obtain the whole real-space dependence of the spin-spin Hamiltonian in one turn, where the algorithm loops over the diagonal
of the matrix $D$ above. The most difficult part here is the $he$ term, which features an $E_k - E_{k'}$ denominator; we handle it
by performing an incremental on-the-fly $LDU$ decomposition of the matrix $(1 / (E_k - E_{k'}))_{k,k'}$, which yields $D$ of the
size $\min(n_v, n_c)$ (the number of valence or conduction band states, whichever is lower).

Another challenge in the double integration over the Brillouin zone, in the case of the interband $he$ term in zero-gap HgTe,
is the singularity that appears at $k, k' \rightarrow 0$. We follow the procedure elaborated previously \cite{Sliwa_2021}, which
involves a shift of the $k$ grid by different $\vartheta$ values. The number of $k$ and $\vartheta$ values employed here insures
convergence of the results.

As an ending remark, the fourth order perturbation to the Hamiltonian for a pair of transition metal impurities, after averaging over orbitals and
neglecting the spin-orbit interaction, assumes the Heisenberg form. This is so because the electron's spin $s = 1/2$ in the underlying theory;
however, other forms beyond the approach implemented here --- like biquadratic of four-spin couplings --- are possible and can be found by examining
ab initio the dependence of the system energy on the $q$-vector of a frozen magnon, as pioneered by Liechtensten et al.\ \cite{Liechtenstein:1984_JPF}.

\section{Numerical values of the exchange integrals}
\begin{figure*}[tb]
\includegraphics[width=0.92\textwidth]{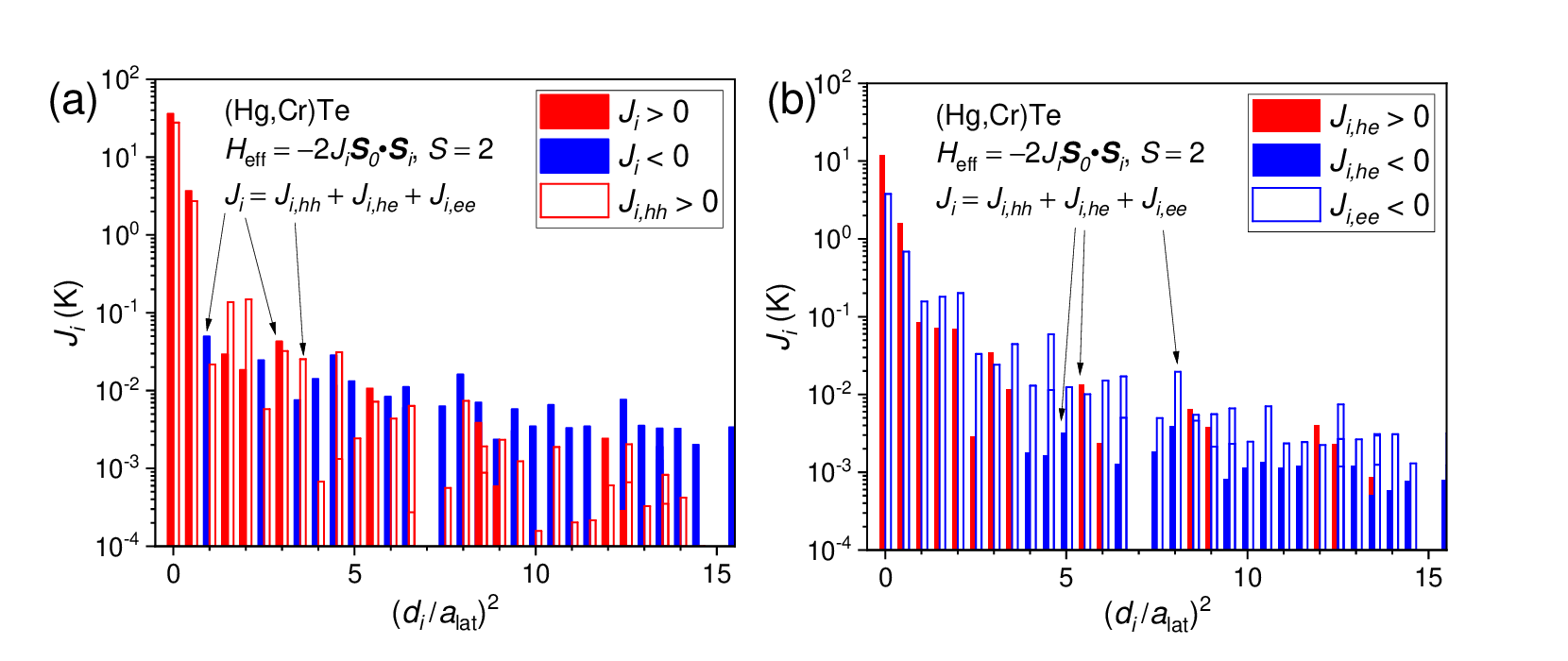}
\caption{Exchange integrals $J_{i}$ in a logarithmic scale for spin pairs at distances $d_i$ corresponding
to the first thirty cation coordination spheres $1 \leq i \leq 30$,
$d_i = (i/2)^{1/2}a_{\text{lat}}$ in zinc-blende topological (Hg,Cr)Te.
(a) The total exchange energies and the superexchange ($hh$) contribution; (b) the remaining contributions:
the electron-hole (Bloembergen-Rowland) and two-electron terms ($he$ an $ee$, respectively).
Red and blue colors correspond to the FM and AFM sign of the interaction, respectively.
As demonstrated elsewhere in the context of (Hg,Mn)Te \cite{Sliwa_2021},
the self-interaction terms at $d_i = 0$ do not contribute
to spin-spin exchange energies but account for
a sizable overestimation of the interband
contribution $he$ to the Curie temperature $T_{\text{C}}$,
if evaluated in terms of the interband Van Vleck susceptibility \cite{Yu:2010_S}.}
\label{fig: HgCrTe}
\end{figure*}
%
\begin{figure*}
\includegraphics[width=0.92\textwidth]{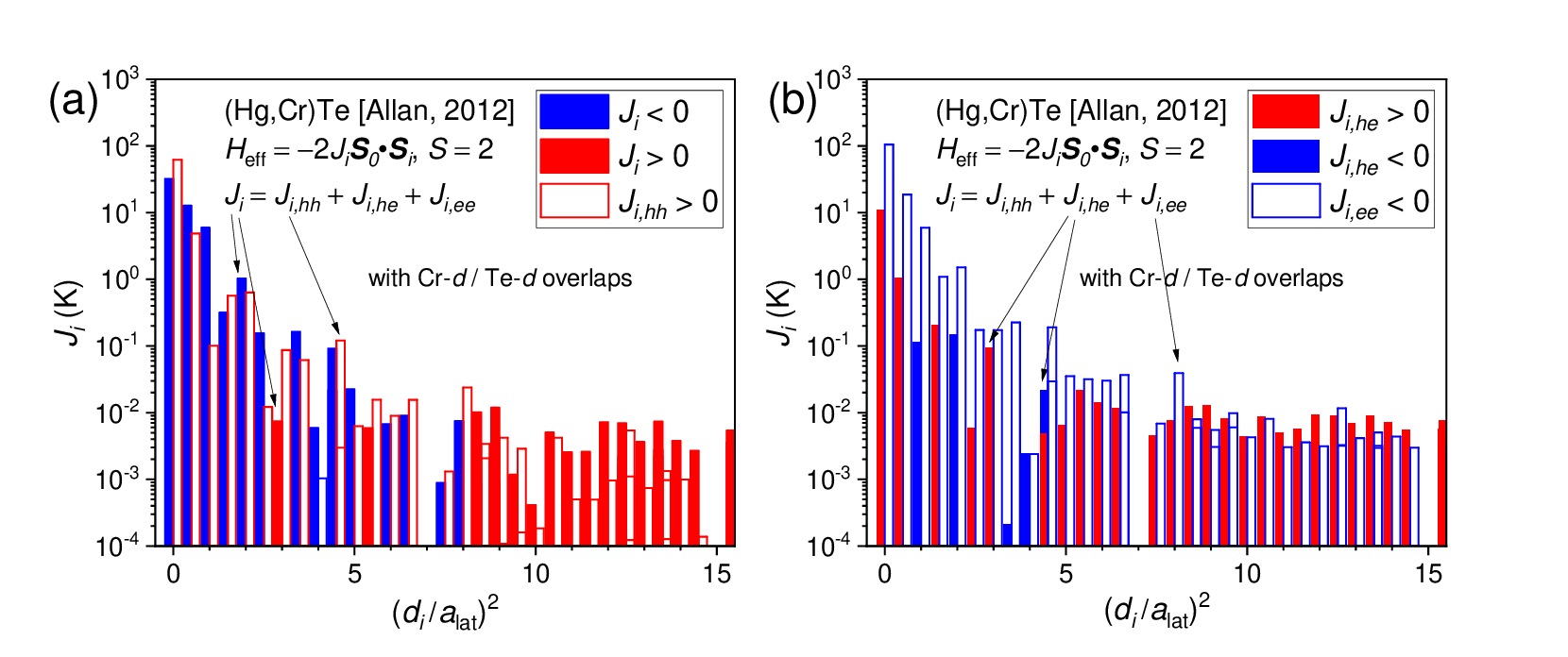}
\caption{Exchange integrals $J_{i}$ in zinc-blende topological (Hg,Cr)Te according to the band-structure
model of Ref.\ \onlinecite{Allan:2012_PRB} (2048 $k$-points, 64 $\theta$-points).}
\label{fig: HgCrTe_Allan}
\end{figure*}
\begin{figure*}
\centerline{\includegraphics[width=0.92\textwidth]{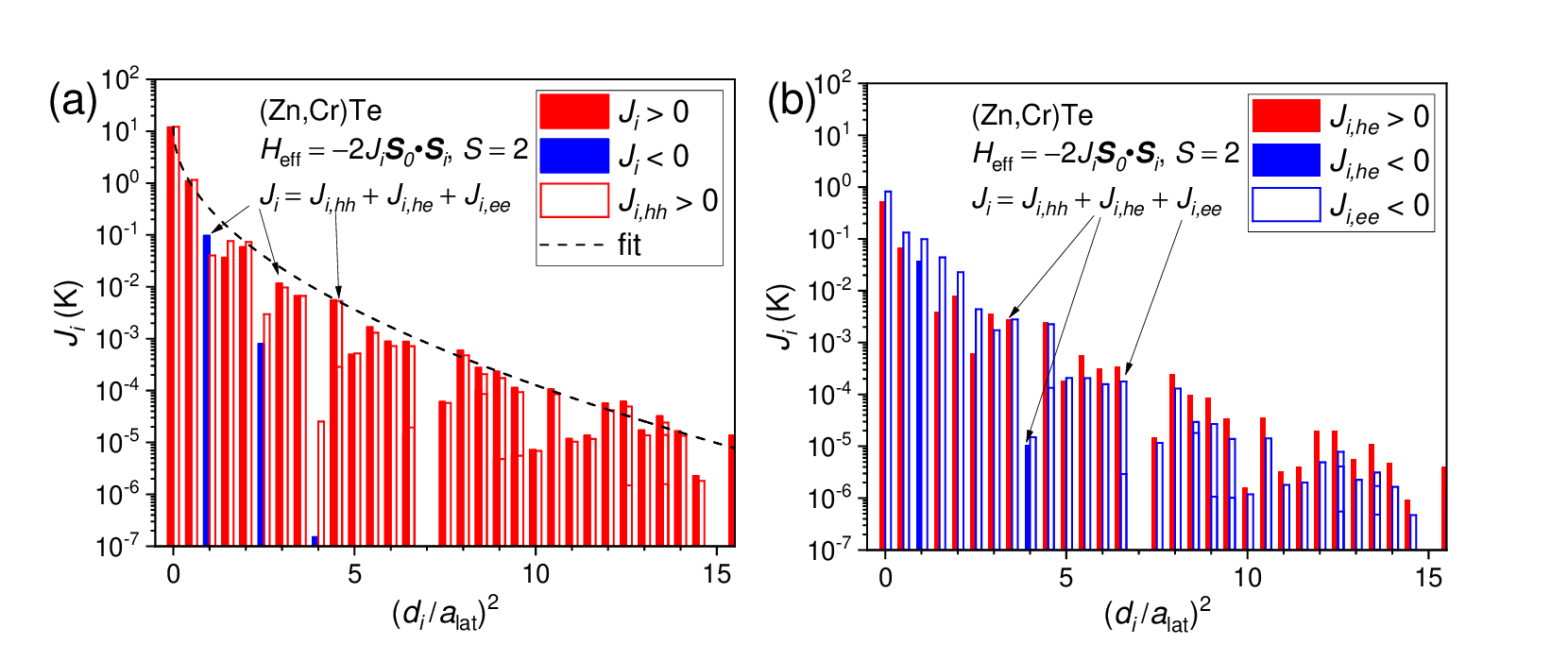}}
\caption{Same as in Fig.\ref{fig: HgCrTe} but for topologically trivial zinc-blende (Zn,Cr)Te.}
\label{fig: ZnCrTe}
\end{figure*}
\begin{figure*}
\includegraphics[width=0.92\textwidth]{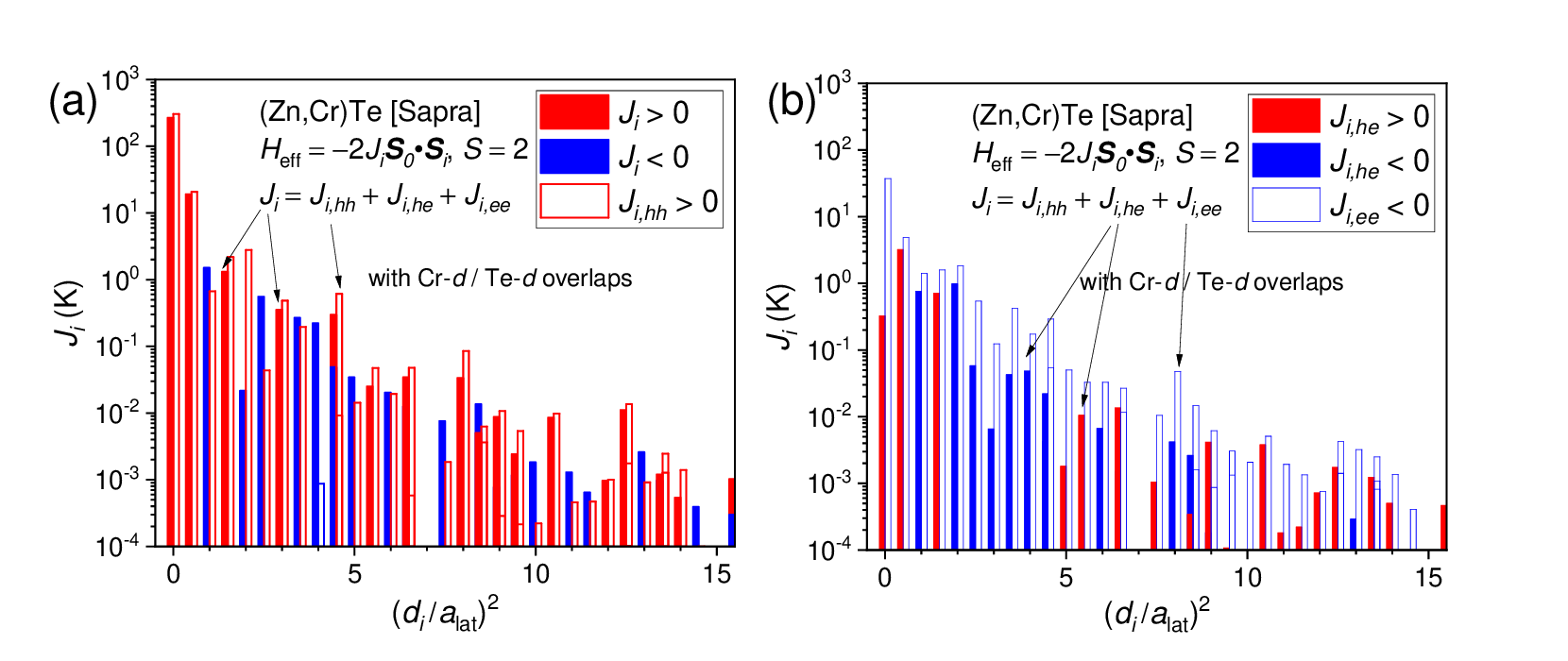}
\caption{Exchange integrals $J_{i}$ in zinc-blende topologically trivial (Zn,Cr)Te according to the band-structure
model of Ref.\ \onlinecite{Sapra:2002_PRB} (without second neighbor interactions; 16384 $k$-points).}
\label{fig: ZnCrTe_Sapra}
\end{figure*}
\begin{figure*}
\centerline{\includegraphics[width=0.92\textwidth]{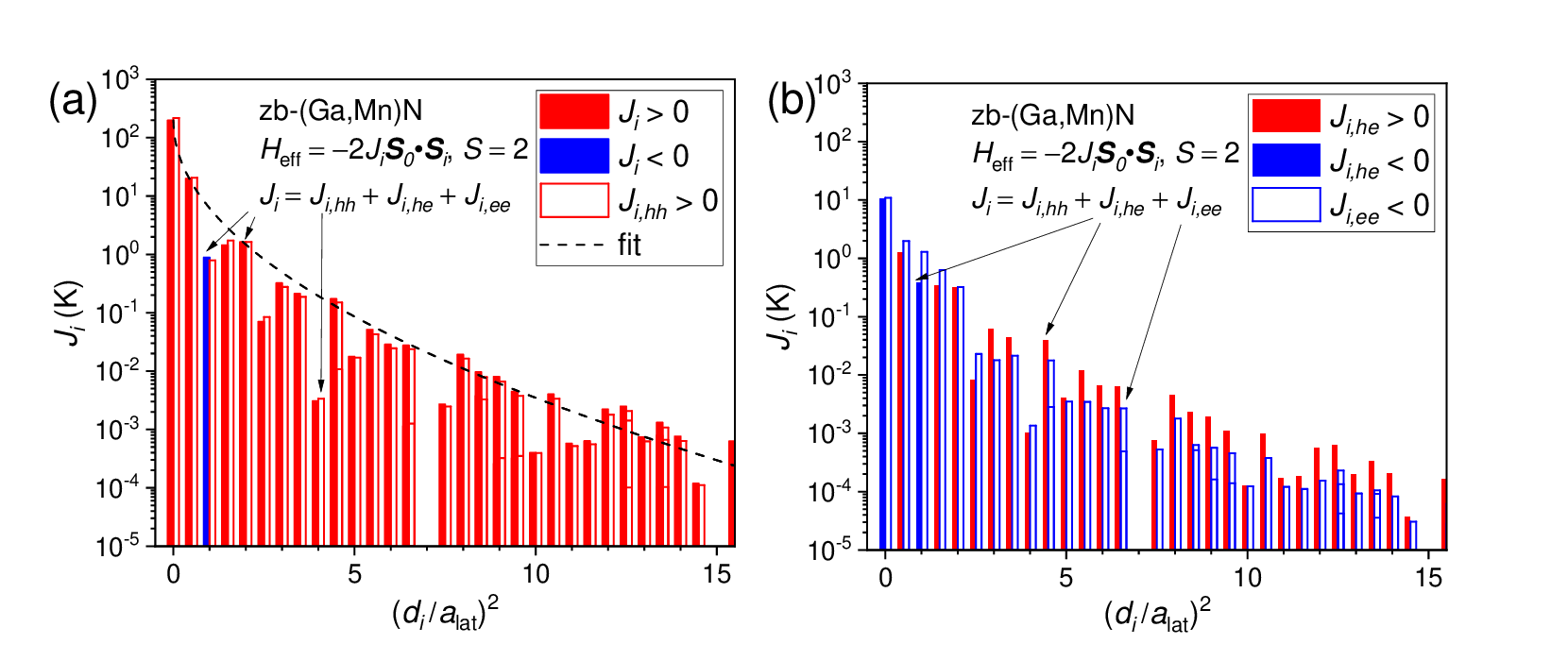}}
\caption{Same as in Fig.\ref{fig: HgCrTe} but for for topologically trivial zinc-blende (Ga,Mn)N.}
\label{fig: zb-GaMnN}
\end{figure*}

We report our results in the convention according to which the Hamiltonian $H_{ij}$ for a pair $(i,j)$ is
\begin{equation}
  H_{ij} = -2 J_{ij} \mathbf{S}_i \cdot \mathbf{S}_j,
\end{equation}
where $\mathbf{S}_i$ and $\mathbf{S}_j$ are the quantum operators for $S = 2$ spins. We present, in
Figs.\ \ref{fig: HgCrTe}--\ \ref{fig: zb-GaMnN} for (Hg,Cr)Te, (Zn,Cr)Te, and zb-(Ga,Mn)N, respectively
the determined values
of $J_{ij} \equiv J_i$ vs. the spin pair distance $R_{ij} \equiv d_i$, where $d_i =(i/\sqrt{2})a_{\text{lat}}$ denote the positions of the sequential cation
coordination spheres and the lattice parameters $a_{\text{lat}} =0.646$ and $0.610$\,nm for HgTe and ZnTe, respectively.  Since the magnitudes of $J_{ij}$ decay exponentially with $d_i$, only the first few $J_{ij}$ values are significant and, therefore, shown in Tables \ref{tab: hgte}--\ref{tab: wz-gan} for the same systems and wz-(Ga,Mn)N.

\begin{figure*}
\centerline{\includegraphics[width=0.92\textwidth]{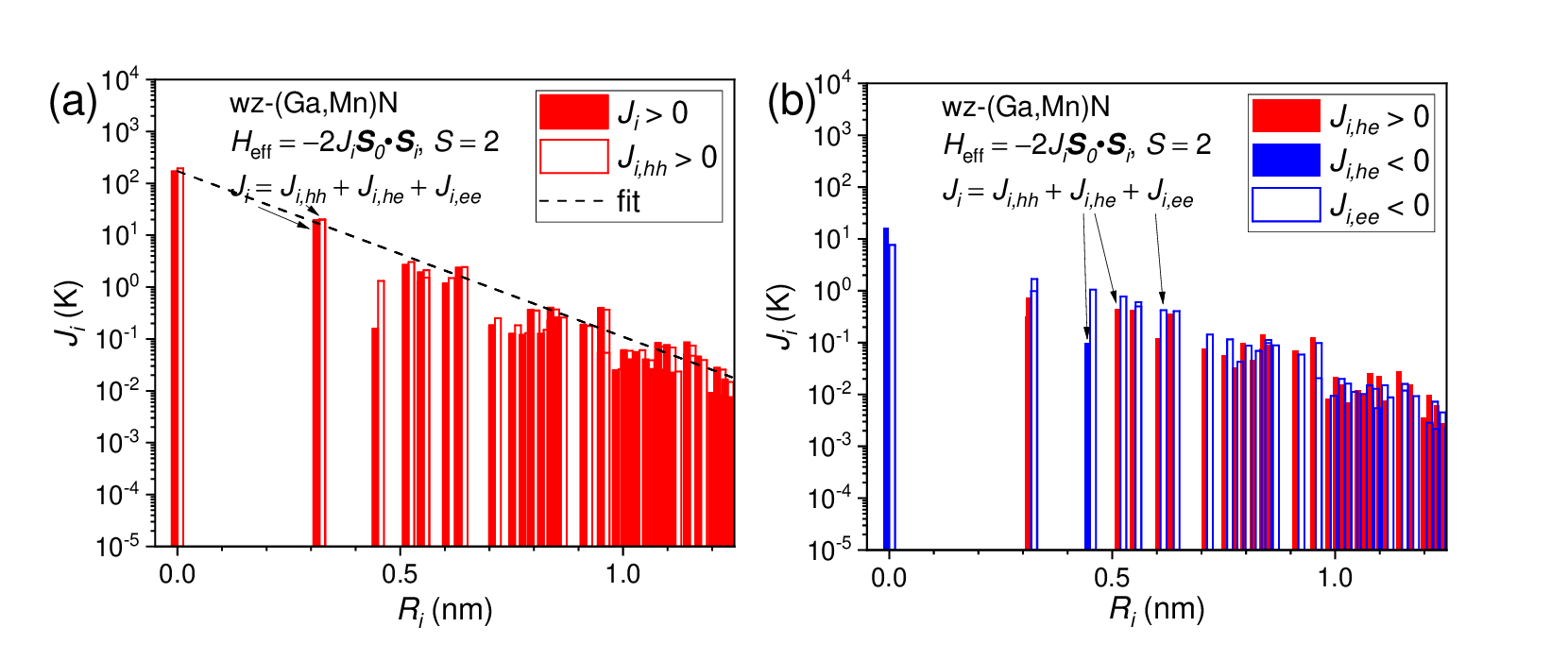}}
\caption{Exchange integrals $J_{i}$ in a logarithmic scale for spin pairs at distances $R_i < 1 \, \mathrm{nm}$ in topologically-trivial wurtzite (Ga,Mn)N.
See the caption of Fig.\ref{fig: HgCrTe} for the legend.}
\label{fig: wz-GaN-hh}
\end{figure*}

\begin{table}
\caption{$(\mathrm{Hg},\mathrm{Cr})\mathrm{Te}$: Numerical values of the total pair exchange integrals $J_i$
and contributions $hh$, $he$, and $ee$ for the nearest cation coordination zones
located at distances $d_i = \sqrt{i/2}a_{\text{lat}}$ and containing $z_i$ cations.
Positive and negative signs of $J$ correspond to FM and AFM coupling, respectively.
The computations have been performed for 16384 $k$-points and 512 $\theta$-points
(see, Sec.\,\ref{sec: theory} and Ref.\,\onlinecite{Sliwa_2021}).}
\label{tab: hgte}
\begin{tabular}{r|r|rrr|r}
\vspace{0.2cm}
$i$&       $J_i (\mathrm{K})$&
                   $J_{i,hh} (\mathrm{K})$&
                   $J_{i,he} (\mathrm{K})$&
                   $J_{i,ee} (\mathrm{K})$&  $z_i$\\
\hline
  1&      $3.6631$&
          $2.7219$&
          $1.6255$&
         $-0.6843$&        12\\
  2&     $-0.0496$&
          $0.0217$&
          $0.0857$&
         $-0.1570$&         6\\
  3&      $0.0292$&
          $0.1371$&
          $0.0730$&
         $-0.1809$&        24\\
  4&      $0.0185$&
          $0.1489$&
          $0.0707$&
         $-0.2011$&        12\\
\hline
\end{tabular}
\end{table}

\begin{table}
\caption{Same as in Table\,\ref{tab: hgte} but for zb-(Zn,Cr)Te (16384 $k$-points).}
\begin{tabular}{r|r|rrr|r}
\vspace{0.2cm}
$i$&       $J_i (\mathrm{K})$&
                   $J_{i,hh} (\mathrm{K})$&
                   $J_{i,he} (\mathrm{K})$&
                   $J_{i,ee} (\mathrm{K})$&  $z_i$\\
\hline
  1&                $1.0936$&
                    $1.1579$&
                    $0.0687$&
                   $-0.1330$&     12\\
  2&               $-0.0967$&
                    $0.0402$&
                   $-0.0384$&
                   $-0.0985$&      6\\
  3&                $0.0366$&
                    $0.0765$&
                    $0.0040$&
                   $-0.0438$&     24\\
  4&                $0.0588$&
                    $0.0734$&
                    $0.0082$&
                   $-0.0228$&     12\\
\hline
\end{tabular}
\label{tab: znte}
\end{table}

\begin{table}
\caption{Same as in Table\,\ref{tab: hgte} but for zb-(Ga,Mn)N (16384 $k$-points).}
\begin{tabular}{r|r|rrr|r}
\vspace{0.2cm}
$i$&       $J_i (\mathrm{K})$&
                   $J_{i,hh} (\mathrm{K})$&
                   $J_{i,he} (\mathrm{K})$&
                   $J_{i,ee} (\mathrm{K})$&  $z_i$\\
\hline
  1&          $19.879$&
              $20.579$&
              $ 1.289$&
              $-1.989$&       12\\
  2&          $-0.884$&
               $0.788$&
              $-0.386$&
              $-1.286$&        6\\
  3&           $1.440$&
               $1.725$&
               $0.345$&
              $-0.630$&       24\\
  4&           $1.629$&
               $1.625$&
               $0.326$&
              $-0.321$&       12\\
\hline
\end{tabular}
\label{tab: zb-gan}
\end{table}

\begin{table}[tb]
\caption{Same as Table\,\ref{tab: hgte} but for wz-(Ga,Mn)N
(1152 $k$-points per unit cell of the reciprocal lattice).}
\begin{tabular}{r|r|r|rrr|r}
\vspace{0.2cm}
$i$& $R_i (\text{\AA})$& $J_i (\mathrm{K})$&
  $J_{i,hh} (\mathrm{K})$& $J_{i,he} (\mathrm{K})$& $J_{i,ee} (\mathrm{K})$& $z_i$\\
\hline
  1&    $3.180$&       $19.498$&
                       $20.166$&
                        $0.325$&
                       $-0.988$&         6\\
  2&    $3.189$&       $18.898$&
                       $19.859$&
                        $0.729$&
                       $-1.690$&         6\\
  3&    $4.503$&        $0.158$&
                        $1.309$&
                       $-0.101$&
                       $-1.050$&         6\\
  4&    $5.185$&        $2.704$&
                        $3.034$&
                        $0.449$&
                       $-0.779$&         2\\
  5&    $5.518$&        $1.272$&
                        $1.519$&
                        $0.250$&
                       $-0.497$&        12\\
  6&    $5.524$&        $1.944$&
                        $2.113$&
                        $0.430$&
                       $-0.599$&         6\\
  7&    $6.087$&        $1.186$&
                        $1.487$&
                        $0.120$&
                       $-0.421$&        12\\
  8&    $6.378$&        $2.376$&
                        $2.420$&
                        $0.361$&
                       $-0.405$&         6\\
\hline
\end{tabular}
\label{tab: wz-gan}
\end{table}

Another relevant quantity is the Curie-Weiss parameter,
\begin{equation}
\Theta_0 = \frac{2}{3}\sum_{i\geq 1} z_iJ_iS(S+1),
\end{equation}
where $z_i$ is the number of cations in the $i$-th coordination sphere. In terms of $\Theta_0$ and according to the high-temperature
expansion \cite{Spalek:1986_PRB}, Curie-Weiss temperature [equal to  Curie temperature $T_{\text{C}}$ in the mean-field approximation (MFA)], is given by $\Theta_{\text{CW}} = x\Theta_0$, if a dependence of the band structure parameters on fractional magnetic cation content $x$ can be neglected. The values of $\Theta_{\text{CW}}$ for $x = 0.1$ are shown in Table~\ref{tab: Theta} for the studied systems.

\begin{table}
\caption{Curie-Weiss temperatures $\Theta_{\text{CW}}$ equivalent to Curie temperatures  $T_{\text{C}}$ within the mean-field
approximation compared to $T_{\text{C}}$ values obtained by Monte-Carlo (MC) simulations and the percolation (perc) theory
(see Sec.\,\ref{sec: TC}) for the systems under study with 10\% cations substituted by magnetic impurities.}
\begin{tabular}{c|c c c}
\vspace{0.2cm}
compound& $\Theta_{\text{CW}}$\,(K)& $T_{\text{C}}^{\text{MC}}$\,(K) & $T_{\text{C}}^{\text{perc}}$\,(K)\\
\hline
zb-Hg$_{0.9}$Cr$_{0.1}$Te& $15.3$&--&--\\
zb-Hg$_{0.9}$Cr$_{0.1}$Te [Allan]& $-86.0$&--&--\\
zb-Zn$_{0.9}$Cr$_{0.1}$Te& $5.90$& $0.75$& $1.50$\\
zb-Zn$_{0.9}$Cr$_{0.1}$Te [Sapra]& $92.0$& --& --\\
zb-Ga$_{0.9}$Mn$_{0.1}$N& $123.8$& $28$& $30.1$\\
wz-Ga$_{0.9}$Mn$_{0.1}$N& $129.1$& $32$& $31.3$\\
\hline
\end{tabular}
\label{tab: Theta}
\end{table}

Several conclusions emerge from the results displayed in Figs.~\ref{fig: HgCrTe}--\ref{fig: wz-GaN-hh} and Tables~\ref{tab: hgte}--\ref{tab: Theta}.
First, as shown previously \cite{Sliwa_2021}, within the Van Vleck susceptiblity model \cite{Yu:2010_S}, $T_{\text{C}}^{(VV)} = x\Theta_0^{(VV)}$, where $\Theta_0^{(VV)}$ involves the summation only over $J_{i,he}$ and includes the self-interaction $J_{0,he}$ term with $z_0 = 1$. Since this spurious $J_{0,he}$ term is quite large, its inclusion with a simultaneous disregarding of superexchange (the $hh$ term), leads to the improper conclusion about the dominant role of the Van Vleck mechanism. Second, judging from the $\Theta_0$ values, FM interactions are about an order of magnitude stronger in both zb- and wz-(Ga,Mn)N compared to (Hg,Cr)Te and (Zn,Cr)Te. Within our model, this fact results from a short lattice constant of GaN leading to sizable $pd$ hybridization and a rather large magnitude of $J_1$, as found by {\em ab initio} studies and experimentally \cite{Bonanni:2011_PRB}. Third, AFM $ee$ term is surprisingly large in the case of (Hg,Cr)Te and (Zn,Cr)Te pointing to a possible competition between spin-glass and FM ordering at low temperatures in those systems. The presence of two competing terms indicates that theoretical conclusions on the magnetic ground state and corresponding ordering temperature may sensitively depend on the assumed values of the $d$-shell energies ($E_d, U, J$) and tight-binding parameters.
The computations performed for two tight-binding models presented in Figs.~\ref{fig: HgCrTe}--\ref{fig: ZnCrTe_Sapra} for (Hg,Cr)Te and (Zn,Cr)Te confirm such a strong sensitivity of $J_i$ to details of the band-structure representation,
and indicate importance of the hybridization between the magnetic $d$-shells of Cr and the $d$ orbitals of Te.


In the situation outlined above, the results of recent first principles computations, carried out for (Hg,Cr)Te and (Cd,Cr)Te \cite{Cuono:2023_arXiv} using the hybrid functional approach,
are particularly relevant. Surprisingly, those data have demonstrated a dependence of the interaction sign on the setting of the functional mixing parameter $a_{\text{HSE}}$: the interaction tended to be FM
in (Hg,Cr)Te for $a_{\text{HSE}} = 0.5$, but AFM for $a_{\text{HSE}} = 0.32$ and $0.25$. Altogether, both tight-binding and ab initio results point to the fact that spin-spin coupling in (Hg,Cr)Te is on the borderline
between FM and AFM regime, which opens a door for magnetism manipulations by e.g. strain, pressure and electric field.

\section{Curie temperatures from Monte Carlo simulations and percolation theory}
\label{sec: TC}

Issues that can be encountered while attempting to find, by means of Monte Carlo simulations,
the Curie temperature of a site-diluted system, were recollected elsewhere \cite{Sliwa_2022}.
We assume that a fraction ($x = 0.1$) of randomly-chosen cation sites are occupied by local spins,
represented in the simulation by unit vectors $\mathbf{m}_i$ and pairwise coupled by isotropic Heisenberg interaction with $S (2 S + 1) J_{ij}$,
where $S = 2$ and $J_{ij}$ values are given in Tables \ref{tab: hgte}--\ref{tab: wz-gan}.
In the zinc-blende case, we limit the interaction to the neighbor pairs at the distance
$R_{ij} \leq 2\sqrt{2} a_{\text{lat}}$; in the wurtzite variant, $R_{ij} \leq 1.0$\,nm.

Periodic boundary conditions have been assumed, and --- for the smallest simulated
sizes --- the couplings with images of each spin in neighboring supercells are included by summing
within the truncation distance. System sizes range from $L = 4$ (256 disorder realizations)
to $L = 24$ (16 disorder realizations). In the wurtzite case, the proportions of the lattice block
are 3:3:2 (and $L$ refers to the linear system size along the $c$-axis). The numbers of temperatures in the
simulation is kept constant at $N_T = 24$. For each realization, $4+4$ burnin/measurement cycles are
performed, with the number of Monte Carlo steps in each cycle increasing linearly with the linear system size,
starting at 2000 steps ($L = 4$).

\begin{figure}
\centerline{\includegraphics[width=0.75\columnwidth]{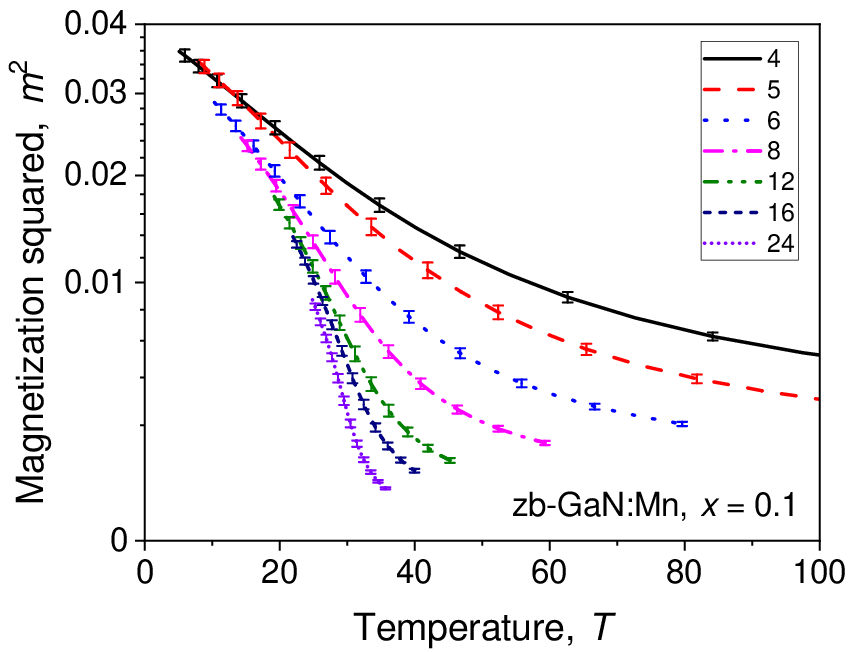}}
\caption{Examples of temperature dependence of magnetization square in zb-Ga$_{1-x}$Mn$_x$N, $x = 0.1$ obtained by the Monte-Carlo simulations.}
\label{fig: zb-GaN-m2}
\end{figure}

\begin{figure}
\centerline{\includegraphics[width=0.75\columnwidth]{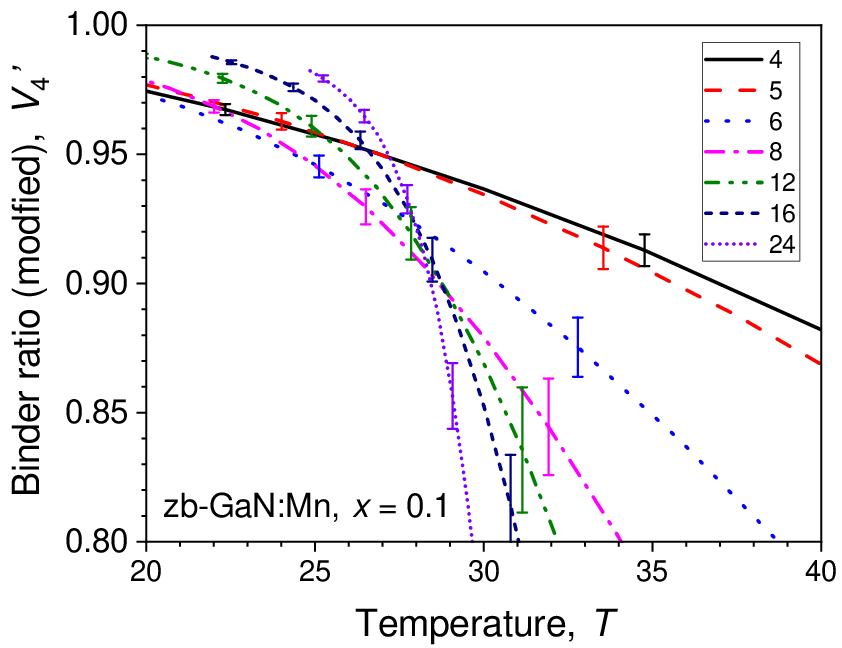}}
\caption{Examples of temperature dependencies of the modified Binder ratio for zb-Ga$_{1-x}$Mn$_x$N, $x = 0.1$ obtained by Monte-Carlo simulations.}
\label{fig: zb-GaN-v4alt}
\end{figure}

Figures~\ref{fig: zb-GaN-m2} and \ref{fig: zb-GaN-v4alt} present examples of
temperature dependencies of
magnetization square $[\langle\mathbf{m}\rangle^2]$ and the modified Binder cumulant for disordered
magnetic systems \cite{Sliwa_2022},
\begin{equation}
V'_4\{m\} = \frac{1}{2}\frac{[5\langle\mathbf{m}^2\rangle^2 -3\langle\mathbf{m}^4\rangle]}{[\langle\mathbf{m}^2\rangle^2]},
\end{equation}
obtained by Monte-Carlo simulations for zb-Ga$_{0.9}$Mn$_{0.1}$N with various system sizes $L$. Here, the square
brackets denote an average over disorder realizations.
The crossing point of $V'_4\{m\}$ curves determines Curie temperature  $T_{\text{C}}$.
The same procedure has been successful in the case of wz-Ga$_{0.9}$Mn$_{0.1}$N and zb-Zn$_{0.9}$Cr$_{0.1}$Te, and the
obtained $T_{\text{C}}$ values are displayed in Table~\ref{tab: Theta}.
However, because of competitions between ferromagnetic and antiferromagnetic
interactions, as shown in Fig.~\ref{fig: HgCrTe}, Monte Carlo simulations have not been conclusive for zb-Hg$_{0.9}$Cr$_{0.1}$Te,
pointing to a possibility of spin-glass freezing in that system.

It is interesting to compare the Monte-Carlo results to the expectations of the MFA and percolation theory for dilute
ferromagnetic systems. For instance, in the case of zinc-blende Ga$_{0.9}$Mn$_{0.1}$N, according to Table~\ref{tab: Theta}
 $T_{\text{C}}^{\text{MFA}} = 124$\,K, whereas $T_{\text{C}}^{\text{MC}} = 28$\,K. This significant difference
was already noted in the context of {\em ab initio} studies of DFSs with short-range exchange interactions \cite{Sato:2010_RMP}.
However, in the case of, for instance (Ga,Mn)As, where long range carrier-mediated interactions dominate, the difference between $T_{\text{C}}^{\text{MFA}}$ and $T_{\text{C}}^{\text{MC}}$ is much smaller, typically below 50\% \cite{Brey:2003_PRB}.

\begin{figure}
\centerline{\includegraphics[width=0.98\columnwidth]{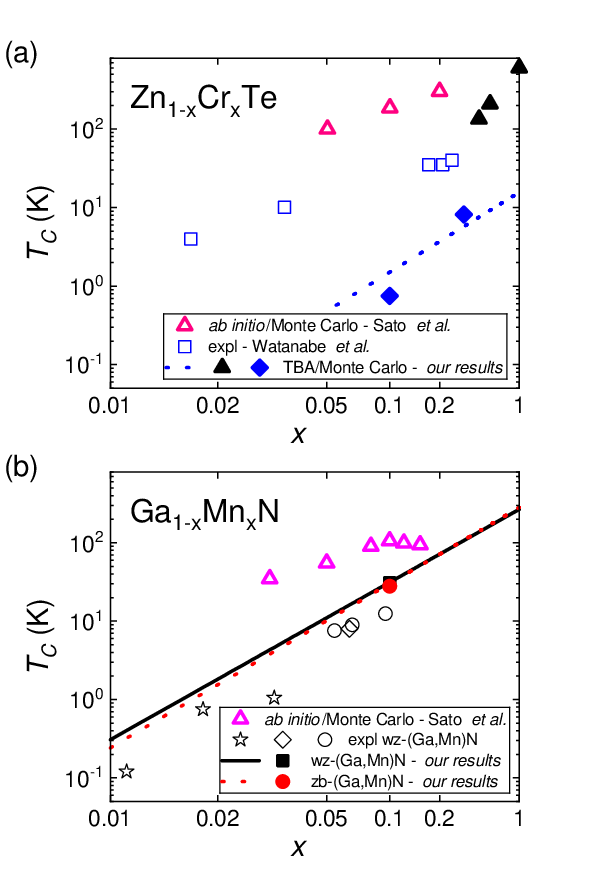}}
\caption{Curie temperatures $T_{\text{C}}$ in (a) Zn$_{1-x}$Cr$_x$Te and (b) Ga$_{1-x}$Mn$_x$N. Present results are shown by
solid points obtained by tight-binding computations (Figs.\ \ref{fig: ZnCrTe}--\ref{fig: wz-GaN-hh}) and Monte Carlo simulations.
The fits of $J(r)$ served to determine $T_{\text{C}}(x)$ from
the percolation theory, as shown by dotted and solid lines for zinc-blende and wurtzite systems, respectively.
Solid triangles: alternative parameterization data (Fig.\ \ref{fig: ZnCrTe_Sapra}).
Hollow triangles: previous {\em ab initio} and Monte Carlo results \cite{Sato:2010_RMP}; experimental data: squares
\cite{Watanabe:2019_APL}, diamond \cite{Sarigiannidou:2006_PRB}, stars \cite{Sawicki:2012_PRB}, circles \cite{Stefanowicz:2013_PRB}.}
\label{fig: percol}
\end{figure}

The percolation theory \cite{Korenblit_1973} was developed for continuous systems with exchange coupling $J(r)$ decaying exponentially with the spin pair distance $r$,
\begin{equation}
  J(r) = J_0 \exp\left(-r / b\right),
\label{eq: J0}
\end{equation}
for which $T_{\text{C}}$ can be written in the form \cite{Korenblit_1973},
\begin{equation}
T_{\text{C}}^{\text{perc}}\approx [S (2 S + 1)] J_0 \exp\left(-\frac{0.89}{b \, (xN_0)^{1/3}}\right),
\label{eq: perc}
\end{equation}
where $N_0$ is the cation concentration of magnetic ions.
The values of $T_{\text{C}}$ computed with these parameters and (\ref{eq: perc}) are shown in Fig.\ \ref{fig: percol} as a function of $x$ in comparison to the Monte Carlo and experimental results.

By fitting Eq.\,\ref{eq: J0} to $J_i$ data displayed in Figs.~\ref{fig: ZnCrTe}, \ref{fig: zb-GaMnN}, and \ref{fig: wz-GaN-hh} we obtain,
$J_0 \approx 12.0, 197, 172 \,\mathrm{K}$ and $b \approx 0.168, 0.130$, and  0.136\,nm
for zb-(Zn,Cr)Te, zb-(Ga,Mn)N, and wz-(Ga,Mn)N, respectively. We omit fitting the data for (Hg,Cr)Te, as a departure from (\ref{eq: J0}) is evident from oscillations of $J$. Similarly, we were unsuccessful fitting the alternative $J(r)$ for (Zn,Cr)Te (Fig.\ \ref{fig: ZnCrTe_Sapra}).
As shown in Table~\ref{tab: Theta}, the
resulting magnitudes of $T_{\text{C}}^{\text{perc}}$ agree with the $T_{\text{C}}$ values determined
by Monte-Carlo simulations in both zb-(Ga,Mn)N and wz-(Ga,Mn)N. This agreement allows us to evaluate $T_{\text{C}}(x)$ from the percolation formula (\ref{eq: perc}),
avoiding computationally expensive Monte-Carlo simulations for many $x$ values and disorder realizations.
In Fig.~\ref{fig: percol}, the values of $T_{\text{C}}(x)$ we have obtained in this way are compared to experimental data \cite{Watanabe:2019_APL,Sarigiannidou:2006_PRB,Sawicki:2012_PRB,Stefanowicz:2013_PRB} and earlier {\em ab initio} results
\cite{Sato:2010_RMP}. As seen, the present theory predicts much smaller values of $T_{\text{C}}$ than the {\em ab initio}
method which, within the local functional approximation, underestimates the localization degree of transition metal $d$ orbitals in semiconductors.
Our $T_{\text{C}}$ values are lower than experimental points in the case of Zn$_{1-x}$Cr$_x$Te at low $x$, which may point out to some aggregation of Cr ions in the studied layers,
sensitivity to band-structure modelling, or significance of the departure from (\ref{eq: J0}).

\section{$sp$-$d$ exchange integrals and the quantum topological Hall effects}
In tetrahedrally coordinated DMSs, exchange coupling between band carriers near the Brillouin zone center and cation-substitutional magnetic ions,
${\cal{H}}_{sp-d}^{(i)} = -{\cal{J}}\mathbf{s}\cdot\mathbf{S}_i$, is described by two
exchange integrals \cite{Kacman:2001_SST}, $\alpha = \langle S|{\cal{J}}|S\rangle$ and $\beta = \langle X|{\cal{J}}|X\rangle$, where here $S$ and $X$ are the periodic part
of the Bloch functions (Kohn-Luttinger amplitudes) that
transform as atomic $s$ and $p_x$ wave functions under the point symmetry group operations. Typically, $\alpha$ and $\beta$ originate from the FM intra-atomic $s$-$d$
potential exchange and the $p$-$d$ hybridization, respectively.  Furthermore, in many cases the molecular-field and virtual-crystal approximations hold allowing
a straightforward determination of $\alpha$ and $\beta$ from band splittings once macroscopic magnetization is known.

Making use of {\em NIST Atomic Spectra Database Levels}, we obtain  ${\cal{J}}_{\text{$4s$-$3d$}} = 0.372$\,eV for Cr$^{1{+}}$ ions,
which constitutes an upper limit for the exchange energy $N_0\alpha$ in Cr-doped compounds. For comparison, ${\cal{J}}_{\text{$4s$-$3d$}} =0.391$\,eV for Mn$^{1{+}}$, close to experimental
\cite{Dobrowolska:1981_SST,Bauer:1985_PRB} and {\em ab initio} \cite{Autieri:2021_PRB}
values for (Hg,Mn)Te, $N_0\alpha = 0.4\pm 0.1$ and $0.35\pm 0.05$\,eV, respectively. In view of this discussion,
we expect
\begin{equation}
  N_0\alpha = 0.3\pm 0.1 \,\mathrm{eV}
\end{equation}
for Hg$_{1-x}$Cr$_x$Te.

However, in the case of wide-gap II-VI Mn-contained DMSs, in which the conduction-band wave function
becomes significantly spread on anions, $N_0\alpha$ values are much reduced, e.g., $N_0\alpha = 0.19$\,eV
in the case of (Zn,Mn)Te \cite{Twardowski:1984_SSC}. An interesting situation occurs in III-V DMS with Mn$^{3+}$ ions, in which ferromagnetic $s$-$d$ coupling is compensated
by antiferromagnetic exchange with holes tightly bound by Mn$^{2}$ ions \cite{Sliwa:2008_PRB}, resulting in $N_0\alpha= 0.0 \pm 0.1$\,eV in wz-(Ga,Mn)N \cite{Suffczynski:2011_PRB}.

Using notation and parameter values introduced in Secs.\,\ref{sec: bands}, the $p$-$d$ exchange energy
for transition metal ions with the high-spin ($S = 2$) $d^{4}$ configuration is given by \cite{Kacman:2001_SST},
\begin{equation}
  N_0\beta = V_{pd}^2\left(\frac{1}{3}\frac{1}{E_{d}} + \frac{2}{15}\frac{1}{E_{d}+U} - \frac{7}{15}\frac{1}{E_{d} + U + J}\right),
\label{eq: beta}
\end{equation}
where ($P_{\mathrm{Te},p}$ projects the wavefunctions at $\Gamma$ onto the $p$-symmetry orbitals of tellurium)
\begin{equation}
  V_{pd} = \frac{4}{3}\left(V_{pd\sigma} - \frac{2}{\sqrt{3}}V_{pd\pi}\right)
    \left| \left< \psi_\Gamma \middle| P_{\mathrm{Te},p} \middle| \psi_\Gamma \right> \right|,
\end{equation}
which results in
\begin{equation}
  N_0\beta = -1.8\pm0.5 \,\mathrm{eV}
\end{equation}
for Hg$_{1-x}$Cr$_x$Te in the small $x$ value limit and $E_d = -0.3 \pm0.1$\,eV,
with an about 50\% enhancement in the alternative model which includes the $d$-$d$ hybridization.
Large magnitudes of $N_0\alpha =0.3$ and $N_0\beta = -1.8$\,eV  mean
that full polarization of Cr spins in Hg$_{0.99}$Cr$_{0.01}$Te will change
$E_g$ by about 21\,meV, strongly affecting optical and transport properties.

In the same way, we can evaluate a magnitude of the upward
shift $\Delta E_v$ of the Hg$_{1-x}$Cr$_x$ valence band top
in respect to HgTe, introduced by $p$-$d$ hybridization,
\begin{eqnarray}
  \lefteqn{\Delta E_v = xN_0W_{pd}} \nonumber \\
  & = & -xV_{pd}^2\left(\frac{1}{3}\frac{1}{E_{d}} + \frac{1}{5}\frac{1}{E_{d}+U} + \frac{7}{15}\frac{1}{E_{d} + U + J}\right).
\end{eqnarray}
This equation leads to
\begin{equation}
N_0W _{pd} = 1.2\pm0.5\,\mathrm{eV},
\end{equation}
for Hg$_{1-x}$Cr$_x$Te in the small $x$ limit.  This value of $N_0W _{pd}$ implies that $p$-$d$ hybridization
give a sizable contribution to
the gap change. In particular, this fact is expected to enlarge the topological region, $E_g < 0$, to $x \approx 10$\% in Hg$_{1-x}$Cr$_x$Te
compare to Hg$_{1-x}$Mn$_x$Te, where it extends to $x=7$\%.

Particularly interesting is the case of Hg$_{1-x}$Cr$_x$Te topological QWs. We assume that our evaluations of $\alpha$ and $\beta$ magnitudes are correct
and list out expected phenomena brought about by cation-substitutional randomly distributed Cr ions.
In the paramagnetic phase, guided by the Mn case \cite{Shamim:2020_SA}, we expect that the range of QW thicknesses corresponding to the
topological phase shrinks with $x$, and the trivial phase occurs at any thicknesses for $x\gtrsim 0.07$ \cite{Sawicki:1983_Pr}.
In the topological phase, two new effects of paramagnetic impurities upon the quantum spin
Hall effect have been recently identified \cite{Dietl:2023_PRL,Dietl:2023_PRB}:
\begin{enumerate}
\item The formation of bound magnetic polarons by holes residing on residual acceptor impurities.
The associated spin-splitting $\Delta$ of acceptor states diminishes spin-flip Kondo backscattering
of edge electrons by acceptor holes at low temperatures, $k_BT \lesssim \Delta(T)$,
where $\Delta(T)$ scales with $\chi(T){\cal{J}}_{sp-d}^2$, where $\chi(T)$ is magnetic susceptibility of
localized spins and ${\cal{J}}_{sp-d}^2$ is a weighted combination of $\alpha$ and $\beta$ \cite{Dietl:2023_PRB}. This
model is experimentally corroborated by  a recovery of the conductance quantization at low temperatures
in topological Hg$_{1-x}$Mn$_x$Te QWs \cite{Shamim:2021_NC}.
\item Precessional dephasing of edge electron spins and momenta by a dense cloud of randomly oriented
magnetic impurity spins. It has been argued that the constraint imposed by spin-momentum locking
on the efficiency of backscattering  by localized spins \cite{Tanaka:2011_PRL} is relaxed by a flow of spin momenta to the bath of interacting
magnetic impurities \cite{Dietl:2023_PRB}. This effect is relatively weakly dependent on temperature,
as it scales with $T\chi(T){\cal{J}}_{sp-d}'^2$, where ${\cal{J}}_{sp-d}'^2$ is another
combination of $\alpha$ and $\beta$ \cite{Dietl:2023_PRB}.
\end{enumerate}
Interestingly, due to larger magnitudes of $\beta$ and $\chi$ (enhanced by ferromagnetic components in $J_{ij}$),
both effects are expected to be substantially stronger in Hg$_{1-x}$Cr$_x$Te compared to Hg$_{1-x}$Mn$_x$Te.

Can one observe the anomalous quantum Hall effect in Hg$_{1-x}$Cr$_x$Te quantum wells (QWs)? As we noted,
there is a competition of FM and AFM interactions, so that either FM or spin-glass
phase is expected at low temperatures. Furthermore, it was demonstrated
that the formation of a single chiral edge channels occurs for spin polarized magnetic ions along the growth direction, if
$p =-\alpha/\beta \gtrsim 0.25$ \cite{Liu:2008_PRLb}, whereas the values quoted above point to
$p = 0.17 \pm 0.1$. These two facts call for experimental verification. In the case of $p < 0.25$,
polarization of Cr spins, either spontaneous or driven by an external magnetic field along the growth direction,
will lead to the closure of the topological gap and the associated colossal drop
of resistance.

We note, however, that the spin splitting of subbands with a heavy-hole character vanishes for the in-plane magnetization direction due to a competition between spin-orbit and $p$-$d$ interactions \cite{Peyla:1993_PRB}. This means that one can change the parameter $p$ by tilting the magnetization direction. In order to verify this expectation we have adapted the eight bands' $k \cdot p$ model for the Hg$_{1-x}$Cr$_x$Te quantum well surrounded by 30\,nm-thick Hg$_{1-y}$Cd$_{y}$Te barriers \cite{Novik:2005_PRB,Dietl:2023_PRB}. We abandon the axial approximation and employ the values of $N_0\alpha$, $N_0\beta$, and $N_0W_{pd}$ quoted above. The electron envelope function in the growth direction is taken as a linear combination of 51 Fourier components \footnote{Boundary terms at the well-barrier interfaces are neglected.}, whose contributions are obtained by diagonalization of the resulting $408 \times 408$ Hamiltonian ${\cal{H}}_{k_x,k_y}$ assuming periodic boundary conditions in the $z$-direction \cite{Dietl:2023_PRB}.

\begin{figure}[tb]
\hspace*{-0.5cm}
\hbox to \hsize{
\includegraphics[scale=0.5, angle=0]{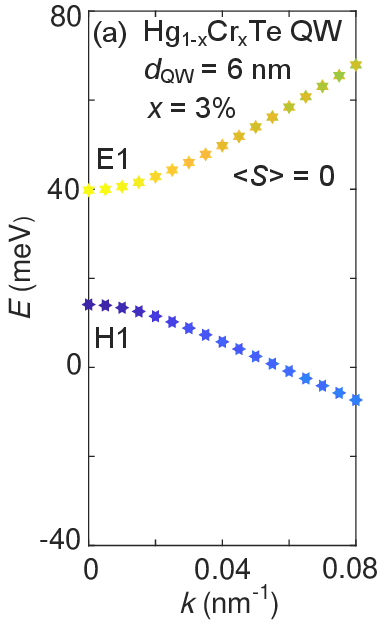}
\includegraphics[scale=0.5, angle=0]{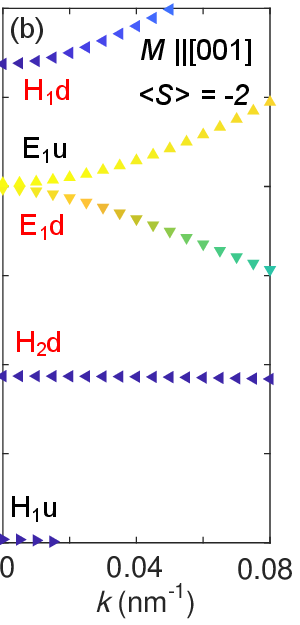}
\includegraphics[scale=0.5, angle=0]{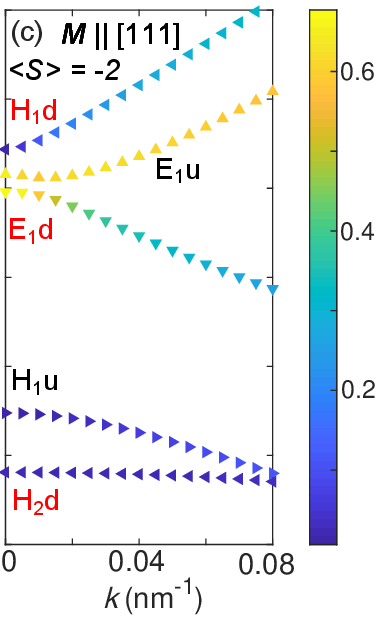}}
	\caption{Computed subband structure in unstrained Hg$_{0.97}$Cr$_{0.03}$Te/Hg$_{0.3}$Cd$_{0.7}$Te quantum wells; labels H and E refer to subbands originating from the heavy-hole and electron-like bands in non-topological semiconductors; up and right oriented triangles correspond to spin-up orientation for subbands with the electron and hole character, respectively. (a) Paramagnetic case, i.e., average Cr spin $\langle S\rangle =0$; (b) $\langle S\rangle =2$ and magnetization is oriented along the growth direction; (c) $\langle S\rangle =2$ and magnetization is oriented along  the [111] directions. Colors describe the participation of the $s_{1/2,\pm1/2}$ Kohn-Luttinger amplitude in the electronic wave function; Hg$_{0.97}$Cr$_{0.03}$Te QW thickness $d_{\text{QW}} = 6$\,nm.}
\label{fig:QAHE}
\end{figure}

Figure~\ref{fig:QAHE} shows the subband structure for unstrained 6\,nm-thick Hg$_{0.97}$Cr$_{0.03}$Te/Hg$_{0.3}$Cd$_{0.7}$Te QW in the absence of an external magnetic field. In the case of ferromagnetic (Hg,Cr)Te in a zinc-blende structure, for which cubic magnetic anisotropy is expected, the magnetization easy axis will assume either the $\langle100\rangle$ or $\langle111\rangle$ crystallographic direction. As seen in Fig.~\ref{fig:QAHE}, for the easy axis along the growth direction, $\mathbf{M} \| [001]$, the presence of non-zero magnetization tends to close the gap, in agreement with the value $p < 0.25$. In contrast, for $\mathbf{M} \| \langle111\rangle$, the topological gap persists and, moreover, the inverted band structure is present for the spin-down channel only. We conclude that the  quantum anomalous Hall effect is expected under these conditions.

To substantiate the above conclusion, we have calculated Berry's curvature $\Omega^{(x,y)}(\mathbf{k})$ for the is 306 occupied valence subbands, i.e., for the Fermi level in the gap. Its integral over the 2D Brillouin zone provides $2\pi{\cal{C}}$, where the Chern number ${\cal{C}}$  is directly related to Hall conductivity, $\sigma_{xy} = (e^2/h){\cal{C}}$. For the fcc lattice and (001) orientation of the QW, the 2D Brillouin zone is a square spanned on  the $[11]$ and $[\bar{1}1]$ of the length $G = 2\sqrt{2}\pi/a_0$, where $a_0$ is the lattice constant. The computation
involves two steps. First,  to ensure the periodicity of ${\cal{H}}_k$, we  introduce a $\mathbf{k}$ vector regularization, according to
\begin{eqnarray}
k_i & \rightarrow & (G/2\pi) \sin(2\pi k_i/G) \\
k_i k_j & \rightarrow & (G/\pi)^2 \sin(\pi k_i/G) \sin(\pi k_jG),
\end{eqnarray}
where the components $k_1$ and $k_2$ of the vectors in the Brillouin zone are related to $k_x$ and $k_y$ and their combinations in the $kp$ Hamiltonian by,
\begin{eqnarray}
k_{x,y} & = & (k_1 \mp k_2)/\sqrt{2} \\
k_{x,y}^2 & = & (k_1 \mp k_2)^2/2 \\
k_x k_y & = & (k_1^2 - k_2^2)/2.
\end{eqnarray}

Second, we use the placket method introduced by Fukui {\em et al.} \cite{Fukui:2005_JPSJ} in a numerically efficient implementation elaborated by Brzezicki \cite{Brzezicki},
to calculate the total (a trace) of the Berry curvature over the occupied bands $E_n(\mathbf{k})$, $E_n(\mathbf{k}) < E_F$, as \footnote{If the presence of degenerate bands cannot be excluded, the determinants of the submatrices corresponding to each of the mutually different eigenergies must be computed instead. Then the imaginary part of the trace can be obtained as the logarithm of the product of all the determinants (normalized to the unit circle), and is the Abelian curvature of the adiabatic transport of full-dimension orthonormal sets of energy eigenstates. Considering the principal branch of the logarithm \cite{Fukui:2005_JPSJ} eliminates ambiguity of the result.}
\begin{equation}
(\Delta k)^2 \Omega^{(x,y)}(\mathbf{k}) = \mathop{\mathrm{Im}} \left\{ \mathop{\mathrm{Tr}} \left[ \prod_{i = 1}^4 V_i V_i^\dagger \right] \right\},
\end{equation}
where $\Delta k$ is the length of the placket side; $V_i$ are matrices of eigenvectors whose columns correspond to the 306 eigenenergies $E_n(k_{i,1}, k_{i,2}) < E_F$ at the four wave vectors determining the placket position, $[k_1, k_2]$, $[k_1 + \Delta k, k_2]$, $[k_1 + \Delta k, k_2 + \Delta k]$, $[k_1, k_2 + \Delta k]$.

\begin{figure}[tb]
\includegraphics[width=8 cm]{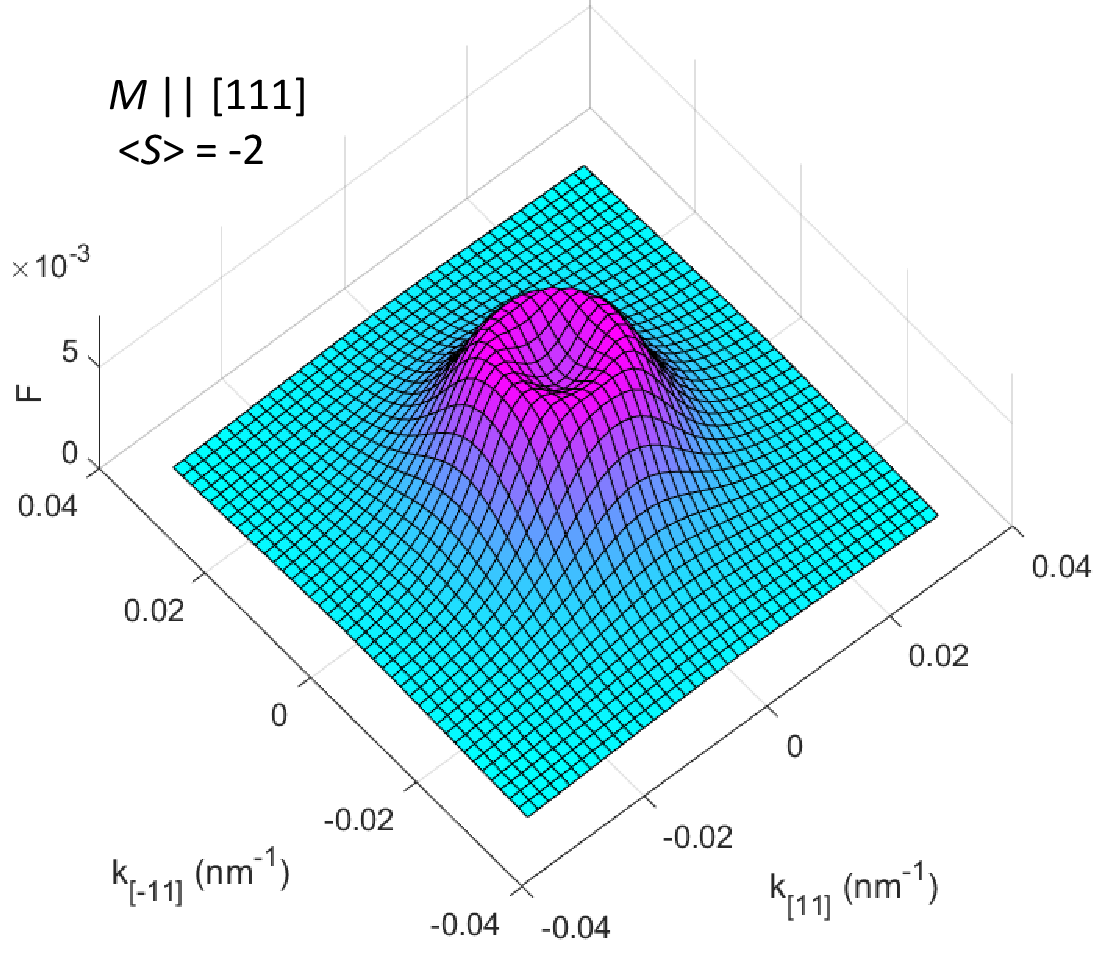}
	\caption{Berry curvature integrated over the placket area $(\Delta k)^2$,
$F(\mathbf{k})$, computed for the valence band subbands in a magnetized
Hg$_{0.97}$Cr$_{0.03}$Te/Hg$_{0.3}$Cd$_{0.7}$Te (001) QW quantum well,
whose electronic structure is presented in Fig.~\ref{fig:QAHE}(c). The
data are shown for the Brillouin zone center (a side of the square is $G/400$), that provides the dominant
contribution. The  Chern number ${\cal{C}} = 1.0$, obtained by summing
up $F$ values over the whole 2D Brillouin zone, indicates the presence of
the quantum anomalous Hall effect. }
\label{fig:Berry}
\end{figure}

As shown in Fig.~\ref{fig:Berry}, non-zero Berry curvature weights for spin-polarized Cr ions, $S_{\text{av}} = -2$, reside in the Brillouin zone center, i.e., in the gap region.  By computing $\Omega^{(x,y)}(\mathbf{k})(\Delta k)^2 / 2 \pi$ values with adaptive $\Delta k$ magnitudes over the whole 2D Brillouin zone, we obtained the Chern number  ${\cal{C}}= 1.0$. We also checked that ${\cal{C}}= 0.00$ if $S_{\text{av}} = -0.1$, at which the band structure is no longer inverted. These outcomes confirm the prediction of the quantized Hall resistance.

Finally, we mention that in the case of (Zn,Cr)Te and (Ga,Mn)N the relevant donor level introduced
by TM ions resides in the gap, i.e., $E_d >0$.
In qualitative agreement with Eq.\,\ref{eq: beta}, the FM sign of $\beta$ was observed
in these two systems \cite{Twardowski:1987_PRB,Suffczynski:2011_PRB}.
However, as mentioned when discussing $N_0\alpha$ case, the Mn$^{3}$ state can be regarded as the Mn$^{2{+}}$
 acceptor with a tightly bound hole.
The resulting bare coupling between a band hole and Mn$^{2+}$ ion is then AFM. However, the experimentally proven
existence of the hole bound state in the gap (Mn$^{3+/4+}$ complex with $E_d =1.1$\,eV) means that band hole can be trapped by Mn,
meaning that the molecular-field approximation does not hold.
A non-perturbative approach demonstrates that the apparent $p$-$d$ coupling is FM under these conditions \cite{Dietl:2008_PRB}, in agreement
with experimental findings \cite{Suffczynski:2011_PRB}.

\section{Conclusions and outlook}

We have presented the theoretical results for $d$-$d$ and $p$-$d$ exchange
interactions in diluted magnetic insulators for the $S = 2$ case, i.e., for
Cr in HgTe and ZnTe as well as for Mn in zinc-blende and wurtzite GaN. Our
approach not only addresses some weak points of earlier theories of
ferromagnetic superexchange but takes into account the interband
Bloembergen-Rowland-Van Vleck and two-electron contributions.
These additional terms are of a lesser importance in (Ga,Mn)N but play a
significant role in (Zn,Cr)Te and, particularly, in topological
(Hg,Cr)Te, where they introduce antiferromagnetic coupling for certain
pairs of Cr atoms.  We have found that the presence of the competing
interactions makes that the theoretical results become rather sensitive to
the adopted tight-binding model of the host band structure. Because of
this competition, and in agreement with experimental data, Curie
temperatures $T_{\text{C}}$ in (Zn,Cr)Te are expected to be lower compared
to (Ga,Mn)N with the identical concentration of magnetic ions. At the
same time, we cannot exclude the presence of a spin-glass phase rather
than ferromagnetism in topological (Hg,Cr)Te. If this would be the case, a
relatively large magnitude of the $p$-$d$ exchange integral $\beta$ that we
predict for (Hg,Cr)Te should stabilize the quantum spin Hall effect by the
formation of bound magnetic polarons that weaken Kondo backscattering of
edge electrons by holes trapped to residual acceptor impurities. The large
magnitude of $\beta$ means also that the quantum anomalous Hall effect
can be observed in topological (Hg,Cr)Te for the magnetization vector
tilted away from the direction perpendicular to the quantum well plane,
as confirmed by the Chern number determination.

Compared to ferromagnetic Bi-Sb chalcogenides, it is harder to introduce
Cr and V to HgTe and related systems. However, once obtained, they should
show lower areal density of native defects, as according to gating
characteristics, the concentration
of the in-gap localized states is almost two orders of magnitude
smaller in HgTe quantum wells \cite{Bendias:2018_NL,Yahniuk:2021_arXiv}
than in (Bi,Sb,Cr,V)$_2$Te$_3$ layers
\cite{Okazaki:2022_NP,Fijalkowski:2021_NC,Rodenbach:2023_arXiv}.
As the exchange gap is typically in the dozen meV range (see
Fig.~\ref{fig:QAHE}), we assign the thermally activated
conductivity $\sigma_{xx}$  in (Bi,Sb,Cr,V)$_2$Te$_3$ layers
\cite{Okazaki:2022_NP,Fijalkowski:2021_NC,Rodenbach:2023_arXiv} to the
Efros-Shklovskii hopping between in-gap states. A lower concentration of
such states will allow for the operation
of resistance standards at higher temperature.

\section*{Acknowledgments}

We acknowledge J. A. Majewski for sharing with us a code implementing formulas for
superexchange (in the $d^5$ and $d^4$ cases) and W. Brzezicki for the Chern number formula.  This research was supported by the Foundation for Polish Science through the IRA Programme co-financed by EU within SG OP and EFSE (project ``MagTop'' no. FENG.02.01-IP.05-0028/23),
 by funds from the state budget allocated by the Minister
of Science (Poland) as part of the Polish Metrology II programme project no.
PM-II/SP/0012/2024/02, and by
Interdisciplinary Centre for Mathematical and Computational Modelling, University of Warsaw (ICM UW)
under computational allocations numbers G93-1595 and G93-1601.

\appendix

\begin{widetext}

\section{Transformation matrix to $T_2 \oplus E$}
\label{app: rot_d}

With the threefold wurtzite $c$-axis as the $z$ (quantization) axis, a symmetry-invariant decomposition
of the Hilbert space for $L = 2$ into $3+2$ dimensions ($T_2 \oplus E$) is parameterized below
by $(\theta, \phi)$ [cf.\ Eq.\ 68 of Ref.\ \onlinecite{Goodenough_1963}]. The threefold symmetry
acts by cycling the $T_2$ vectors $(\psi_{T_2, i})_{i = 1, 2, 3}$.
Under cubic symmetry, $\cos(\theta) = 1 / \sqrt{3}$; the deviation from this value is a material parameter.
On the other hand, due to the wurtzite mirror symmetry, $\phi$ is not variable, but it takes different and
conventions-dependent values for the two inequivalent cation positions \cite{Yang_1995} (here, $\phi = -\pi/2$ for $t_2$
and $\phi = \pi/2$ for $t_4$).
\begin{eqnarray}
  \left< yz \middle| \psi_{T_2,1} \right> & = & \sqrt{2/3} \cos(\theta) \sin(\phi) \\
  \left< xz \middle| \psi_{T_2,1} \right> & = & -\sqrt{2/3} \cos(\theta) \cos(\phi) \\
  \left< xy \middle| \psi_{T_2,1} \right> & = & -\sqrt{2/3} \sin(\theta) \sin(2\phi) \\
  \left< 3z^2-r^2 \middle| \psi_{T_2,1} \right> & = & \sqrt{1/3} \\
  \left< x^2 - y^2 \middle| \psi_{T_2,1} \right> & = & \sqrt{2/3} \sin(\theta) \cos(2\phi) \\
  \left< yz \middle| \psi_{T_2,2} \right> & = & -\cos\theta \left[3 \cos(\phi) + \sqrt{3} \sin(\phi)\right] / (3 \sqrt{2}) \\
  \left< xz \middle| \psi_{T_2,2} \right> & = & \cos\theta \left[\cos(\phi) / \sqrt{6} - \sin(\phi) / \sqrt{2}\right] \\
  \left< xy \middle| \psi_{T_2,2} \right> & = & -\sin\theta \left[\cos(2\phi) / \sqrt{2} - \sin(2\phi) / \sqrt{6}\right] \\
  \left< 3z^2-r^2 \middle| \psi_{T_2,2} \right> & = & \sqrt{1/3} \\
  \left< x^2 - y^2 \middle| \psi_{T_2,2} \right> & = & -\sin\theta \left[\sqrt{3} \cos(2\phi) + 3 \sin(2 \phi)\right] / (3 \sqrt{2}) \\
  \left< yz \middle| \psi_{T_2,3} \right> & = & \cos\theta \left[\cos(\phi) / \sqrt{2} - \sin(\phi) / \sqrt{6}\right] \\
  \left< xz \middle| \psi_{T_2,3} \right> & = & \cos\theta \left[\cos(\phi) / \sqrt{6} + \sin(\phi) / \sqrt{2}\right] \\
  \left< xy \middle| \psi_{T_2,3} \right> & = & \sin\theta \left[\cos(2\phi) / \sqrt{2} + \sin(2\phi) / \sqrt{6}\right] \\
  \left< 3z^2-r^2 \middle| \psi_{T_2,3} \right> & = & \sqrt{1/3} \\
  \left< x^2 - y^2 \middle| \psi_{T_2,3} \right> & = & -\sin\theta \left[\cos(2\phi) / \sqrt{6} - \sin(2 \phi) / \sqrt{2}\right] \\
  \left< yz \middle| \psi_{E,1} \right> & = & -\sin\theta \left[\cos(\phi) + \sin(\phi) \right] / \sqrt{2}\\
  \left< xz \middle| \psi_{E,1} \right> & = & \sin\theta \left[\cos(\phi) / \sqrt{6} + \sin(\phi) / \sqrt{2}\right] \\
  \left< xy \middle| \psi_{E,1} \right> & = & \cos\theta \left[\cos(\phi) - \sin(\phi) \right] / \sqrt{2} \\
  \left< 3z^2-r^2 \middle| \psi_{E,1} \right> & = & 0 \\
  \left< x^2 - y^2 \middle| \psi_{E,1} \right> & = & \cos\theta \left[\cos(2\phi) + \sin(2 \phi)\right]  / \sqrt{2} \\
  \left< yz \middle| \psi_{E,2} \right> & = & \sin\theta \left[\cos(\phi) - \sin(\phi)\right] / \sqrt{2} \\
  \left< xz \middle| \psi_{E,2} \right> & = & \sin\theta \left[\cos(\phi) + \sin(\phi)\right] / \sqrt{2} \\
  \left< xy \middle| \psi_{E,2} \right> & = & -\cos\theta \left[\cos(2\phi) + \sin(2\phi)\right] / \sqrt{2} \\
  \left< 3z^2-r^2 \middle| \psi_{E,2} \right> & = & 0 \\
  \left< x^2 - y^2 \middle| \psi_{E,2} \right> & = & \cos\theta \left[\cos(2\phi) - \sin(2 \phi)\right] / \sqrt{2}
\end{eqnarray}
\end{widetext}

\section{Definitions of the correlation lengths}

The definition of the correlation length varies from a crystal structure to another. For wurtzite, we assume that
the correlator decays exponentially with distance ($\mathbf{r} = x_1 \mathbf{a}_1 + x_2 \mathbf{a}_2 + x_3 \mathbf{a}_3$) as:
\begin{equation}
  C(\mathbf{r}) = \exp\left(-\frac{(|x_1|+|x_2|+|x_1+x_2|)}{\xi_a}\right) \exp\left(-\frac{|x_3|}{\xi_c}\right).
\end{equation}
We estimate $\xi_c$ as follows: a simulation is performed in a lattice block of size $L_1 \times L_2 \times L_3$,
$L_1 / L_3 = L_2 / L_3 = 3 / 2$, and the susceptibility $\chi_{k = k_{min,c}}$ at
momentum $k_{min,c} = (k_1, k_2, k_3) = \left(0, 0, \frac{4 \pi}{L_3}\right)$,
together with magnetization squared (i.e.\ the susceptibility at zero momentum, $k_1 = k_2 = k_3 = 0$), are both evaluated.
We define $\xi_{wz,c} \approx \xi_c$ as
\begin{equation}
  \xi_{wz,c} \approx \frac{1}{3 \sin{\frac{4 \pi}{3 L_3}}} \sqrt{\frac{\chi_{k = 0}}{\chi_{k = k_{min,c}}} - 1}.
\end{equation}
Analogously, $\xi_{wz,a} \approx \xi_a$ can be defined as
\begin{equation}
  \xi_{wz,a} \approx \frac{7}{18 \sin{\frac{2 \pi}{L_1}}} \sqrt{\frac{\chi_{k = 0}}{\chi_{k = k_{min,a}}} - 1},
\end{equation}
with $k_{min,a} = \left(\frac{6 \pi}{L_1}, 0, 0\right)$.

\bibliography{HgCrTe_etc_td,spin_interactions_cdmnte_and_hgmnte_24May2023}

\begin{thebibliography}{77}%
\makeatletter
\providecommand \@ifxundefined [1]{%
 \@ifx{#1\undefined}
}%
\providecommand \@ifnum [1]{%
 \ifnum #1\expandafter \@firstoftwo
 \else \expandafter \@secondoftwo
 \fi
}%
\providecommand \@ifx [1]{%
 \ifx #1\expandafter \@firstoftwo
 \else \expandafter \@secondoftwo
 \fi
}%
\providecommand \natexlab [1]{#1}%
\providecommand \enquote  [1]{``#1''}%
\providecommand \bibnamefont  [1]{#1}%
\providecommand \bibfnamefont [1]{#1}%
\providecommand \citenamefont [1]{#1}%
\providecommand \href@noop [0]{\@secondoftwo}%
\providecommand \href [0]{\begingroup \@sanitize@url \@href}%
\providecommand \@href[1]{\@@startlink{#1}\@@href}%
\providecommand \@@href[1]{\endgroup#1\@@endlink}%
\providecommand \@sanitize@url [0]{\catcode `\\12\catcode `\$12\catcode
  `\&12\catcode `\#12\catcode `\^12\catcode `\_12\catcode `\%12\relax}%
\providecommand \@@startlink[1]{}%
\providecommand \@@endlink[0]{}%
\providecommand \url  [0]{\begingroup\@sanitize@url \@url }%
\providecommand \@url [1]{\endgroup\@href {#1}{\urlprefix }}%
\providecommand \urlprefix  [0]{URL }%
\providecommand \Eprint [0]{\href }%
\providecommand \doibase [0]{http://dx.doi.org/}%
\providecommand \selectlanguage [0]{\@gobble}%
\providecommand \bibinfo  [0]{\@secondoftwo}%
\providecommand \bibfield  [0]{\@secondoftwo}%
\providecommand \translation [1]{[#1]}%
\providecommand \BibitemOpen [0]{}%
\providecommand \bibitemStop [0]{}%
\providecommand \bibitemNoStop [0]{.\EOS\space}%
\providecommand \EOS [0]{\spacefactor3000\relax}%
\providecommand \BibitemShut  [1]{\csname bibitem#1\endcsname}%
\let\auto@bib@innerbib\@empty
\bibitem [{\citenamefont {Dietl}\ and\ \citenamefont
  {Ohno}(2014)}]{Dietl:2014_RMP}%
  \BibitemOpen
  \bibfield  {author} {\bibinfo {author} {\bibfnamefont {T.}~\bibnamefont
  {Dietl}}\ and\ \bibinfo {author} {\bibfnamefont {H.}~\bibnamefont {Ohno}},\
  }\bibfield  {title} {\enquote {\bibinfo {title} {Dilute ferromagnetic
  semiconductors: Physics and spintronic structures},}\ }\href {\doibase
  10.1103/RevModPhys.86.187} {\bibfield  {journal} {\bibinfo  {journal} {Rev.
  Mod. Phys.}\ }\textbf {\bibinfo {volume} {86}},\ \bibinfo {pages} {187--251}
  (\bibinfo {year} {2014})}\BibitemShut {NoStop}%
\bibitem [{\citenamefont {Jungwirth}\ \emph {et~al.}(2014)\citenamefont
  {Jungwirth}, \citenamefont {Wunderlich}, \citenamefont {Nov\'ak},
  \citenamefont {Olejn\'ik}, \citenamefont {Gallagher}, \citenamefont
  {Campion}, \citenamefont {Edmonds}, \citenamefont {Rushforth}, \citenamefont
  {Ferguson},\ and\ \citenamefont {N\ifmmode~\check{e}\else
  \v{e}\fi{}mec}}]{Jungwirth:2014_RMP}%
  \BibitemOpen
  \bibfield  {author} {\bibinfo {author} {\bibfnamefont {T.}~\bibnamefont
  {Jungwirth}}, \bibinfo {author} {\bibfnamefont {J.}~\bibnamefont
  {Wunderlich}}, \bibinfo {author} {\bibfnamefont {V.}~\bibnamefont {Nov\'ak}},
  \bibinfo {author} {\bibfnamefont {K.}~\bibnamefont {Olejn\'ik}}, \bibinfo
  {author} {\bibfnamefont {B.~L.}\ \bibnamefont {Gallagher}}, \bibinfo {author}
  {\bibfnamefont {R.~P.}\ \bibnamefont {Campion}}, \bibinfo {author}
  {\bibfnamefont {K.~W.}\ \bibnamefont {Edmonds}}, \bibinfo {author}
  {\bibfnamefont {A.~W.}\ \bibnamefont {Rushforth}}, \bibinfo {author}
  {\bibfnamefont {A.~J.}\ \bibnamefont {Ferguson}}, \ and\ \bibinfo {author}
  {\bibfnamefont {P.}~\bibnamefont {N\ifmmode~\check{e}\else \v{e}\fi{}mec}},\
  }\bibfield  {title} {\enquote {\bibinfo {title} {Spin-dependent phenomena and
  device concepts explored in {(Ga,Mn)As}},}\ }\href {\doibase
  10.1103/RevModPhys.86.855} {\bibfield  {journal} {\bibinfo  {journal} {Rev.
  Mod. Phys.}\ }\textbf {\bibinfo {volume} {86}},\ \bibinfo {pages} {855--896}
  (\bibinfo {year} {2014})}\BibitemShut {NoStop}%
\bibitem [{\citenamefont {Sztenkiel}\ \emph {et~al.}(2016)\citenamefont
  {Sztenkiel}, \citenamefont {Foltyn}, \citenamefont {Mazur}, \citenamefont
  {Adhikari}, \citenamefont {Kosiel}, \citenamefont {Gas}, \citenamefont
  {Zgirski}, \citenamefont {Kruszka}, \citenamefont {Jakie\l{}a}, \citenamefont
  {Li}, \citenamefont {Piotrowska}, \citenamefont {Bonanni}, \citenamefont
  {Sawicki},\ and\ \citenamefont {Dietl}}]{Sztenkiel:2016_NC}%
  \BibitemOpen
  \bibfield  {author} {\bibinfo {author} {\bibfnamefont {D.}~\bibnamefont
  {Sztenkiel}}, \bibinfo {author} {\bibfnamefont {M.}~\bibnamefont {Foltyn}},
  \bibinfo {author} {\bibfnamefont {G.~P.}\ \bibnamefont {Mazur}}, \bibinfo
  {author} {\bibfnamefont {R.}~\bibnamefont {Adhikari}}, \bibinfo {author}
  {\bibfnamefont {K.}~\bibnamefont {Kosiel}}, \bibinfo {author} {\bibfnamefont
  {K.}~\bibnamefont {Gas}}, \bibinfo {author} {\bibfnamefont {M.}~\bibnamefont
  {Zgirski}}, \bibinfo {author} {\bibfnamefont {R.}~\bibnamefont {Kruszka}},
  \bibinfo {author} {\bibfnamefont {R.}~\bibnamefont {Jakie\l{}a}}, \bibinfo
  {author} {\bibfnamefont {Tian}\ \bibnamefont {Li}}, \bibinfo {author}
  {\bibfnamefont {A.}~\bibnamefont {Piotrowska}}, \bibinfo {author}
  {\bibfnamefont {A.}~\bibnamefont {Bonanni}}, \bibinfo {author} {\bibfnamefont
  {M.}~\bibnamefont {Sawicki}}, \ and\ \bibinfo {author} {\bibfnamefont
  {T.}~\bibnamefont {Dietl}},\ }\bibfield  {title} {\enquote {\bibinfo {title}
  {Stretching magnetism with an electric field in a nitride semiconductor},}\
  }\href {\doibase 10.1038/ncomms13232} {\bibfield  {journal} {\bibinfo
  {journal} {Nat. Commun.}\ }\textbf {\bibinfo {volume} {7}},\ \bibinfo {pages}
  {13232} (\bibinfo {year} {2016})}\BibitemShut {NoStop}%
\bibitem [{\citenamefont {Mogi}\ \emph {et~al.}(2022)\citenamefont {Mogi},
  \citenamefont {Okamura}, \citenamefont {Kawamura}, \citenamefont {Yoshimi},
  \citenamefont {Yasuda}, \citenamefont {Tsukazaki}, \citenamefont {Takahashi},
  \citenamefont {Morimoto}, \citenamefont {Nagaosa}, \citenamefont {Kawasaki},
  \citenamefont {Takahashi},\ and\ \citenamefont {Tokura}}]{Mogi:2022_NP}%
  \BibitemOpen
  \bibfield  {author} {\bibinfo {author} {\bibfnamefont {M.}~\bibnamefont
  {Mogi}}, \bibinfo {author} {\bibfnamefont {Y.}~\bibnamefont {Okamura}},
  \bibinfo {author} {\bibfnamefont {M.}~\bibnamefont {Kawamura}}, \bibinfo
  {author} {\bibfnamefont {R.}~\bibnamefont {Yoshimi}}, \bibinfo {author}
  {\bibfnamefont {K.}~\bibnamefont {Yasuda}}, \bibinfo {author} {\bibfnamefont
  {A.}~\bibnamefont {Tsukazaki}}, \bibinfo {author} {\bibfnamefont {K.~S.}\
  \bibnamefont {Takahashi}}, \bibinfo {author} {\bibfnamefont {T.}~\bibnamefont
  {Morimoto}}, \bibinfo {author} {\bibfnamefont {N.}~\bibnamefont {Nagaosa}},
  \bibinfo {author} {\bibfnamefont {M.}~\bibnamefont {Kawasaki}}, \bibinfo
  {author} {\bibfnamefont {Y.}~\bibnamefont {Takahashi}}, \ and\ \bibinfo
  {author} {\bibfnamefont {Y.}~\bibnamefont {Tokura}},\ }\bibfield  {title}
  {\enquote {\bibinfo {title} {Experimental signature of the parity anomaly in
  a semi-magnetic topological insulator},}\ }\href {\doibase
  10.1038/s41567-021-01490-y} {\bibfield  {journal} {\bibinfo  {journal} {Nat.
  Phys.}\ }\textbf {\bibinfo {volume} {18}},\ \bibinfo {pages} {390--394}
  (\bibinfo {year} {2022})}\BibitemShut {NoStop}%
\bibitem [{\citenamefont {Yu}\ \emph {et~al.}(2010)\citenamefont {Yu},
  \citenamefont {Zhang}, \citenamefont {Zhang}, \citenamefont {Zhang},
  \citenamefont {Dai},\ and\ \citenamefont {Fang}}]{Yu:2010_S}%
  \BibitemOpen
  \bibfield  {author} {\bibinfo {author} {\bibfnamefont {Rui}\ \bibnamefont
  {Yu}}, \bibinfo {author} {\bibfnamefont {Wei}\ \bibnamefont {Zhang}},
  \bibinfo {author} {\bibfnamefont {Hai-Jun}\ \bibnamefont {Zhang}}, \bibinfo
  {author} {\bibfnamefont {Shou-Cheng}\ \bibnamefont {Zhang}}, \bibinfo
  {author} {\bibfnamefont {Xi}~\bibnamefont {Dai}}, \ and\ \bibinfo {author}
  {\bibfnamefont {Zhong}\ \bibnamefont {Fang}},\ }\bibfield  {title} {\enquote
  {\bibinfo {title} {Quantized anomalous {H}all effect in magnetic topological
  insulators},}\ }\href {\doibase 10.1126/science.1187485} {\bibfield
  {journal} {\bibinfo  {journal} {Science}\ }\textbf {\bibinfo {volume}
  {329}},\ \bibinfo {pages} {61--64} (\bibinfo {year} {2010})}\BibitemShut
  {NoStop}%
\bibitem [{\citenamefont {Chang}\ \emph {et~al.}(2013)\citenamefont {Chang},
  \citenamefont {Zhang}, \citenamefont {Feng}, \citenamefont {Shen},
  \citenamefont {Zhang}, \citenamefont {Guo}, \citenamefont {Li}, \citenamefont
  {Ou}, \citenamefont {Wei}, \citenamefont {Wang}, \citenamefont {Ji},
  \citenamefont {Feng}, \citenamefont {Ji}, \citenamefont {Chen}, \citenamefont
  {Jia}, \citenamefont {Dai}, \citenamefont {Fang}, \citenamefont {Zhang},
  \citenamefont {He}, \citenamefont {Wang}, \citenamefont {Lu}, \citenamefont
  {Ma},\ and\ \citenamefont {Xue}}]{Chang:2013_S}%
  \BibitemOpen
  \bibfield  {author} {\bibinfo {author} {\bibfnamefont {Cui-Zu}\ \bibnamefont
  {Chang}}, \bibinfo {author} {\bibfnamefont {Jinsong}\ \bibnamefont {Zhang}},
  \bibinfo {author} {\bibfnamefont {Xiao}\ \bibnamefont {Feng}}, \bibinfo
  {author} {\bibfnamefont {Jie}\ \bibnamefont {Shen}}, \bibinfo {author}
  {\bibfnamefont {Zuocheng}\ \bibnamefont {Zhang}}, \bibinfo {author}
  {\bibfnamefont {Minghua}\ \bibnamefont {Guo}}, \bibinfo {author}
  {\bibfnamefont {Kang}\ \bibnamefont {Li}}, \bibinfo {author} {\bibfnamefont
  {Yunbo}\ \bibnamefont {Ou}}, \bibinfo {author} {\bibfnamefont {Pang}\
  \bibnamefont {Wei}}, \bibinfo {author} {\bibfnamefont {Li-Li}\ \bibnamefont
  {Wang}}, \bibinfo {author} {\bibfnamefont {Zhong-Qing}\ \bibnamefont {Ji}},
  \bibinfo {author} {\bibfnamefont {Yang}\ \bibnamefont {Feng}}, \bibinfo
  {author} {\bibfnamefont {Shuaihua}\ \bibnamefont {Ji}}, \bibinfo {author}
  {\bibfnamefont {Xi}~\bibnamefont {Chen}}, \bibinfo {author} {\bibfnamefont
  {Jinfeng}\ \bibnamefont {Jia}}, \bibinfo {author} {\bibfnamefont
  {Xi}~\bibnamefont {Dai}}, \bibinfo {author} {\bibfnamefont {Zhong}\
  \bibnamefont {Fang}}, \bibinfo {author} {\bibfnamefont {Shou-Cheng}\
  \bibnamefont {Zhang}}, \bibinfo {author} {\bibfnamefont {Ke}~\bibnamefont
  {He}}, \bibinfo {author} {\bibfnamefont {Yayu}\ \bibnamefont {Wang}},
  \bibinfo {author} {\bibfnamefont {Li}~\bibnamefont {Lu}}, \bibinfo {author}
  {\bibfnamefont {Xu-Cun}\ \bibnamefont {Ma}}, \ and\ \bibinfo {author}
  {\bibfnamefont {Qi-Kun}\ \bibnamefont {Xue}},\ }\bibfield  {title} {\enquote
  {\bibinfo {title} {Experimental observation of the quantum anomalous {H}all
  effect in a magnetic topological insulator},}\ }\href {\doibase
  10.1126/science.1234414} {\bibfield  {journal} {\bibinfo  {journal}
  {Science}\ }\textbf {\bibinfo {volume} {340}},\ \bibinfo {pages} {167--170}
  (\bibinfo {year} {2013})}\BibitemShut {NoStop}%
\bibitem [{\citenamefont {Ke}\ \emph {et~al.}(2018)\citenamefont {Ke},
  \citenamefont {Wang},\ and\ \citenamefont {Xue}}]{Ke:2018_ARCMP}%
  \BibitemOpen
  \bibfield  {author} {\bibinfo {author} {\bibfnamefont {He}~\bibnamefont
  {Ke}}, \bibinfo {author} {\bibfnamefont {Yayu}\ \bibnamefont {Wang}}, \ and\
  \bibinfo {author} {\bibfnamefont {Qi-Kun}\ \bibnamefont {Xue}},\ }\bibfield
  {title} {\enquote {\bibinfo {title} {Topological materials: quantum anomalous
  {H}all system},}\ }\href {\doibase 10.1146/annurev-conmatphys-033117-054144}
  {\bibfield  {journal} {\bibinfo  {journal} {Annu. Rev. Cond. Mat. Phys.}\
  }\textbf {\bibinfo {volume} {9}},\ \bibinfo {pages} {329--344} (\bibinfo
  {year} {2018})}\BibitemShut {NoStop}%
\bibitem [{\citenamefont {Tokura}\ \emph {et~al.}(2019)\citenamefont {Tokura},
  \citenamefont {Yasuda},\ and\ \citenamefont {Tsukazaki}}]{Tokura:2019_NRP}%
  \BibitemOpen
  \bibfield  {author} {\bibinfo {author} {\bibfnamefont {Y.}~\bibnamefont
  {Tokura}}, \bibinfo {author} {\bibfnamefont {K.}~\bibnamefont {Yasuda}}, \
  and\ \bibinfo {author} {\bibfnamefont {A.}~\bibnamefont {Tsukazaki}},\
  }\bibfield  {title} {\enquote {\bibinfo {title} {Magnetic topological
  insulators},}\ }\href {\doibase 10.1038/s42254-018-0011-5} {\bibfield
  {journal} {\bibinfo  {journal} {Nat. Rev. Phys.}\ }\textbf {\bibinfo {volume}
  {1}},\ \bibinfo {pages} {126--143} (\bibinfo {year} {2019})}\BibitemShut
  {NoStop}%
\bibitem [{\citenamefont {Bernevig}\ \emph {et~al.}(2022)\citenamefont
  {Bernevig}, \citenamefont {Felser},\ and\ \citenamefont
  {Beidenkopf}}]{Bernevig:2022_N}%
  \BibitemOpen
  \bibfield  {author} {\bibinfo {author} {\bibfnamefont {B.~A.}\ \bibnamefont
  {Bernevig}}, \bibinfo {author} {\bibfnamefont {C.}~\bibnamefont {Felser}}, \
  and\ \bibinfo {author} {\bibfnamefont {H.}~\bibnamefont {Beidenkopf}},\
  }\bibfield  {title} {\enquote {\bibinfo {title} {Progress and prospects in
  magnetic topological materials},}\ }\href {\doibase
  10.1038/s41586-021-04105-x} {\bibfield  {journal} {\bibinfo  {journal}
  {Nature}\ }\textbf {\bibinfo {volume} {603}},\ \bibinfo {pages} {41--51}
  (\bibinfo {year} {2022})}\BibitemShut {NoStop}%
\bibitem [{\citenamefont {Chang}\ \emph {et~al.}(2023)\citenamefont {Chang},
  \citenamefont {Liu},\ and\ \citenamefont {MacDonald}}]{Chang:2023_RMP}%
  \BibitemOpen
  \bibfield  {author} {\bibinfo {author} {\bibfnamefont {Cui-Zu}\ \bibnamefont
  {Chang}}, \bibinfo {author} {\bibfnamefont {Chao-Xing}\ \bibnamefont {Liu}},
  \ and\ \bibinfo {author} {\bibfnamefont {A.~H.}\ \bibnamefont {MacDonald}},\
  }\bibfield  {title} {\enquote {\bibinfo {title} {Quantum anomalous {H}all
  effect},}\ }\href {\doibase 10.1103/RevModPhys.95.011002} {\bibfield
  {journal} {\bibinfo  {journal} {Rev. Mod. Phys.}\ }\textbf {\bibinfo {volume}
  {95}},\ \bibinfo {pages} {011002} (\bibinfo {year} {2023})}\BibitemShut
  {NoStop}%
\bibitem [{\citenamefont {Goetz}\ \emph {et~al.}(2018)\citenamefont {Goetz},
  \citenamefont {Fijalkowski}, \citenamefont {Pesel}, \citenamefont {Hartl},
  \citenamefont {Schreyeck}, \citenamefont {Winnerlein}, \citenamefont
  {Grauer}, \citenamefont {Scherer}, \citenamefont {Brunner}, \citenamefont
  {Gould}, \citenamefont {Ahlers},\ and\ \citenamefont
  {Molenkamp}}]{Goetz:2018_APL}%
  \BibitemOpen
  \bibfield  {author} {\bibinfo {author} {\bibfnamefont {M.}~\bibnamefont
  {Goetz}}, \bibinfo {author} {\bibfnamefont {K.~M.}\ \bibnamefont
  {Fijalkowski}}, \bibinfo {author} {\bibfnamefont {E.}~\bibnamefont {Pesel}},
  \bibinfo {author} {\bibfnamefont {M.}~\bibnamefont {Hartl}}, \bibinfo
  {author} {\bibfnamefont {S.}~\bibnamefont {Schreyeck}}, \bibinfo {author}
  {\bibfnamefont {M.}~\bibnamefont {Winnerlein}}, \bibinfo {author}
  {\bibfnamefont {S.}~\bibnamefont {Grauer}}, \bibinfo {author} {\bibfnamefont
  {H.}~\bibnamefont {Scherer}}, \bibinfo {author} {\bibfnamefont
  {K.}~\bibnamefont {Brunner}}, \bibinfo {author} {\bibfnamefont
  {C.}~\bibnamefont {Gould}}, \bibinfo {author} {\bibfnamefont {F.~J.}\
  \bibnamefont {Ahlers}}, \ and\ \bibinfo {author} {\bibfnamefont {L.~W.}\
  \bibnamefont {Molenkamp}},\ }\bibfield  {title} {\enquote {\bibinfo {title}
  {Precision measurement of the quantized anomalous {Hall} resistance at zero
  magnetic field},}\ }\href {\doibase 10.1063/1.5009718} {\bibfield  {journal}
  {\bibinfo  {journal} {Appl. Phys. Lett.}\ }\textbf {\bibinfo {volume}
  {112}},\ \bibinfo {pages} {072102} (\bibinfo {year} {2018})}\BibitemShut
  {NoStop}%
\bibitem [{\citenamefont {Fox}\ \emph {et~al.}(2018)\citenamefont {Fox},
  \citenamefont {Rosen}, \citenamefont {Yang}, \citenamefont {Jones},
  \citenamefont {Elmquist}, \citenamefont {Kou}, \citenamefont {Pan},
  \citenamefont {Wang},\ and\ \citenamefont {Goldhaber-Gordon}}]{Fox:2018_PRB}%
  \BibitemOpen
  \bibfield  {author} {\bibinfo {author} {\bibfnamefont {E.~J.}\ \bibnamefont
  {Fox}}, \bibinfo {author} {\bibfnamefont {I.~T.}\ \bibnamefont {Rosen}},
  \bibinfo {author} {\bibfnamefont {Yanfei}\ \bibnamefont {Yang}}, \bibinfo
  {author} {\bibfnamefont {G.~R.}\ \bibnamefont {Jones}}, \bibinfo {author}
  {\bibfnamefont {R.~E.}\ \bibnamefont {Elmquist}}, \bibinfo {author}
  {\bibfnamefont {Xufeng}\ \bibnamefont {Kou}}, \bibinfo {author}
  {\bibfnamefont {Lei}\ \bibnamefont {Pan}}, \bibinfo {author} {\bibfnamefont
  {Kang~L.}\ \bibnamefont {Wang}}, \ and\ \bibinfo {author} {\bibfnamefont
  {D.}~\bibnamefont {Goldhaber-Gordon}},\ }\bibfield  {title} {\enquote
  {\bibinfo {title} {Part-per-million quantization and current-induced
  breakdown of the quantum anomalous {Hall} effect},}\ }\href {\doibase
  10.1103/PhysRevB.98.075145} {\bibfield  {journal} {\bibinfo  {journal} {Phys.
  Rev. B}\ }\textbf {\bibinfo {volume} {98}},\ \bibinfo {pages} {075145}
  (\bibinfo {year} {2018})}\BibitemShut {NoStop}%
\bibitem [{\citenamefont {Okazaki}\ \emph {et~al.}(2022)\citenamefont
  {Okazaki}, \citenamefont {Oe}, \citenamefont {Kawamura}, \citenamefont
  {Yoshimi}, \citenamefont {Nakamura}, \citenamefont {Takada}, \citenamefont
  {Mogi}, \citenamefont {Takahashi}, \citenamefont {Tsukazaki}, \citenamefont
  {Kawasaki}, \citenamefont {Tokura},\ and\ \citenamefont
  {Kaneko}}]{Okazaki:2022_NP}%
  \BibitemOpen
  \bibfield  {author} {\bibinfo {author} {\bibfnamefont {Y.}~\bibnamefont
  {Okazaki}}, \bibinfo {author} {\bibfnamefont {T.}~\bibnamefont {Oe}},
  \bibinfo {author} {\bibfnamefont {M.}~\bibnamefont {Kawamura}}, \bibinfo
  {author} {\bibfnamefont {R.}~\bibnamefont {Yoshimi}}, \bibinfo {author}
  {\bibfnamefont {S.}~\bibnamefont {Nakamura}}, \bibinfo {author}
  {\bibfnamefont {S.}~\bibnamefont {Takada}}, \bibinfo {author} {\bibfnamefont
  {M.}~\bibnamefont {Mogi}}, \bibinfo {author} {\bibfnamefont {K.~S.}\
  \bibnamefont {Takahashi}}, \bibinfo {author} {\bibfnamefont {A.}~\bibnamefont
  {Tsukazaki}}, \bibinfo {author} {\bibfnamefont {M.}~\bibnamefont {Kawasaki}},
  \bibinfo {author} {\bibfnamefont {Y.}~\bibnamefont {Tokura}}, \ and\ \bibinfo
  {author} {\bibfnamefont {N.-H.}\ \bibnamefont {Kaneko}},\ }\bibfield  {title}
  {\enquote {\bibinfo {title} {Quantum anomalous {Hall} effect with a permanent
  magnet defines a quantum resistance standard},}\ }\href {\doibase
  10.1038/s41567-021-01424-8} {\bibfield  {journal} {\bibinfo  {journal} {Nat.
  Phys.}\ }\textbf {\bibinfo {volume} {18}},\ \bibinfo {pages} {25} (\bibinfo
  {year} {2022})}\BibitemShut {NoStop}%
\bibitem [{\citenamefont {Rodenbach}\ \emph {et~al.}(2023)\citenamefont
  {Rodenbach}, \citenamefont {{Ngoc Thanh Mai Tran}}, \citenamefont
  {Underwood}, \citenamefont {Panna}, \citenamefont {Andersen}, \citenamefont
  {Barcikowski}, \citenamefont {Payagala}, \citenamefont {Zhang}, \citenamefont
  {Tai}, \citenamefont {Wang}, \citenamefont {Elmquist}, \citenamefont
  {Jarrett}, \citenamefont {Newell}, \citenamefont {Rigosi},\ and\
  \citenamefont {Goldhaber-Gordon}}]{Rodenbach:2023_arXiv}%
  \BibitemOpen
  \bibfield  {author} {\bibinfo {author} {\bibfnamefont {L.~K.}\ \bibnamefont
  {Rodenbach}}, \bibinfo {author} {\bibnamefont {{Ngoc Thanh Mai Tran}}},
  \bibinfo {author} {\bibfnamefont {J.~M.}\ \bibnamefont {Underwood}}, \bibinfo
  {author} {\bibfnamefont {A.~R.}\ \bibnamefont {Panna}}, \bibinfo {author}
  {\bibfnamefont {M.~P.}\ \bibnamefont {Andersen}}, \bibinfo {author}
  {\bibfnamefont {Z.~S.}\ \bibnamefont {Barcikowski}}, \bibinfo {author}
  {\bibfnamefont {S.~U.}\ \bibnamefont {Payagala}}, \bibinfo {author}
  {\bibfnamefont {{Peng}}\ \bibnamefont {Zhang}}, \bibinfo {author}
  {\bibfnamefont {{Lixuan}}\ \bibnamefont {Tai}}, \bibinfo {author}
  {\bibfnamefont {{Kang}~L.}\ \bibnamefont {Wang}}, \bibinfo {author}
  {\bibfnamefont {R.~E.}\ \bibnamefont {Elmquist}}, \bibinfo {author}
  {\bibfnamefont {D.~G.}\ \bibnamefont {Jarrett}}, \bibinfo {author}
  {\bibfnamefont {D.~B.}\ \bibnamefont {Newell}}, \bibinfo {author}
  {\bibfnamefont {A.~F.}\ \bibnamefont {Rigosi}}, \ and\ \bibinfo {author}
  {\bibfnamefont {D.}~\bibnamefont {Goldhaber-Gordon}},\ }\bibfield  {title}
  {\enquote {\bibinfo {title} {Realization of the quantum ampere using the
  quantum anomalous {Hall} and {Josephson} effects},}\ }\href {\doibase
  10.48550/arXiv.2308.00200} {\bibfield  {journal} {\bibinfo  {journal}
  {arXiv:2308.00200}\ } (\bibinfo {year} {2023}),\
  10.48550/arXiv.2308.00200}\BibitemShut {NoStop}%
\bibitem [{\citenamefont {Peixoto}\ \emph {et~al.}(2020)\citenamefont
  {Peixoto}, \citenamefont {Bentmann}, \citenamefont {R\"{u}{\ss}mann},
  \citenamefont {Tcakaev}, \citenamefont {Winnerlein}, \citenamefont
  {Schreyeck}, \citenamefont {Schatz}, \citenamefont {Vidal}, \citenamefont
  {Stier}, \citenamefont {Zabolotnyy}, \citenamefont {Green}, \citenamefont
  {Min}, \citenamefont {Fornari}, \citenamefont {H.}, \citenamefont {Vasili},
  \citenamefont {Gargiani}, \citenamefont {Valvidares}, \citenamefont {Barla},
  \citenamefont {Buck}, \citenamefont {Hoesch}, \citenamefont {Diekmann},
  \citenamefont {Rohlf}, \citenamefont {Kall\"ane}, \citenamefont {Rossnagel},
  \citenamefont {Gould}, \citenamefont {Brunner}, \citenamefont {Bl\"ugel},
  \citenamefont {Hinkov}, \citenamefont {Molenkamp},\ and\ \citenamefont
  {Reinert}}]{Peixoto:2020_QM}%
  \BibitemOpen
  \bibfield  {author} {\bibinfo {author} {\bibfnamefont {T.~R.~F.}\
  \bibnamefont {Peixoto}}, \bibinfo {author} {\bibfnamefont {H.}~\bibnamefont
  {Bentmann}}, \bibinfo {author} {\bibfnamefont {P.}~\bibnamefont
  {R\"{u}{\ss}mann}}, \bibinfo {author} {\bibfnamefont {A.-V.}\ \bibnamefont
  {Tcakaev}}, \bibinfo {author} {\bibfnamefont {M.}~\bibnamefont {Winnerlein}},
  \bibinfo {author} {\bibfnamefont {S.}~\bibnamefont {Schreyeck}}, \bibinfo
  {author} {\bibfnamefont {S.}~\bibnamefont {Schatz}}, \bibinfo {author}
  {\bibfnamefont {R.~C.}\ \bibnamefont {Vidal}}, \bibinfo {author}
  {\bibfnamefont {F.}~\bibnamefont {Stier}}, \bibinfo {author} {\bibfnamefont
  {V.}~\bibnamefont {Zabolotnyy}}, \bibinfo {author} {\bibfnamefont {R.~J.}\
  \bibnamefont {Green}}, \bibinfo {author} {\bibfnamefont {Chul~Hee}\
  \bibnamefont {Min}}, \bibinfo {author} {\bibfnamefont {C.~I.}\ \bibnamefont
  {Fornari}}, \bibinfo {author} {\bibfnamefont {Maa{\ss}}\ \bibnamefont {H.}},
  \bibinfo {author} {\bibfnamefont {H.~B.}\ \bibnamefont {Vasili}}, \bibinfo
  {author} {\bibfnamefont {P.}~\bibnamefont {Gargiani}}, \bibinfo {author}
  {\bibfnamefont {M.}~\bibnamefont {Valvidares}}, \bibinfo {author}
  {\bibfnamefont {A.}~\bibnamefont {Barla}}, \bibinfo {author} {\bibfnamefont
  {J.}~\bibnamefont {Buck}}, \bibinfo {author} {\bibfnamefont {M.}~\bibnamefont
  {Hoesch}}, \bibinfo {author} {\bibfnamefont {F.}~\bibnamefont {Diekmann}},
  \bibinfo {author} {\bibfnamefont {S.}~\bibnamefont {Rohlf}}, \bibinfo
  {author} {\bibfnamefont {M.}~\bibnamefont {Kall\"ane}}, \bibinfo {author}
  {\bibfnamefont {K.}~\bibnamefont {Rossnagel}}, \bibinfo {author}
  {\bibfnamefont {Ch.}\ \bibnamefont {Gould}}, \bibinfo {author} {\bibfnamefont
  {K.}~\bibnamefont {Brunner}}, \bibinfo {author} {\bibfnamefont
  {S.}~\bibnamefont {Bl\"ugel}}, \bibinfo {author} {\bibfnamefont
  {V.}~\bibnamefont {Hinkov}}, \bibinfo {author} {\bibfnamefont {L.~W.}\
  \bibnamefont {Molenkamp}}, \ and\ \bibinfo {author} {\bibfnamefont
  {F.}~\bibnamefont {Reinert}},\ }\bibfield  {title} {\enquote {\bibinfo
  {title} {Non-local effect of impurity states on the exchange coupling
  mechanism in magnetic topological insulators},}\ }\href {\doibase
  10.1038/s41535-020-00288-0} {\bibfield  {journal} {\bibinfo  {journal} {npj
  Quant. Mater.}\ }\textbf {\bibinfo {volume} {5}},\ \bibinfo {pages} {87}
  (\bibinfo {year} {2020})}\BibitemShut {NoStop}%
\bibitem [{\citenamefont {\'Sliwa}\ \emph {et~al.}(2021)\citenamefont
  {\'Sliwa}, \citenamefont {Autieri}, \citenamefont {Majewski},\ and\
  \citenamefont {Dietl}}]{Sliwa_2021}%
  \BibitemOpen
  \bibfield  {author} {\bibinfo {author} {\bibfnamefont {C.}~\bibnamefont
  {\'Sliwa}}, \bibinfo {author} {\bibfnamefont {C.}~\bibnamefont {Autieri}},
  \bibinfo {author} {\bibfnamefont {J.~A.}\ \bibnamefont {Majewski}}, \ and\
  \bibinfo {author} {\bibfnamefont {T.}~\bibnamefont {Dietl}},\ }\bibfield
  {title} {\enquote {\bibinfo {title} {Superexchange dominates in magnetic
  topological insulators},}\ }\href {\doibase 10.1103/PhysRevB.104.L220404}
  {\bibfield  {journal} {\bibinfo  {journal} {Phys. Rev. B}\ }\textbf {\bibinfo
  {volume} {104}},\ \bibinfo {pages} {L220404} (\bibinfo {year}
  {2021})}\BibitemShut {NoStop}%
\bibitem [{\citenamefont {Watanabe}\ \emph {et~al.}(2019)\citenamefont
  {Watanabe}, \citenamefont {Yoshimi}, \citenamefont {Kawamura}, \citenamefont
  {Mogi}, \citenamefont {Tsukazaki}, \citenamefont {Yu}, \citenamefont
  {Nakajima}, \citenamefont {Takahashi}, \citenamefont {Kawasaki},\ and\
  \citenamefont {Tokura}}]{Watanabe:2019_APL}%
  \BibitemOpen
  \bibfield  {author} {\bibinfo {author} {\bibfnamefont {R.}~\bibnamefont
  {Watanabe}}, \bibinfo {author} {\bibfnamefont {R.}~\bibnamefont {Yoshimi}},
  \bibinfo {author} {\bibfnamefont {M.}~\bibnamefont {Kawamura}}, \bibinfo
  {author} {\bibfnamefont {M.}~\bibnamefont {Mogi}}, \bibinfo {author}
  {\bibfnamefont {A.}~\bibnamefont {Tsukazaki}}, \bibinfo {author}
  {\bibfnamefont {X.~Z.}\ \bibnamefont {Yu}}, \bibinfo {author} {\bibfnamefont
  {K.}~\bibnamefont {Nakajima}}, \bibinfo {author} {\bibfnamefont {K.~S.}\
  \bibnamefont {Takahashi}}, \bibinfo {author} {\bibfnamefont {M.}~\bibnamefont
  {Kawasaki}}, \ and\ \bibinfo {author} {\bibfnamefont {Y.}~\bibnamefont
  {Tokura}},\ }\bibfield  {title} {\enquote {\bibinfo {title} {Quantum
  anomalous {H}all effect driven by magnetic proximity coupling in
  all-telluride based heterostructure},}\ }\href {\doibase 10.1063/1.5111891}
  {\bibfield  {journal} {\bibinfo  {journal} {Appl. Phys. Lett.}\ }\textbf
  {\bibinfo {volume} {115}},\ \bibinfo {pages} {102403} (\bibinfo {year}
  {2019})}\BibitemShut {NoStop}%
\bibitem [{\citenamefont {Escribano}\ \emph {et~al.}(2022)\citenamefont
  {Escribano}, \citenamefont {Maiani}, \citenamefont {Leijnse}, \citenamefont
  {Flensberg}, \citenamefont {Oreg}, \citenamefont {{Levy Yeyati}},
  \citenamefont {Prada},\ and\ \citenamefont {{Seoane
  Souto}}}]{Escribano:2022_QM}%
  \BibitemOpen
  \bibfield  {author} {\bibinfo {author} {\bibfnamefont {S.~D.}\ \bibnamefont
  {Escribano}}, \bibinfo {author} {\bibfnamefont {A.}~\bibnamefont {Maiani}},
  \bibinfo {author} {\bibfnamefont {M.}~\bibnamefont {Leijnse}}, \bibinfo
  {author} {\bibfnamefont {K.}~\bibnamefont {Flensberg}}, \bibinfo {author}
  {\bibfnamefont {Y.}~\bibnamefont {Oreg}}, \bibinfo {author} {\bibfnamefont
  {A.}~\bibnamefont {{Levy Yeyati}}}, \bibinfo {author} {\bibfnamefont
  {E.}~\bibnamefont {Prada}}, \ and\ \bibinfo {author} {\bibfnamefont
  {R.}~\bibnamefont {{Seoane Souto}}},\ }\bibfield  {title} {\enquote {\bibinfo
  {title} {Semiconductor-ferromagnet-superconductor planar heterostructures for
  {1D} topological superconductivity},}\ }\href {\doibase
  10.1038/s41535-022-00489-9} {\bibfield  {journal} {\bibinfo  {journal} {npj
  Quant. Mater.}\ }\textbf {\bibinfo {volume} {7}},\ \bibinfo {pages} {81}
  (\bibinfo {year} {2022})}\BibitemShut {NoStop}%
\bibitem [{\citenamefont {Goodenough}(1963)}]{Goodenough_1963}%
  \BibitemOpen
  \bibfield  {author} {\bibinfo {author} {\bibfnamefont {John~B.}\ \bibnamefont
  {Goodenough}},\ }\href@noop {} {\emph {\bibinfo {title} {Magnetism and the
  Chemical Bond}}},\ Interscience Monographs on Chemistry\ (\bibinfo
  {publisher} {Interscience Publishers},\ \bibinfo {year} {1963})\BibitemShut
  {NoStop}%
\bibitem [{\citenamefont {Blinowski}\ \emph {et~al.}(1996)\citenamefont
  {Blinowski}, \citenamefont {Kacman},\ and\ \citenamefont
  {Majewski}}]{Blinowski_1996}%
  \BibitemOpen
  \bibfield  {author} {\bibinfo {author} {\bibfnamefont {J.}~\bibnamefont
  {Blinowski}}, \bibinfo {author} {\bibfnamefont {P.}~\bibnamefont {Kacman}}, \
  and\ \bibinfo {author} {\bibfnamefont {J.~A.}\ \bibnamefont {Majewski}},\
  }\bibfield  {title} {\enquote {\bibinfo {title} {Ferromagnetic superexchange
  in {Cr}-based diluted magnetic semiconductors},}\ }\href {\doibase
  10.1103/PhysRevB.53.9524} {\bibfield  {journal} {\bibinfo  {journal} {Phys.
  Rev. B}\ }\textbf {\bibinfo {volume} {53}},\ \bibinfo {pages} {9524--9527}
  (\bibinfo {year} {1996})}\BibitemShut {NoStop}%
\bibitem [{\citenamefont {Kuroda}\ \emph
  {et~al.}(2007{\natexlab{a}})\citenamefont {Kuroda}, \citenamefont
  {Nishizawa}, \citenamefont {Takita}, \citenamefont {Mitome}, \citenamefont
  {Bando}, \citenamefont {Osuch},\ and\ \citenamefont
  {Dietl}}]{Kuroda:2007_NM}%
  \BibitemOpen
  \bibfield  {author} {\bibinfo {author} {\bibfnamefont {S.}~\bibnamefont
  {Kuroda}}, \bibinfo {author} {\bibfnamefont {N.}~\bibnamefont {Nishizawa}},
  \bibinfo {author} {\bibfnamefont {K.}~\bibnamefont {Takita}}, \bibinfo
  {author} {\bibfnamefont {M.}~\bibnamefont {Mitome}}, \bibinfo {author}
  {\bibfnamefont {Y.}~\bibnamefont {Bando}}, \bibinfo {author} {\bibfnamefont
  {K.}~\bibnamefont {Osuch}}, \ and\ \bibinfo {author} {\bibfnamefont
  {T.}~\bibnamefont {Dietl}},\ }\bibfield  {title} {\enquote {\bibinfo {title}
  {Origin and control of high-temperature ferromagnetism in semiconductors},}\
  }\href {\doibase 10.1038/nmat1910} {\bibfield  {journal} {\bibinfo  {journal}
  {Nat. Mater.}\ }\textbf {\bibinfo {volume} {6}},\ \bibinfo {pages} {440}
  (\bibinfo {year} {2007}{\natexlab{a}})}\BibitemShut {NoStop}%
\bibitem [{\citenamefont {Sarigiannidou}\ \emph {et~al.}(2006)\citenamefont
  {Sarigiannidou}, \citenamefont {Wilhelm}, \citenamefont {Monroy},
  \citenamefont {Galera}, \citenamefont {Bellet-Amalric}, \citenamefont
  {Rogalev}, \citenamefont {Goulon}, \citenamefont {Cibert},\ and\
  \citenamefont {Mariette}}]{Sarigiannidou:2006_PRB}%
  \BibitemOpen
  \bibfield  {author} {\bibinfo {author} {\bibfnamefont {E.}~\bibnamefont
  {Sarigiannidou}}, \bibinfo {author} {\bibfnamefont {F.}~\bibnamefont
  {Wilhelm}}, \bibinfo {author} {\bibfnamefont {E.}~\bibnamefont {Monroy}},
  \bibinfo {author} {\bibfnamefont {R.~M.}\ \bibnamefont {Galera}}, \bibinfo
  {author} {\bibfnamefont {E.}~\bibnamefont {Bellet-Amalric}}, \bibinfo
  {author} {\bibfnamefont {A.}~\bibnamefont {Rogalev}}, \bibinfo {author}
  {\bibfnamefont {J.}~\bibnamefont {Goulon}}, \bibinfo {author} {\bibfnamefont
  {J.}~\bibnamefont {Cibert}}, \ and\ \bibinfo {author} {\bibfnamefont
  {H.}~\bibnamefont {Mariette}},\ }\bibfield  {title} {\enquote {\bibinfo
  {title} {Intrinsic ferromagnetism in wurtzite {(Ga,Mn)N} semiconductor},}\
  }\href {\doibase 10.1103/PhysRevB.74.041306} {\bibfield  {journal} {\bibinfo
  {journal} {Phys. Rev. B}\ }\textbf {\bibinfo {volume} {74}},\ \bibinfo
  {pages} {041306} (\bibinfo {year} {2006})}\BibitemShut {NoStop}%
\bibitem [{\citenamefont {Sawicki}\ \emph {et~al.}(2012)\citenamefont
  {Sawicki}, \citenamefont {Devillers}, \citenamefont {Ga\l\k{e}ski},
  \citenamefont {Simserides}, \citenamefont {Dobkowska}, \citenamefont {Faina},
  \citenamefont {Grois}, \citenamefont {Navarro-Quezada}, \citenamefont
  {Trohidou}, \citenamefont {Majewski}, \citenamefont {Dietl},\ and\
  \citenamefont {Bonanni}}]{Sawicki:2012_PRB}%
  \BibitemOpen
  \bibfield  {author} {\bibinfo {author} {\bibfnamefont {M.}~\bibnamefont
  {Sawicki}}, \bibinfo {author} {\bibfnamefont {T.}~\bibnamefont {Devillers}},
  \bibinfo {author} {\bibfnamefont {S.}~\bibnamefont {Ga\l\k{e}ski}}, \bibinfo
  {author} {\bibfnamefont {C.}~\bibnamefont {Simserides}}, \bibinfo {author}
  {\bibfnamefont {S.}~\bibnamefont {Dobkowska}}, \bibinfo {author}
  {\bibfnamefont {B.}~\bibnamefont {Faina}}, \bibinfo {author} {\bibfnamefont
  {A.}~\bibnamefont {Grois}}, \bibinfo {author} {\bibfnamefont
  {A.}~\bibnamefont {Navarro-Quezada}}, \bibinfo {author} {\bibfnamefont
  {K.~N.}\ \bibnamefont {Trohidou}}, \bibinfo {author} {\bibfnamefont {J.~A.}\
  \bibnamefont {Majewski}}, \bibinfo {author} {\bibfnamefont {T.}~\bibnamefont
  {Dietl}}, \ and\ \bibinfo {author} {\bibfnamefont {A.}~\bibnamefont
  {Bonanni}},\ }\bibfield  {title} {\enquote {\bibinfo {title} {Origin of
  low-temperature magnetic ordering in
  {$\mathrm{Ga}_{1-x}\mathrm{Mn}_{x}\mathrm{N}$}},}\ }\href {\doibase
  10.1103/PhysRevB.85.205204} {\bibfield  {journal} {\bibinfo  {journal} {Phys.
  Rev. B}\ }\textbf {\bibinfo {volume} {85}},\ \bibinfo {pages} {205204}
  (\bibinfo {year} {2012})}\BibitemShut {NoStop}%
\bibitem [{\citenamefont {Stefanowicz}\ \emph {et~al.}(2013)\citenamefont
  {Stefanowicz}, \citenamefont {Kunert}, \citenamefont {Simserides},
  \citenamefont {Majewski}, \citenamefont {Stefanowicz}, \citenamefont {Kruse},
  \citenamefont {Figge}, \citenamefont {Li}, \citenamefont {Jakie\l{}a},
  \citenamefont {Trohidou}, \citenamefont {Bonanni}, \citenamefont {Hommel},
  \citenamefont {Sawicki},\ and\ \citenamefont {Dietl}}]{Stefanowicz:2013_PRB}%
  \BibitemOpen
  \bibfield  {author} {\bibinfo {author} {\bibfnamefont {S.}~\bibnamefont
  {Stefanowicz}}, \bibinfo {author} {\bibfnamefont {G.}~\bibnamefont {Kunert}},
  \bibinfo {author} {\bibfnamefont {C.}~\bibnamefont {Simserides}}, \bibinfo
  {author} {\bibfnamefont {J.~A.}\ \bibnamefont {Majewski}}, \bibinfo {author}
  {\bibfnamefont {W.}~\bibnamefont {Stefanowicz}}, \bibinfo {author}
  {\bibfnamefont {C.}~\bibnamefont {Kruse}}, \bibinfo {author} {\bibfnamefont
  {S.}~\bibnamefont {Figge}}, \bibinfo {author} {\bibfnamefont {Tian}\
  \bibnamefont {Li}}, \bibinfo {author} {\bibfnamefont {R.}~\bibnamefont
  {Jakie\l{}a}}, \bibinfo {author} {\bibfnamefont {K.~N.}\ \bibnamefont
  {Trohidou}}, \bibinfo {author} {\bibfnamefont {A.}~\bibnamefont {Bonanni}},
  \bibinfo {author} {\bibfnamefont {D.}~\bibnamefont {Hommel}}, \bibinfo
  {author} {\bibfnamefont {M.}~\bibnamefont {Sawicki}}, \ and\ \bibinfo
  {author} {\bibfnamefont {T.}~\bibnamefont {Dietl}},\ }\bibfield  {title}
  {\enquote {\bibinfo {title} {Phase diagram and critical behavior of the
  random ferromagnet {$\mathrm{Ga}_{1-x}\mathrm{Mn}_{x}\mathrm{N}$}},}\ }\href
  {\doibase 10.1103/PhysRevB.88.081201} {\bibfield  {journal} {\bibinfo
  {journal} {Phys. Rev. B}\ }\textbf {\bibinfo {volume} {88}},\ \bibinfo
  {pages} {081201(R)} (\bibinfo {year} {2013})}\BibitemShut {NoStop}%
\bibitem [{\citenamefont {Simserides}\ \emph {et~al.}(2014)\citenamefont
  {Simserides}, \citenamefont {Majewski}, \citenamefont {Trohidou},\ and\
  \citenamefont {Dietl}}]{Simserides_2014}%
  \BibitemOpen
  \bibfield  {author} {\bibinfo {author} {\bibfnamefont {C.}~\bibnamefont
  {Simserides}}, \bibinfo {author} {\bibfnamefont {J.A.}\ \bibnamefont
  {Majewski}}, \bibinfo {author} {\bibfnamefont {K.N.}\ \bibnamefont
  {Trohidou}}, \ and\ \bibinfo {author} {\bibfnamefont {T.}~\bibnamefont
  {Dietl}},\ }\bibfield  {title} {\enquote {\bibinfo {title} {Theory of
  ferromagnetism driven by superexchange in dilute magnetic semiconductors},}\
  }\href {\doibase 10.1051/epjconf/20147501003} {\bibfield  {journal} {\bibinfo
   {journal} {EPJ Web of Conferences}\ }\textbf {\bibinfo {volume} {75}},\
  \bibinfo {pages} {01003} (\bibinfo {year} {2014})}\BibitemShut {NoStop}%
\bibitem [{\citenamefont {Bloembergen}\ and\ \citenamefont
  {Rowland}(1955)}]{Bloembergen:1955_PR}%
  \BibitemOpen
  \bibfield  {author} {\bibinfo {author} {\bibfnamefont {N.}~\bibnamefont
  {Bloembergen}}\ and\ \bibinfo {author} {\bibfnamefont {T.~J.}\ \bibnamefont
  {Rowland}},\ }\bibfield  {title} {\enquote {\bibinfo {title} {Nuclear spin
  exchange in solids: {${\mathrm{Tl}}^{203}$ and ${\mathrm{Tl}}^{205}$}
  magnetic resonance in thallium and thallic oxide},}\ }\href {\doibase
  10.1103/PhysRev.97.1679} {\bibfield  {journal} {\bibinfo  {journal} {Phys.
  Rev.}\ }\textbf {\bibinfo {volume} {97}},\ \bibinfo {pages} {1679--1698}
  (\bibinfo {year} {1955})}\BibitemShut {NoStop}%
\bibitem [{\citenamefont {Lewiner}\ \emph {et~al.}(1980)\citenamefont
  {Lewiner}, \citenamefont {Gaj},\ and\ \citenamefont
  {Bastard}}]{Lewiner:1980_JPCol}%
  \BibitemOpen
  \bibfield  {author} {\bibinfo {author} {\bibfnamefont {C.}~\bibnamefont
  {Lewiner}}, \bibinfo {author} {\bibfnamefont {J.}~\bibnamefont {Gaj}}, \ and\
  \bibinfo {author} {\bibfnamefont {G.}~\bibnamefont {Bastard}},\ }\bibfield
  {title} {\enquote {\bibinfo {title} {Indirect exchange interaction in
  {$\mathrm{Hg}_{1-x}\mathrm{Mn}_x\mathrm{Te}$} and
  {$\mathrm{Cd}_{1-x}\mathrm{Mn}_x\mathrm{Te}$} alloys},}\ }\href {\doibase
  10.1051/jphyscol:1980549} {\bibfield  {journal} {\bibinfo  {journal} {J.
  Phys. Colloq. (Paris)}\ }\textbf {\bibinfo {volume} {41 (C5)}},\ \bibinfo
  {pages} {289--292} (\bibinfo {year} {1980})}\BibitemShut {NoStop}%
\bibitem [{\citenamefont {Larson}\ \emph {et~al.}(1988)\citenamefont {Larson},
  \citenamefont {Hass}, \citenamefont {Ehrenreich},\ and\ \citenamefont
  {Carlsson}}]{Larson_1988}%
  \BibitemOpen
  \bibfield  {author} {\bibinfo {author} {\bibfnamefont {B.~E.}\ \bibnamefont
  {Larson}}, \bibinfo {author} {\bibfnamefont {K.~C.}\ \bibnamefont {Hass}},
  \bibinfo {author} {\bibfnamefont {H.}~\bibnamefont {Ehrenreich}}, \ and\
  \bibinfo {author} {\bibfnamefont {A.~E.}\ \bibnamefont {Carlsson}},\
  }\bibfield  {title} {\enquote {\bibinfo {title} {Theory of exchange
  interactions and chemical trends in diluted magnetic semiconductors},}\
  }\href {\doibase 10.1103/PhysRevB.37.4137} {\bibfield  {journal} {\bibinfo
  {journal} {Phys. Rev. B}\ }\textbf {\bibinfo {volume} {37}},\ \bibinfo
  {pages} {4137--4154} (\bibinfo {year} {1988})}\BibitemShut {NoStop}%
\bibitem [{\citenamefont {Dietl}\ \emph {et~al.}(2001)\citenamefont {Dietl},
  \citenamefont {Ohno},\ and\ \citenamefont {Matsukura}}]{Dietl:2001_PRB}%
  \BibitemOpen
  \bibfield  {author} {\bibinfo {author} {\bibfnamefont {T.}~\bibnamefont
  {Dietl}}, \bibinfo {author} {\bibfnamefont {H.}~\bibnamefont {Ohno}}, \ and\
  \bibinfo {author} {\bibfnamefont {F.}~\bibnamefont {Matsukura}},\ }\bibfield
  {title} {\enquote {\bibinfo {title} {Hole-mediated ferromagnetism in
  tetrahedrally coordinated semiconductors},}\ }\href {\doibase
  10.1103/PhysRevB.63.195205} {\bibfield  {journal} {\bibinfo  {journal} {Phys.
  Rev. B}\ }\textbf {\bibinfo {volume} {63}},\ \bibinfo {pages} {195205}
  (\bibinfo {year} {2001})}\BibitemShut {NoStop}%
\bibitem [{\citenamefont {Kacman}(2001)}]{Kacman:2001_SST}%
  \BibitemOpen
  \bibfield  {author} {\bibinfo {author} {\bibfnamefont {P.}~\bibnamefont
  {Kacman}},\ }\bibfield  {title} {\enquote {\bibinfo {title} {Spin
  interactions in diluted magnetic semiconductors and magnetic semiconductor
  structures},}\ }\href {\doibase 10.1088/0268-1242/16/4/201} {\bibfield
  {journal} {\bibinfo  {journal} {Semicon. Sci. Technol.}\ }\textbf {\bibinfo
  {volume} {16}},\ \bibinfo {pages} {R25--R39} (\bibinfo {year}
  {2001})}\BibitemShut {NoStop}%
\bibitem [{\citenamefont {\'Sliwa}\ and\ \citenamefont
  {Dietl}(2018)}]{Sliwa:2018_PRB}%
  \BibitemOpen
  \bibfield  {author} {\bibinfo {author} {\bibfnamefont {C.}~\bibnamefont
  {\'Sliwa}}\ and\ \bibinfo {author} {\bibfnamefont {T.}~\bibnamefont
  {Dietl}},\ }\bibfield  {title} {\enquote {\bibinfo {title} {Thermodynamic
  perturbation theory for noninteracting quantum particles with application to
  spin-spin interactions in solids},}\ }\href {\doibase
  10.1103/PhysRevB.98.035105} {\bibfield  {journal} {\bibinfo  {journal} {Phys.
  Rev. B}\ }\textbf {\bibinfo {volume} {98}},\ \bibinfo {pages} {035105}
  (\bibinfo {year} {2018})}\BibitemShut {NoStop}%
\bibitem [{\citenamefont {Korenblit}\ \emph {et~al.}(1973)\citenamefont
  {Korenblit}, \citenamefont {Shender},\ and\ \citenamefont
  {Shklovsky}}]{Korenblit_1973}%
  \BibitemOpen
  \bibfield  {author} {\bibinfo {author} {\bibfnamefont {I.~Ya.}\ \bibnamefont
  {Korenblit}}, \bibinfo {author} {\bibfnamefont {E.~F.}\ \bibnamefont
  {Shender}}, \ and\ \bibinfo {author} {\bibfnamefont {B.~I.}\ \bibnamefont
  {Shklovsky}},\ }\bibfield  {title} {\enquote {\bibinfo {title} {Percolation
  approach to the phase transition in very dilute ferromagnetic alloys},}\
  }\href {\doibase 10.1016/0375-9601(73)90219-3} {\bibfield  {journal}
  {\bibinfo  {journal} {Phys. Lett. A}\ }\textbf {\bibinfo {volume} {46}},\
  \bibinfo {pages} {275--276} (\bibinfo {year} {1973})}\BibitemShut {NoStop}%
\bibitem [{\citenamefont {Sato}\ \emph {et~al.}(2010)\citenamefont {Sato},
  \citenamefont {Bergqvist}, \citenamefont {Kudrnovsk\'y}, \citenamefont
  {Dederichs}, \citenamefont {Eriksson}, \citenamefont {Turek}, \citenamefont
  {Sanyal}, \citenamefont {Bouzerar}, \citenamefont {Katayama-Yoshida},
  \citenamefont {Dinh}, \citenamefont {Fukushima}, \citenamefont {Kizaki},\
  and\ \citenamefont {Zeller}}]{Sato:2010_RMP}%
  \BibitemOpen
  \bibfield  {author} {\bibinfo {author} {\bibfnamefont {K.}~\bibnamefont
  {Sato}}, \bibinfo {author} {\bibfnamefont {L.}~\bibnamefont {Bergqvist}},
  \bibinfo {author} {\bibfnamefont {J.}~\bibnamefont {Kudrnovsk\'y}}, \bibinfo
  {author} {\bibfnamefont {P.~H.}\ \bibnamefont {Dederichs}}, \bibinfo {author}
  {\bibfnamefont {O.}~\bibnamefont {Eriksson}}, \bibinfo {author}
  {\bibfnamefont {I.}~\bibnamefont {Turek}}, \bibinfo {author} {\bibfnamefont
  {B.}~\bibnamefont {Sanyal}}, \bibinfo {author} {\bibfnamefont
  {G.}~\bibnamefont {Bouzerar}}, \bibinfo {author} {\bibfnamefont
  {H.}~\bibnamefont {Katayama-Yoshida}}, \bibinfo {author} {\bibfnamefont
  {V.~A.}\ \bibnamefont {Dinh}}, \bibinfo {author} {\bibfnamefont
  {T.}~\bibnamefont {Fukushima}}, \bibinfo {author} {\bibfnamefont
  {H.}~\bibnamefont {Kizaki}}, \ and\ \bibinfo {author} {\bibfnamefont
  {R.}~\bibnamefont {Zeller}},\ }\bibfield  {title} {\enquote {\bibinfo {title}
  {First-principles theory of dilute magnetic semiconductors},}\ }\href
  {\doibase 10.1103/RevModPhys.82.1633} {\bibfield  {journal} {\bibinfo
  {journal} {Rev. Mod. Phys.}\ }\textbf {\bibinfo {volume} {82}},\ \bibinfo
  {pages} {1633--1690} (\bibinfo {year} {2010})}\BibitemShut {NoStop}%
\bibitem [{\citenamefont {Bonanni}\ \emph {et~al.}(2021)\citenamefont
  {Bonanni}, \citenamefont {Dietl},\ and\ \citenamefont
  {Ohno}}]{Bonanni:2021_HB}%
  \BibitemOpen
  \bibfield  {author} {\bibinfo {author} {\bibfnamefont {A.}~\bibnamefont
  {Bonanni}}, \bibinfo {author} {\bibfnamefont {T.}~\bibnamefont {Dietl}}, \
  and\ \bibinfo {author} {\bibfnamefont {H.}~\bibnamefont {Ohno}},\ }\bibfield
  {title} {\enquote {\bibinfo {title} {Dilute magnetic materials},}\ }in\
  \href@noop {} {\emph {\bibinfo {booktitle} {Handbook of Magnetism and
  Magnetic Materials}}},\ \bibinfo {editor} {edited by\ \bibinfo {editor}
  {\bibfnamefont {M.}~\bibnamefont {Coey}}\ and\ \bibinfo {editor}
  {\bibfnamefont {S.}~\bibnamefont {Parkin}}}\ (\bibinfo  {publisher}
  {Springer, Berlin},\ \bibinfo {year} {2021})\BibitemShut {NoStop}%
\bibitem [{\citenamefont {Dietl}(2008)}]{Dietl:2008_PRB}%
  \BibitemOpen
  \bibfield  {author} {\bibinfo {author} {\bibfnamefont {T.}~\bibnamefont
  {Dietl}},\ }\bibfield  {title} {\enquote {\bibinfo {title} {Hole states in
  wide band-gap diluted magnetic semiconductors and oxides},}\ }\href {\doibase
  10.1103/PhysRevB.77.085208} {\bibfield  {journal} {\bibinfo  {journal} {Phys.
  Rev. B}\ }\textbf {\bibinfo {volume} {77}},\ \bibinfo {pages} {085208}
  (\bibinfo {year} {2008})}\BibitemShut {NoStop}%
\bibitem [{\citenamefont {Mac}\ \emph {et~al.}(1996)\citenamefont {Mac},
  \citenamefont {Twardowski},\ and\ \citenamefont {Demianiuk}}]{Mac:1996_PRB}%
  \BibitemOpen
  \bibfield  {author} {\bibinfo {author} {\bibfnamefont {W.}~\bibnamefont
  {Mac}}, \bibinfo {author} {\bibfnamefont {A.}~\bibnamefont {Twardowski}}, \
  and\ \bibinfo {author} {\bibfnamefont {M.}~\bibnamefont {Demianiuk}},\
  }\bibfield  {title} {\enquote {\bibinfo {title} {s,p-d exchange interaction
  in {Cr}-based diluted magnetic semiconductors},}\ }\href {\doibase
  10.1103/PhysRevB.54.5528} {\bibfield  {journal} {\bibinfo  {journal} {Phys.
  Rev. B}\ }\textbf {\bibinfo {volume} {54}},\ \bibinfo {pages} {5528--5535}
  (\bibinfo {year} {1996})}\BibitemShut {NoStop}%
\bibitem [{\citenamefont {Suffczy\'{n}ski}\ \emph {et~al.}(2011)\citenamefont
  {Suffczy\'{n}ski}, \citenamefont {Grois}, \citenamefont {Pacuski},
  \citenamefont {Golnik}, \citenamefont {Gaj}, \citenamefont {Navarro-Quezada},
  \citenamefont {Faina}, \citenamefont {Devillers},\ and\ \citenamefont
  {Bonanni}}]{Suffczynski:2011_PRB}%
  \BibitemOpen
  \bibfield  {author} {\bibinfo {author} {\bibfnamefont {J.}~\bibnamefont
  {Suffczy\'{n}ski}}, \bibinfo {author} {\bibfnamefont {A.}~\bibnamefont
  {Grois}}, \bibinfo {author} {\bibfnamefont {W.}~\bibnamefont {Pacuski}},
  \bibinfo {author} {\bibfnamefont {A.}~\bibnamefont {Golnik}}, \bibinfo
  {author} {\bibfnamefont {J.~A.}\ \bibnamefont {Gaj}}, \bibinfo {author}
  {\bibfnamefont {A.}~\bibnamefont {Navarro-Quezada}}, \bibinfo {author}
  {\bibfnamefont {B.}~\bibnamefont {Faina}}, \bibinfo {author} {\bibfnamefont
  {T.}~\bibnamefont {Devillers}}, \ and\ \bibinfo {author} {\bibfnamefont
  {A.}~\bibnamefont {Bonanni}},\ }\bibfield  {title} {\enquote {\bibinfo
  {title} {Effects of $s$,$p$-$d$ and $s$-$p$ exchange interactions probed by
  exciton magnetospectroscopy in {(Ga,Mn)N}},}\ }\href {\doibase
  10.1103/PhysRevB.83.094421} {\bibfield  {journal} {\bibinfo  {journal} {Phys.
  Rev. B}\ }\textbf {\bibinfo {volume} {83}},\ \bibinfo {pages} {094421}
  (\bibinfo {year} {2011})}\BibitemShut {NoStop}%
\bibitem [{\citenamefont {Liu}\ \emph {et~al.}(2008{\natexlab{a}})\citenamefont
  {Liu}, \citenamefont {Qi}, \citenamefont {Dai}, \citenamefont {Fang},\ and\
  \citenamefont {Zhang}}]{Liu:2008_PRL}%
  \BibitemOpen
  \bibfield  {author} {\bibinfo {author} {\bibfnamefont {Chao-Xing}\
  \bibnamefont {Liu}}, \bibinfo {author} {\bibfnamefont {Xiao-Liang}\
  \bibnamefont {Qi}}, \bibinfo {author} {\bibfnamefont {Xi}~\bibnamefont
  {Dai}}, \bibinfo {author} {\bibfnamefont {Zhong}\ \bibnamefont {Fang}}, \
  and\ \bibinfo {author} {\bibfnamefont {Shou-Cheng}\ \bibnamefont {Zhang}},\
  }\bibfield  {title} {\enquote {\bibinfo {title} {{Quantum Anomalous Hall
  Effect in ${\mathrm{Hg}}_{1\ensuremath{-}y}{\mathrm{Mn}}_{y}\mathrm{Te}$
  Quantum Wells}},}\ }\href {\doibase 10.1103/PhysRevLett.101.146802}
  {\bibfield  {journal} {\bibinfo  {journal} {Phys. Rev. Lett.}\ }\textbf
  {\bibinfo {volume} {101}},\ \bibinfo {pages} {146802} (\bibinfo {year}
  {2008}{\natexlab{a}})}\BibitemShut {NoStop}%
\bibitem [{\citenamefont {Kuroda}\ \emph
  {et~al.}(2007{\natexlab{b}})\citenamefont {Kuroda}, \citenamefont
  {Nishizawa}, \citenamefont {Takita}, \citenamefont {Mitome}, \citenamefont
  {Bando}, \citenamefont {Osuch},\ and\ \citenamefont {Dietl}}]{Kuroda_2007}%
  \BibitemOpen
  \bibfield  {author} {\bibinfo {author} {\bibfnamefont {S.}~\bibnamefont
  {Kuroda}}, \bibinfo {author} {\bibfnamefont {N.}~\bibnamefont {Nishizawa}},
  \bibinfo {author} {\bibfnamefont {K.}~\bibnamefont {Takita}}, \bibinfo
  {author} {\bibfnamefont {M.}~\bibnamefont {Mitome}}, \bibinfo {author}
  {\bibfnamefont {Y.}~\bibnamefont {Bando}}, \bibinfo {author} {\bibfnamefont
  {K.}~\bibnamefont {Osuch}}, \ and\ \bibinfo {author} {\bibfnamefont
  {T.}~\bibnamefont {Dietl}},\ }\bibfield  {title} {\enquote {\bibinfo {title}
  {Origin and control of high-temperature ferromagnetism in semiconductors},}\
  }\href {\doibase 10.1038/nmat1910} {\bibfield  {journal} {\bibinfo  {journal}
  {Nat. Mater.}\ }\textbf {\bibinfo {volume} {6}},\ \bibinfo {pages} {440--446}
  (\bibinfo {year} {2007}{\natexlab{b}})}\BibitemShut {NoStop}%
\bibitem [{\citenamefont {Graf}\ \emph {et~al.}(2003)\citenamefont {Graf},
  \citenamefont {Goennenwein},\ and\ \citenamefont {Brandt}}]{Graf_2003}%
  \BibitemOpen
  \bibfield  {author} {\bibinfo {author} {\bibfnamefont {Tobias}\ \bibnamefont
  {Graf}}, \bibinfo {author} {\bibfnamefont {Sebastian T.~B.}\ \bibnamefont
  {Goennenwein}}, \ and\ \bibinfo {author} {\bibfnamefont {Martin~S.}\
  \bibnamefont {Brandt}},\ }\bibfield  {title} {\enquote {\bibinfo {title}
  {Prospects for carrier-mediated ferromagnetism in {GaN}},}\ }\href {\doibase
  10.1002/pssb.200301880} {\bibfield  {journal} {\bibinfo  {journal} {physica
  status solidi (b)}\ }\textbf {\bibinfo {volume} {239}},\ \bibinfo {pages}
  {277--290} (\bibinfo {year} {2003})}\BibitemShut {NoStop}%
\bibitem [{\citenamefont {Han}\ \emph {et~al.}(2005)\citenamefont {Han},
  \citenamefont {Wessels},\ and\ \citenamefont {Ulmer}}]{Han_2005}%
  \BibitemOpen
  \bibfield  {author} {\bibinfo {author} {\bibfnamefont {B.}~\bibnamefont
  {Han}}, \bibinfo {author} {\bibfnamefont {B.~W.}\ \bibnamefont {Wessels}}, \
  and\ \bibinfo {author} {\bibfnamefont {M.~P.}\ \bibnamefont {Ulmer}},\
  }\bibfield  {title} {\enquote {\bibinfo {title} {Optical investigation of
  electronic states of {$\mathrm{Mn}^{4+}$} ions in {$p$-type} {GaN}},}\ }\href
  {\doibase 10.1063/1.1853525} {\bibfield  {journal} {\bibinfo  {journal}
  {Appl. Phys. Lett.}\ }\textbf {\bibinfo {volume} {86}},\ \bibinfo {pages}
  {042505} (\bibinfo {year} {2005})}\BibitemShut {NoStop}%
\bibitem [{\citenamefont {Hwang}\ \emph {et~al.}(2005)\citenamefont {Hwang},
  \citenamefont {Ishida}, \citenamefont {Kobayashi}, \citenamefont {Hirata},
  \citenamefont {Takubo}, \citenamefont {Mizokawa}, \citenamefont {Fujimori},
  \citenamefont {Okamoto}, \citenamefont {Mamiya}, \citenamefont {Saito},
  \citenamefont {Muramatsu}, \citenamefont {Ott}, \citenamefont {Tanaka},
  \citenamefont {Kondo},\ and\ \citenamefont {Munekata}}]{Hwang_2005}%
  \BibitemOpen
  \bibfield  {author} {\bibinfo {author} {\bibfnamefont {J.~I.}\ \bibnamefont
  {Hwang}}, \bibinfo {author} {\bibfnamefont {Y.}~\bibnamefont {Ishida}},
  \bibinfo {author} {\bibfnamefont {M.}~\bibnamefont {Kobayashi}}, \bibinfo
  {author} {\bibfnamefont {H.}~\bibnamefont {Hirata}}, \bibinfo {author}
  {\bibfnamefont {K.}~\bibnamefont {Takubo}}, \bibinfo {author} {\bibfnamefont
  {T.}~\bibnamefont {Mizokawa}}, \bibinfo {author} {\bibfnamefont
  {A.}~\bibnamefont {Fujimori}}, \bibinfo {author} {\bibfnamefont
  {J.}~\bibnamefont {Okamoto}}, \bibinfo {author} {\bibfnamefont
  {K.}~\bibnamefont {Mamiya}}, \bibinfo {author} {\bibfnamefont
  {Y.}~\bibnamefont {Saito}}, \bibinfo {author} {\bibfnamefont
  {Y.}~\bibnamefont {Muramatsu}}, \bibinfo {author} {\bibfnamefont
  {H.}~\bibnamefont {Ott}}, \bibinfo {author} {\bibfnamefont {A.}~\bibnamefont
  {Tanaka}}, \bibinfo {author} {\bibfnamefont {T.}~\bibnamefont {Kondo}}, \
  and\ \bibinfo {author} {\bibfnamefont {H.}~\bibnamefont {Munekata}},\
  }\bibfield  {title} {\enquote {\bibinfo {title} {High-energy spectroscopic
  study of the {III--V} nitride-based diluted magnetic semiconductor
  {$\mathrm{Ga}_{1-x}\mathrm{Mn}_{x}\mathrm{N}$}},}\ }\href {\doibase
  10.1103/PhysRevB.72.085216} {\bibfield  {journal} {\bibinfo  {journal} {Phys.
  Rev. B}\ }\textbf {\bibinfo {volume} {72}},\ \bibinfo {pages} {085216}
  (\bibinfo {year} {2005})}\BibitemShut {NoStop}%
\bibitem [{\citenamefont {Bertho}\ \emph {et~al.}(1991)\citenamefont {Bertho},
  \citenamefont {Boiron}, \citenamefont {Simon}, \citenamefont {Jouanin},\ and\
  \citenamefont {Priester}}]{Bertho_1991}%
  \BibitemOpen
  \bibfield  {author} {\bibinfo {author} {\bibfnamefont {D.}~\bibnamefont
  {Bertho}}, \bibinfo {author} {\bibfnamefont {D.}~\bibnamefont {Boiron}},
  \bibinfo {author} {\bibfnamefont {A.}~\bibnamefont {Simon}}, \bibinfo
  {author} {\bibfnamefont {C.}~\bibnamefont {Jouanin}}, \ and\ \bibinfo
  {author} {\bibfnamefont {C.}~\bibnamefont {Priester}},\ }\bibfield  {title}
  {\enquote {\bibinfo {title} {Calculation of hydrostatic and uniaxial
  deformation potentials with a self-consistent tight-binding model for
  {Zn}-cation-based {II-VI} compounds},}\ }\href {\doibase
  10.1103/PhysRevB.44.6118} {\bibfield  {journal} {\bibinfo  {journal} {Phys.
  Rev. B}\ }\textbf {\bibinfo {volume} {44}},\ \bibinfo {pages} {6118--6124}
  (\bibinfo {year} {1991})}\BibitemShut {NoStop}%
\bibitem [{\citenamefont {Shi}\ and\ \citenamefont
  {Papaconstantopoulos}(2004)}]{Shi:2004_PRB}%
  \BibitemOpen
  \bibfield  {author} {\bibinfo {author} {\bibfnamefont {Lei}\ \bibnamefont
  {Shi}}\ and\ \bibinfo {author} {\bibfnamefont {Dimitrios~A.}\ \bibnamefont
  {Papaconstantopoulos}},\ }\bibfield  {title} {\enquote {\bibinfo {title}
  {Modifications and extensions to {H}arrison's tight-binding theory},}\ }\href
  {\doibase 10.1103/PhysRevB.70.205101} {\bibfield  {journal} {\bibinfo
  {journal} {Phys. Rev. B}\ }\textbf {\bibinfo {volume} {70}},\ \bibinfo
  {pages} {205101} (\bibinfo {year} {2004})}\BibitemShut {NoStop}%
\bibitem [{\citenamefont {Parmenter}(1973)}]{Parmenter_1973}%
  \BibitemOpen
  \bibfield  {author} {\bibinfo {author} {\bibfnamefont {R.~H.}\ \bibnamefont
  {Parmenter}},\ }\bibfield  {title} {\enquote {\bibinfo {title} {Effect of
  orbital degeneracy on the {A}nderson model of a localized moment in a
  metal},}\ }\href {\doibase 10.1103/PhysRevB.8.1273} {\bibfield  {journal}
  {\bibinfo  {journal} {Phys. Rev. B}\ }\textbf {\bibinfo {volume} {8}},\
  \bibinfo {pages} {1273--1275} (\bibinfo {year} {1973})}\BibitemShut {NoStop}%
\bibitem [{\citenamefont {Winkler}(2003)}]{Winkler_2003}%
  \BibitemOpen
  \bibfield  {author} {\bibinfo {author} {\bibfnamefont {Roland}\ \bibnamefont
  {Winkler}},\ }\href {\doibase 10.1007/b13586} {\emph {\bibinfo {title}
  {Spin-orbit coupling effects in two-dimensional electron and hole systems}}}\
  (\bibinfo  {publisher} {Springer Verlag},\ \bibinfo {address} {Berlin,
  Heidelberg},\ \bibinfo {year} {2003})\BibitemShut {NoStop}%
\bibitem [{\citenamefont {Liechtenstein}\ \emph {et~al.}(1984)\citenamefont
  {Liechtenstein}, \citenamefont {Katsnelson},\ and\ \citenamefont
  {Gubanov}}]{Liechtenstein:1984_JPF}%
  \BibitemOpen
  \bibfield  {author} {\bibinfo {author} {\bibfnamefont {A.~I.}\ \bibnamefont
  {Liechtenstein}}, \bibinfo {author} {\bibfnamefont {M.~I.}\ \bibnamefont
  {Katsnelson}}, \ and\ \bibinfo {author} {\bibfnamefont {V.~A.}\ \bibnamefont
  {Gubanov}},\ }\bibfield  {title} {\enquote {\bibinfo {title} {Exchange
  interactions and spin-wave stiffness in ferromagnetic metals},}\ }\href
  {\doibase 10.1088/0305-4608/14/7/007} {\bibfield  {journal} {\bibinfo
  {journal} {J. Phys. F: Met. Phys.}\ }\textbf {\bibinfo {volume} {14}},\
  \bibinfo {pages} {L125} (\bibinfo {year} {1984})}\BibitemShut {NoStop}%
\bibitem [{\citenamefont {Allan}\ and\ \citenamefont
  {Delerue}(2012)}]{Allan:2012_PRB}%
  \BibitemOpen
  \bibfield  {author} {\bibinfo {author} {\bibfnamefont {Guy}\ \bibnamefont
  {Allan}}\ and\ \bibinfo {author} {\bibfnamefont {Christophe}\ \bibnamefont
  {Delerue}},\ }\bibfield  {title} {\enquote {\bibinfo {title} {Tight-binding
  calculations of the optical properties of {HgTe} nanocrystals},}\ }\href
  {\doibase 10.1103/PhysRevB.86.165437} {\bibfield  {journal} {\bibinfo
  {journal} {Phys. Rev. B}\ }\textbf {\bibinfo {volume} {86}},\ \bibinfo
  {pages} {165437} (\bibinfo {year} {2012})}\BibitemShut {NoStop}%
\bibitem [{\citenamefont {Sapra}\ \emph {et~al.}(2002)\citenamefont {Sapra},
  \citenamefont {Shanthi},\ and\ \citenamefont {Sarma}}]{Sapra:2002_PRB}%
  \BibitemOpen
  \bibfield  {author} {\bibinfo {author} {\bibfnamefont {Sameer}\ \bibnamefont
  {Sapra}}, \bibinfo {author} {\bibfnamefont {N.}~\bibnamefont {Shanthi}}, \
  and\ \bibinfo {author} {\bibfnamefont {D.~D.}\ \bibnamefont {Sarma}},\
  }\bibfield  {title} {\enquote {\bibinfo {title} {Realistic tight-binding
  model for the electronic structure of {II-VI} semiconductors},}\ }\href
  {\doibase 10.1103/PhysRevB.66.205202} {\bibfield  {journal} {\bibinfo
  {journal} {Phys. Rev. B}\ }\textbf {\bibinfo {volume} {66}},\ \bibinfo
  {pages} {205202} (\bibinfo {year} {2002})}\BibitemShut {NoStop}%
\bibitem [{\citenamefont {Spa{\l}ek}\ \emph {et~al.}(1986)\citenamefont
  {Spa{\l}ek}, \citenamefont {Lewicki}, \citenamefont {Tarnawski},
  \citenamefont {Furdyna}, \citenamefont {Galazka},\ and\ \citenamefont
  {Obuszko}}]{Spalek:1986_PRB}%
  \BibitemOpen
  \bibfield  {author} {\bibinfo {author} {\bibfnamefont {J.}~\bibnamefont
  {Spa{\l}ek}}, \bibinfo {author} {\bibfnamefont {A.}~\bibnamefont {Lewicki}},
  \bibinfo {author} {\bibfnamefont {Z.}~\bibnamefont {Tarnawski}}, \bibinfo
  {author} {\bibfnamefont {J.~K.}\ \bibnamefont {Furdyna}}, \bibinfo {author}
  {\bibfnamefont {R.~R.}\ \bibnamefont {Galazka}}, \ and\ \bibinfo {author}
  {\bibfnamefont {Z.}~\bibnamefont {Obuszko}},\ }\bibfield  {title} {\enquote
  {\bibinfo {title} {Magnetic susceptibility of semimagnetic semiconductors:
  The high-temperature regime and the role of superexchange},}\ }\href
  {\doibase 10.1103/PhysRevB.33.3407} {\bibfield  {journal} {\bibinfo
  {journal} {Phys. Rev. B}\ }\textbf {\bibinfo {volume} {33}},\ \bibinfo
  {pages} {3407--3418} (\bibinfo {year} {1986})}\BibitemShut {NoStop}%
\bibitem [{\citenamefont {Bonanni}\ \emph {et~al.}(2011)\citenamefont
  {Bonanni}, \citenamefont {Sawicki}, \citenamefont {Devillers}, \citenamefont
  {Stefanowicz}, \citenamefont {Faina}, \citenamefont {Li}, \citenamefont
  {Winkler}, \citenamefont {Sztenkiel}, \citenamefont {Navarro-Quezada},
  \citenamefont {Rovezzi}, \citenamefont {Jakie\l{}a}, \citenamefont {Grois},
  \citenamefont {Wegscheider}, \citenamefont {Jantsch}, \citenamefont
  {Suffczy\ifmmode~\acute{n}\else \'{n}\fi{}ski}, \citenamefont {D'Acapito},
  \citenamefont {Meingast}, \citenamefont {Kothleitner},\ and\ \citenamefont
  {Dietl}}]{Bonanni:2011_PRB}%
  \BibitemOpen
  \bibfield  {author} {\bibinfo {author} {\bibfnamefont {A.}~\bibnamefont
  {Bonanni}}, \bibinfo {author} {\bibfnamefont {M.}~\bibnamefont {Sawicki}},
  \bibinfo {author} {\bibfnamefont {T.}~\bibnamefont {Devillers}}, \bibinfo
  {author} {\bibfnamefont {W.}~\bibnamefont {Stefanowicz}}, \bibinfo {author}
  {\bibfnamefont {B.}~\bibnamefont {Faina}}, \bibinfo {author} {\bibfnamefont
  {Tian}\ \bibnamefont {Li}}, \bibinfo {author} {\bibfnamefont {T.~E.}\
  \bibnamefont {Winkler}}, \bibinfo {author} {\bibfnamefont {D.}~\bibnamefont
  {Sztenkiel}}, \bibinfo {author} {\bibfnamefont {A.}~\bibnamefont
  {Navarro-Quezada}}, \bibinfo {author} {\bibfnamefont {M.}~\bibnamefont
  {Rovezzi}}, \bibinfo {author} {\bibfnamefont {R.}~\bibnamefont {Jakie\l{}a}},
  \bibinfo {author} {\bibfnamefont {A.}~\bibnamefont {Grois}}, \bibinfo
  {author} {\bibfnamefont {M.}~\bibnamefont {Wegscheider}}, \bibinfo {author}
  {\bibfnamefont {W.}~\bibnamefont {Jantsch}}, \bibinfo {author} {\bibfnamefont
  {J.}~\bibnamefont {Suffczy\ifmmode~\acute{n}\else \'{n}\fi{}ski}}, \bibinfo
  {author} {\bibfnamefont {F.}~\bibnamefont {D'Acapito}}, \bibinfo {author}
  {\bibfnamefont {A.}~\bibnamefont {Meingast}}, \bibinfo {author}
  {\bibfnamefont {G.}~\bibnamefont {Kothleitner}}, \ and\ \bibinfo {author}
  {\bibfnamefont {T.}~\bibnamefont {Dietl}},\ }\bibfield  {title} {\enquote
  {\bibinfo {title} {Experimental probing of exchange interactions between
  localized spins in the dilute magnetic insulator {(Ga,Mn)N}},}\ }\href
  {\doibase 10.1103/PhysRevB.84.035206} {\bibfield  {journal} {\bibinfo
  {journal} {Phys. Rev. B}\ }\textbf {\bibinfo {volume} {84}},\ \bibinfo
  {pages} {035206} (\bibinfo {year} {2011})}\BibitemShut {NoStop}%
\bibitem [{\citenamefont {Cuono}\ \emph {et~al.}(2023)\citenamefont {Cuono},
  \citenamefont {Autieri},\ and\ \citenamefont {Dietl}}]{Cuono:2023_arXiv}%
  \BibitemOpen
  \bibfield  {author} {\bibinfo {author} {\bibfnamefont {Giuseppe}\
  \bibnamefont {Cuono}}, \bibinfo {author} {\bibfnamefont {Carmine}\
  \bibnamefont {Autieri}}, \ and\ \bibinfo {author} {\bibfnamefont {Tomasz}\
  \bibnamefont {Dietl}},\ }\href@noop {} {\enquote {\bibinfo {title} {{CdTe}
  and {HgTe} doped with {V}, {Cr}, and {Mn} --- prospects for the quantum
  anomalous {H}all effect},}\ } (\bibinfo {year} {2023}),\ \Eprint
  {http://arxiv.org/abs/2312.16732} {arXiv:2312.16732 [cond-mat.mtrl-sci]}
  \BibitemShut {NoStop}%
\bibitem [{\citenamefont {\'Sliwa}(2022)}]{Sliwa_2022}%
  \BibitemOpen
  \bibfield  {author} {\bibinfo {author} {\bibfnamefont {C.}~\bibnamefont
  {\'Sliwa}},\ }\href {\doibase 10.48550/ARXIV.2205.00977} {\enquote {\bibinfo
  {title} {Disorder-averaged {B}inder ratio in site-diluted {H}eisenberg
  model},}\ } (\bibinfo {year} {2022}),\ \bibinfo {note}
  {{arXiv:2205.00977}}\BibitemShut {NoStop}%
\bibitem [{\citenamefont {Brey}\ and\ \citenamefont
  {G\'omez-Santos}(2003)}]{Brey:2003_PRB}%
  \BibitemOpen
  \bibfield  {author} {\bibinfo {author} {\bibfnamefont {L.}~\bibnamefont
  {Brey}}\ and\ \bibinfo {author} {\bibfnamefont {G.}~\bibnamefont
  {G\'omez-Santos}},\ }\bibfield  {title} {\enquote {\bibinfo {title} {Magnetic
  properties of {(Ga,Mn)As} from an effective {H}eisenberg {H}amiltonian},}\
  }\href {\doibase 10.1103/PhysRevB.68.115206} {\bibfield  {journal} {\bibinfo
  {journal} {Phys. Rev. B}\ }\textbf {\bibinfo {volume} {68}},\ \bibinfo
  {pages} {115206} (\bibinfo {year} {2003})}\BibitemShut {NoStop}%
\bibitem [{\citenamefont {Dobrowolska}\ and\ \citenamefont
  {Dobrowolski}(1981)}]{Dobrowolska:1981_SST}%
  \BibitemOpen
  \bibfield  {author} {\bibinfo {author} {\bibfnamefont {M.}~\bibnamefont
  {Dobrowolska}}\ and\ \bibinfo {author} {\bibfnamefont {W.}~\bibnamefont
  {Dobrowolski}},\ }\bibfield  {title} {\enquote {\bibinfo {title} {Temperature
  study of interband magnetoabsorption in
  {$\mathrm{Hg}_{1-x}\mathrm{Mn}_x\mathrm{Te}$} mixed crystals},}\ }\href
  {\doibase 10.1088/0022-3719/14/36/012} {\bibfield  {journal} {\bibinfo
  {journal} {J. Phys. C: Solid State Phys.}\ }\textbf {\bibinfo {volume}
  {14}},\ \bibinfo {pages} {5689} (\bibinfo {year} {1981})}\BibitemShut
  {NoStop}%
\bibitem [{\citenamefont {Bauer}\ \emph {et~al.}(1985)\citenamefont {Bauer},
  \citenamefont {Kossut}, \citenamefont {Faymonville},\ and\ \citenamefont
  {Dornhaus}}]{Bauer:1985_PRB}%
  \BibitemOpen
  \bibfield  {author} {\bibinfo {author} {\bibfnamefont {G.}~\bibnamefont
  {Bauer}}, \bibinfo {author} {\bibfnamefont {J.}~\bibnamefont {Kossut}},
  \bibinfo {author} {\bibfnamefont {R.}~\bibnamefont {Faymonville}}, \ and\
  \bibinfo {author} {\bibfnamefont {R.}~\bibnamefont {Dornhaus}},\ }\bibfield
  {title} {\enquote {\bibinfo {title} {Magnetoreflectivity study of the band
  structure of
  {${\mathrm{Hg}}_{1\mathrm{\ensuremath{-}}\mathrm{x}}$${\mathrm{Mn}}_{\mathrm{x}}$Te}
  (0.026\ensuremath{\le}x\ensuremath{\le}0.106)},}\ }\href {\doibase
  10.1103/PhysRevB.31.2040} {\bibfield  {journal} {\bibinfo  {journal} {Phys.
  Rev. B}\ }\textbf {\bibinfo {volume} {31}},\ \bibinfo {pages} {2040--2048}
  (\bibinfo {year} {1985})}\BibitemShut {NoStop}%
\bibitem [{\citenamefont {Autieri}\ \emph {et~al.}(2021)\citenamefont
  {Autieri}, \citenamefont {\'Sliwa}, \citenamefont {Islam}, \citenamefont
  {Cuono},\ and\ \citenamefont {Dietl}}]{Autieri:2021_PRB}%
  \BibitemOpen
  \bibfield  {author} {\bibinfo {author} {\bibfnamefont {C.}~\bibnamefont
  {Autieri}}, \bibinfo {author} {\bibfnamefont {C.}~\bibnamefont {\'Sliwa}},
  \bibinfo {author} {\bibfnamefont {R.}~\bibnamefont {Islam}}, \bibinfo
  {author} {\bibfnamefont {G.}~\bibnamefont {Cuono}}, \ and\ \bibinfo {author}
  {\bibfnamefont {T.}~\bibnamefont {Dietl}},\ }\bibfield  {title} {\enquote
  {\bibinfo {title} {Momentum-resolved spin splitting in {Mn}-doped trivial
  {CdTe} and topological {HgTe} semiconductors},}\ }\href {\doibase
  10.1103/PhysRevB.103.115209} {\bibfield  {journal} {\bibinfo  {journal}
  {Phys. Rev. B}\ }\textbf {\bibinfo {volume} {103}},\ \bibinfo {pages}
  {115209} (\bibinfo {year} {2021})}\BibitemShut {NoStop}%
\bibitem [{\citenamefont {Twardowski}\ \emph {et~al.}(1984)\citenamefont
  {Twardowski}, \citenamefont {Swiderski}, \citenamefont {{von Ortenberg}},\
  and\ \citenamefont {Pauthenet}}]{Twardowski:1984_SSC}%
  \BibitemOpen
  \bibfield  {author} {\bibinfo {author} {\bibfnamefont {A.}~\bibnamefont
  {Twardowski}}, \bibinfo {author} {\bibfnamefont {P.}~\bibnamefont
  {Swiderski}}, \bibinfo {author} {\bibfnamefont {M.}~\bibnamefont {{von
  Ortenberg}}}, \ and\ \bibinfo {author} {\bibfnamefont {R.}~\bibnamefont
  {Pauthenet}},\ }\bibfield  {title} {\enquote {\bibinfo {title}
  {Magnetoabsorption and magnetization of {Zn$_{1-x}$Mn$_x$Te} mixed
  crystals},}\ }\href {\doibase https://doi.org/10.1016/0038-1098(84)90318-1}
  {\bibfield  {journal} {\bibinfo  {journal} {Solid State Commun.}\ }\textbf
  {\bibinfo {volume} {50}},\ \bibinfo {pages} {509--513} (\bibinfo {year}
  {1984})}\BibitemShut {NoStop}%
\bibitem [{\citenamefont {\'Sliwa}\ and\ \citenamefont
  {Dietl}(2008)}]{Sliwa:2008_PRB}%
  \BibitemOpen
  \bibfield  {author} {\bibinfo {author} {\bibfnamefont {Cezary}\ \bibnamefont
  {\'Sliwa}}\ and\ \bibinfo {author} {\bibfnamefont {Tomasz}\ \bibnamefont
  {Dietl}},\ }\bibfield  {title} {\enquote {\bibinfo {title} {Electron-hole
  contribution to the apparent $s-d$ exchange interaction in {III-V} dilute
  magnetic semiconductors},}\ }\href {\doibase 10.1103/PhysRevB.78.165205}
  {\bibfield  {journal} {\bibinfo  {journal} {Phys. Rev. B}\ }\textbf {\bibinfo
  {volume} {78}},\ \bibinfo {pages} {165205} (\bibinfo {year}
  {2008})}\BibitemShut {NoStop}%
\bibitem [{\citenamefont {Shamim}\ \emph {et~al.}(2020)\citenamefont {Shamim},
  \citenamefont {Beugeling}, \citenamefont {B\"ottcher}, \citenamefont
  {Shekhar}, \citenamefont {Budewitz}, \citenamefont {Leubner}, \citenamefont
  {Lunczer}, \citenamefont {Hankiewicz}, \citenamefont {Buhmann},\ and\
  \citenamefont {Molenkamp}}]{Shamim:2020_SA}%
  \BibitemOpen
  \bibfield  {author} {\bibinfo {author} {\bibfnamefont {S.}~\bibnamefont
  {Shamim}}, \bibinfo {author} {\bibfnamefont {W.}~\bibnamefont {Beugeling}},
  \bibinfo {author} {\bibfnamefont {J.}~\bibnamefont {B\"ottcher}}, \bibinfo
  {author} {\bibfnamefont {P.}~\bibnamefont {Shekhar}}, \bibinfo {author}
  {\bibfnamefont {A.}~\bibnamefont {Budewitz}}, \bibinfo {author}
  {\bibfnamefont {P.}~\bibnamefont {Leubner}}, \bibinfo {author} {\bibfnamefont
  {L.}~\bibnamefont {Lunczer}}, \bibinfo {author} {\bibfnamefont {E.~M.}\
  \bibnamefont {Hankiewicz}}, \bibinfo {author} {\bibfnamefont
  {H.}~\bibnamefont {Buhmann}}, \ and\ \bibinfo {author} {\bibfnamefont
  {L.~W.}\ \bibnamefont {Molenkamp}},\ }\bibfield  {title} {\enquote {\bibinfo
  {title} {Emergent quantum {Hall} effects below 50 {mT} in a two-dimensional
  topological insulator},}\ }\href {\doibase 10.1126/sciadv.aba4625} {\bibfield
   {journal} {\bibinfo  {journal} {Adv. Sci.}\ }\textbf {\bibinfo {volume}
  {6}},\ \bibinfo {pages} {eaba4625} (\bibinfo {year} {2020})}\BibitemShut
  {NoStop}%
\bibitem [{\citenamefont {Sawicki}\ \emph {et~al.}(1983)\citenamefont
  {Sawicki}, \citenamefont {Dietl}, \citenamefont {Plesiewicz}, \citenamefont
  {S\k{e}kowski}, \citenamefont {\'Sniadower}, \citenamefont {Baj},\ and\
  \citenamefont {Dmowski}}]{Sawicki:1983_Pr}%
  \BibitemOpen
  \bibfield  {author} {\bibinfo {author} {\bibfnamefont {M.}~\bibnamefont
  {Sawicki}}, \bibinfo {author} {\bibfnamefont {T.}~\bibnamefont {Dietl}},
  \bibinfo {author} {\bibfnamefont {W.}~\bibnamefont {Plesiewicz}}, \bibinfo
  {author} {\bibfnamefont {P.}~\bibnamefont {S\k{e}kowski}}, \bibinfo {author}
  {\bibfnamefont {L.}~\bibnamefont {\'Sniadower}}, \bibinfo {author}
  {\bibfnamefont {M.}~\bibnamefont {Baj}}, \ and\ \bibinfo {author}
  {\bibfnamefont {L.}~\bibnamefont {Dmowski}},\ }\bibfield  {title} {\enquote
  {\bibinfo {title} {Influence of an acceptor state on transport in zero-gap
  {Hg$_{1-x}$Mn$_x$Te}},}\ }in\ \href {\doibase 10.1007/3-540-11996-5_55}
  {\emph {\bibinfo {booktitle} {Application of High Magnetic Fields in
  Semiconductor Physics}}},\ \bibinfo {editor} {edited by\ \bibinfo {editor}
  {\bibfnamefont {G.}~\bibnamefont {Landwehr}}}\ (\bibinfo  {publisher}
  {Springer Berlin Heidelberg},\ \bibinfo {address} {Berlin, Heidelberg},\
  \bibinfo {year} {1983})\ pp.\ \bibinfo {pages} {382--385}\BibitemShut
  {NoStop}%
\bibitem [{\citenamefont {Dietl}(2023{\natexlab{a}})}]{Dietl:2023_PRL}%
  \BibitemOpen
  \bibfield  {author} {\bibinfo {author} {\bibfnamefont {T.}~\bibnamefont
  {Dietl}},\ }\bibfield  {title} {\enquote {\bibinfo {title} {Effects of charge
  dopants in quantum spin {Hall} materials},}\ }\href {\doibase
  10.1103/PhysRevLett.130.086202} {\bibfield  {journal} {\bibinfo  {journal}
  {Phys. Rev. Lett.}\ }\textbf {\bibinfo {volume} {130}},\ \bibinfo {pages}
  {086202} (\bibinfo {year} {2023}{\natexlab{a}})}\BibitemShut {NoStop}%
\bibitem [{\citenamefont {Dietl}(2023{\natexlab{b}})}]{Dietl:2023_PRB}%
  \BibitemOpen
  \bibfield  {author} {\bibinfo {author} {\bibfnamefont {T.}~\bibnamefont
  {Dietl}},\ }\bibfield  {title} {\enquote {\bibinfo {title} {Quantitative
  theory of backscattering in topological {HgTe} and {(Hg,Mn)Te} quantum wells:
  Acceptor states, {Kondo} effect, precessional dephasing, and bound magnetic
  polaron},}\ }\href {\doibase 10.1103/PhysRevB.107.085421} {\bibfield
  {journal} {\bibinfo  {journal} {Phys. Rev. B}\ }\textbf {\bibinfo {volume}
  {107}},\ \bibinfo {pages} {085421} (\bibinfo {year}
  {2023}{\natexlab{b}})}\BibitemShut {NoStop}%
\bibitem [{\citenamefont {Shamim}\ \emph {et~al.}(2021)\citenamefont {Shamim},
  \citenamefont {Beugeling}, \citenamefont {Shekhar}, \citenamefont {Bendias},
  \citenamefont {Lunczer}, \citenamefont {Kleinlein}, \citenamefont {Buhmann},\
  and\ \citenamefont {Molenkamp}}]{Shamim:2021_NC}%
  \BibitemOpen
  \bibfield  {author} {\bibinfo {author} {\bibfnamefont {S.}~\bibnamefont
  {Shamim}}, \bibinfo {author} {\bibfnamefont {W.}~\bibnamefont {Beugeling}},
  \bibinfo {author} {\bibfnamefont {P.}~\bibnamefont {Shekhar}}, \bibinfo
  {author} {\bibfnamefont {K.}~\bibnamefont {Bendias}}, \bibinfo {author}
  {\bibfnamefont {L.}~\bibnamefont {Lunczer}}, \bibinfo {author} {\bibfnamefont
  {J.}~\bibnamefont {Kleinlein}}, \bibinfo {author} {\bibfnamefont
  {H.}~\bibnamefont {Buhmann}}, \ and\ \bibinfo {author} {\bibfnamefont
  {L.~W.}\ \bibnamefont {Molenkamp}},\ }\bibfield  {title} {\enquote {\bibinfo
  {title} {Quantized spin {Hall} conductance in a magnetically doped two
  dimensional topological insulator},}\ }\href {\doibase
  10.1038/s41467-021-23262-1} {\bibfield  {journal} {\bibinfo  {journal} {Nat.
  Commun.}\ }\textbf {\bibinfo {volume} {12}},\ \bibinfo {pages} {3193}
  (\bibinfo {year} {2021})}\BibitemShut {NoStop}%
\bibitem [{\citenamefont {Tanaka}\ \emph {et~al.}(2011)\citenamefont {Tanaka},
  \citenamefont {Furusaki},\ and\ \citenamefont {Matveev}}]{Tanaka:2011_PRL}%
  \BibitemOpen
  \bibfield  {author} {\bibinfo {author} {\bibfnamefont {Y.}~\bibnamefont
  {Tanaka}}, \bibinfo {author} {\bibfnamefont {A.}~\bibnamefont {Furusaki}}, \
  and\ \bibinfo {author} {\bibfnamefont {K.~A.}\ \bibnamefont {Matveev}},\
  }\bibfield  {title} {\enquote {\bibinfo {title} {Conductance of a helical
  edge liquid coupled to a magnetic impurity},}\ }\href {\doibase
  10.1103/PhysRevLett.106.236402} {\bibfield  {journal} {\bibinfo  {journal}
  {Phys. Rev. Lett.}\ }\textbf {\bibinfo {volume} {106}},\ \bibinfo {pages}
  {236402} (\bibinfo {year} {2011})}\BibitemShut {NoStop}%
\bibitem [{\citenamefont {Liu}\ \emph {et~al.}(2008{\natexlab{b}})\citenamefont
  {Liu}, \citenamefont {Qi}, \citenamefont {Dai}, \citenamefont {Fang},\ and\
  \citenamefont {Zhang}}]{Liu:2008_PRLb}%
  \BibitemOpen
  \bibfield  {author} {\bibinfo {author} {\bibfnamefont {Chao-Xing}\
  \bibnamefont {Liu}}, \bibinfo {author} {\bibfnamefont {Xiao-Liang}\
  \bibnamefont {Qi}}, \bibinfo {author} {\bibfnamefont {Xi}~\bibnamefont
  {Dai}}, \bibinfo {author} {\bibfnamefont {Zhong}\ \bibnamefont {Fang}}, \
  and\ \bibinfo {author} {\bibfnamefont {Shou-Cheng}\ \bibnamefont {Zhang}},\
  }\bibfield  {title} {\enquote {\bibinfo {title} {Quantum anomalous {Hall}
  effect in {Hg$_{1-y}$Mn$_y$Te} quantum wells},}\ }\href {\doibase
  10.1103/PhysRevLett.101.146802} {\bibfield  {journal} {\bibinfo  {journal}
  {Phys. Rev. Lett.}\ }\textbf {\bibinfo {volume} {101}},\ \bibinfo {pages}
  {146802} (\bibinfo {year} {2008}{\natexlab{b}})}\BibitemShut {NoStop}%
\bibitem [{\citenamefont {Peyla}\ \emph {et~al.}(1993)\citenamefont {Peyla},
  \citenamefont {Wasiela}, \citenamefont {Merle~d'Aubign\'e}, \citenamefont
  {Ashenford},\ and\ \citenamefont {Lunn}}]{Peyla:1993_PRB}%
  \BibitemOpen
  \bibfield  {author} {\bibinfo {author} {\bibfnamefont {P.}~\bibnamefont
  {Peyla}}, \bibinfo {author} {\bibfnamefont {A.}~\bibnamefont {Wasiela}},
  \bibinfo {author} {\bibfnamefont {Y.}~\bibnamefont {Merle~d'Aubign\'e}},
  \bibinfo {author} {\bibfnamefont {D.~E.}\ \bibnamefont {Ashenford}}, \ and\
  \bibinfo {author} {\bibfnamefont {B.}~\bibnamefont {Lunn}},\ }\bibfield
  {title} {\enquote {\bibinfo {title} {Anisotropy of the zeeman effect in
  {CdTe/${\mathrm{Cd}}_{1\mathrm{\ensuremath{-}}\mathit{x}}$${\mathrm{Mn}}_{\mathit{x}}$Te}
  multiple quantum wells},}\ }\href {\doibase 10.1103/PhysRevB.47.3783}
  {\bibfield  {journal} {\bibinfo  {journal} {Phys. Rev. B}\ }\textbf {\bibinfo
  {volume} {47}},\ \bibinfo {pages} {3783--3789} (\bibinfo {year}
  {1993})}\BibitemShut {NoStop}%
\bibitem [{\citenamefont {Novik}\ \emph {et~al.}(2005)\citenamefont {Novik},
  \citenamefont {Pfeuffer-Jeschke}, \citenamefont {Jungwirth}, \citenamefont
  {Latussek}, \citenamefont {Becker}, \citenamefont {Landwehr}, \citenamefont
  {Buhmann},\ and\ \citenamefont {Molenkamp}}]{Novik:2005_PRB}%
  \BibitemOpen
  \bibfield  {author} {\bibinfo {author} {\bibfnamefont {E.~G.}\ \bibnamefont
  {Novik}}, \bibinfo {author} {\bibfnamefont {A.}~\bibnamefont
  {Pfeuffer-Jeschke}}, \bibinfo {author} {\bibfnamefont {T.}~\bibnamefont
  {Jungwirth}}, \bibinfo {author} {\bibfnamefont {V.}~\bibnamefont {Latussek}},
  \bibinfo {author} {\bibfnamefont {C.~R.}\ \bibnamefont {Becker}}, \bibinfo
  {author} {\bibfnamefont {G.}~\bibnamefont {Landwehr}}, \bibinfo {author}
  {\bibfnamefont {H.}~\bibnamefont {Buhmann}}, \ and\ \bibinfo {author}
  {\bibfnamefont {L.~W.}\ \bibnamefont {Molenkamp}},\ }\bibfield  {title}
  {\enquote {\bibinfo {title} {Band structure of semimagnetic
  {Hg$_{1-y}$Mn$_y$Te} quantum wells},}\ }\href {\doibase
  10.1103/PhysRevB.72.035321} {\bibfield  {journal} {\bibinfo  {journal} {Phys.
  Rev. B}\ }\textbf {\bibinfo {volume} {72}},\ \bibinfo {pages} {035321}
  (\bibinfo {year} {2005})}\BibitemShut {NoStop}%
\bibitem [{Note1()}]{Note1}%
  \BibitemOpen
  \bibinfo {note} {Boundary terms at the well-barrier interfaces are
  neglected.}\BibitemShut {Stop}%
\bibitem [{\citenamefont {Fukui}\ \emph {et~al.}(2005)\citenamefont {Fukui},
  \citenamefont {Hatsugai},\ and\ \citenamefont {Suzuki}}]{Fukui:2005_JPSJ}%
  \BibitemOpen
  \bibfield  {author} {\bibinfo {author} {\bibfnamefont {Takahiro}\
  \bibnamefont {Fukui}}, \bibinfo {author} {\bibfnamefont {Yasuhiro}\
  \bibnamefont {Hatsugai}}, \ and\ \bibinfo {author} {\bibfnamefont {Hiroshi}\
  \bibnamefont {Suzuki}},\ }\bibfield  {title} {\enquote {\bibinfo {title}
  {{C}hern numbers in discretized {B}rillouin zone: Efficient method of
  computing (spin) {H}all conductances},}\ }\href {\doibase
  10.1143/JPSJ.74.1674} {\bibfield  {journal} {\bibinfo  {journal} {J. Phys.
  Soc. Jpn.}\ }\textbf {\bibinfo {volume} {74}},\ \bibinfo {pages} {1674--1677}
  (\bibinfo {year} {2005})}\BibitemShut {NoStop}%
\bibitem [{\citenamefont {Brzezicki}()}]{Brzezicki}%
  \BibitemOpen
  \bibfield  {author} {\bibinfo {author} {\bibfnamefont {W.}~\bibnamefont
  {Brzezicki}},\ }\href@noop {} {}\bibinfo {note} {Unpublished}\BibitemShut
  {NoStop}%
\bibitem [{Note2()}]{Note2}%
  \BibitemOpen
  \bibinfo {note} {If the presence of degenerate bands cannot be excluded, the
  determinants of the submatrices corresponding to each of the mutually
  different eigenergies must be computed instead. Then the imaginary part of
  the trace can be obtained as the logarithm of the product of all the
  determinants (normalized to the unit circle), and is the Abelian curvature of
  the adiabatic transport of full-dimension orthonormal sets of energy
  eigenstates. Considering the principal branch of the logarithm \cite
  {Fukui:2005_JPSJ} eliminates ambiguity of the result.}\BibitemShut {Stop}%
\bibitem [{\citenamefont {Twardowski}\ \emph {et~al.}(1987)\citenamefont
  {Twardowski}, \citenamefont {Swagten}, \citenamefont {de~Jonge},\ and\
  \citenamefont {Demianiuk}}]{Twardowski:1987_PRB}%
  \BibitemOpen
  \bibfield  {author} {\bibinfo {author} {\bibfnamefont {A.}~\bibnamefont
  {Twardowski}}, \bibinfo {author} {\bibfnamefont {H.~J.~M.}\ \bibnamefont
  {Swagten}}, \bibinfo {author} {\bibfnamefont {W.~J.~M.}\ \bibnamefont
  {de~Jonge}}, \ and\ \bibinfo {author} {\bibfnamefont {M.}~\bibnamefont
  {Demianiuk}},\ }\bibfield  {title} {\enquote {\bibinfo {title} {Magnetic
  behavior of the diluted magnetic semiconductor
  {${\mathrm{Zn}}_{1\mathrm{\ensuremath{-}}\mathrm{x}}$${\mathrm{Mn}}_{\mathrm{x}}$Se}},}\
  }\href {\doibase 10.1103/PhysRevB.36.7013} {\bibfield  {journal} {\bibinfo
  {journal} {Phys. Rev. B}\ }\textbf {\bibinfo {volume} {36}},\ \bibinfo
  {pages} {7013--7023} (\bibinfo {year} {1987})}\BibitemShut {NoStop}%
\bibitem [{\citenamefont {Bendias}\ \emph {et~al.}(2018)\citenamefont
  {Bendias}, \citenamefont {Shamim}, \citenamefont {Herrmann}, \citenamefont
  {Budewitz}, \citenamefont {Shekhar}, \citenamefont {Leubner}, \citenamefont
  {Kleinlein}, \citenamefont {Bocquillon}, \citenamefont {Buhmann},\ and\
  \citenamefont {Molenkamp}}]{Bendias:2018_NL}%
  \BibitemOpen
  \bibfield  {author} {\bibinfo {author} {\bibfnamefont {K.}~\bibnamefont
  {Bendias}}, \bibinfo {author} {\bibfnamefont {S.}~\bibnamefont {Shamim}},
  \bibinfo {author} {\bibfnamefont {O.}~\bibnamefont {Herrmann}}, \bibinfo
  {author} {\bibfnamefont {A.}~\bibnamefont {Budewitz}}, \bibinfo {author}
  {\bibfnamefont {P.}~\bibnamefont {Shekhar}}, \bibinfo {author} {\bibfnamefont
  {P.}~\bibnamefont {Leubner}}, \bibinfo {author} {\bibfnamefont
  {J.}~\bibnamefont {Kleinlein}}, \bibinfo {author} {\bibfnamefont
  {E.}~\bibnamefont {Bocquillon}}, \bibinfo {author} {\bibfnamefont
  {H.}~\bibnamefont {Buhmann}}, \ and\ \bibinfo {author} {\bibfnamefont
  {L.~W.}\ \bibnamefont {Molenkamp}},\ }\bibfield  {title} {\enquote {\bibinfo
  {title} {High mobility {HgTe} microstructures for quantum spin {H}all
  studies},}\ }\href {\doibase 10.1021/acs.nanolett.8b01405} {\bibfield
  {journal} {\bibinfo  {journal} {Nano Lett.}\ }\textbf {\bibinfo {volume}
  {18}},\ \bibinfo {pages} {4831--4836} (\bibinfo {year} {2018})}\BibitemShut
  {NoStop}%
\bibitem [{\citenamefont {Yahniuk}\ \emph {et~al.}()\citenamefont {Yahniuk},
  \citenamefont {Kazakov}, \citenamefont {Jouault}, \citenamefont
  {Krishtopenko}, \citenamefont {Kret}, \citenamefont {Grabecki}, \citenamefont
  {Cywi\'nski}, \citenamefont {Mikhailov}, \citenamefont {Dvoretskii},
  \citenamefont {Przybytek}, \citenamefont {Gavrilenko}, \citenamefont {Teppe},
  \citenamefont {Dietl},\ and\ \citenamefont {Knap}}]{Yahniuk:2021_arXiv}%
  \BibitemOpen
  \bibfield  {author} {\bibinfo {author} {\bibfnamefont {I.}~\bibnamefont
  {Yahniuk}}, \bibinfo {author} {\bibfnamefont {A.}~\bibnamefont {Kazakov}},
  \bibinfo {author} {\bibfnamefont {B.}~\bibnamefont {Jouault}}, \bibinfo
  {author} {\bibfnamefont {S.~S.}\ \bibnamefont {Krishtopenko}}, \bibinfo
  {author} {\bibfnamefont {S.}~\bibnamefont {Kret}}, \bibinfo {author}
  {\bibfnamefont {G.}~\bibnamefont {Grabecki}}, \bibinfo {author}
  {\bibfnamefont {G.}~\bibnamefont {Cywi\'nski}}, \bibinfo {author}
  {\bibfnamefont {N.~N.}\ \bibnamefont {Mikhailov}}, \bibinfo {author}
  {\bibfnamefont {S.~A.}\ \bibnamefont {Dvoretskii}}, \bibinfo {author}
  {\bibfnamefont {J.}~\bibnamefont {Przybytek}}, \bibinfo {author}
  {\bibfnamefont {V.~I.}\ \bibnamefont {Gavrilenko}}, \bibinfo {author}
  {\bibfnamefont {F.}~\bibnamefont {Teppe}}, \bibinfo {author} {\bibfnamefont
  {T.}~\bibnamefont {Dietl}}, \ and\ \bibinfo {author} {\bibfnamefont
  {W.}~\bibnamefont {Knap}},\ }\bibfield  {title} {\enquote {\bibinfo {title}
  {{HgTe} quantum wells for {QHE} metrology under soft cryomagnetic conditions:
  permanent magnets and liquid {$^4$He} temperatures},}\ }\href {\doibase
  10.48550/arXiv.2111.07581} {\ 10.48550/arXiv.2111.07581}\BibitemShut
  {NoStop}%
\bibitem [{\citenamefont {Fijalkowski}\ \emph {et~al.}(2021)\citenamefont
  {Fijalkowski}, \citenamefont {Liu}, \citenamefont {Mandal}, \citenamefont
  {Schreyeck}, \citenamefont {Brunner}, \citenamefont {Gould},\ and\
  \citenamefont {Molenkamp}}]{Fijalkowski:2021_NC}%
  \BibitemOpen
  \bibfield  {author} {\bibinfo {author} {\bibfnamefont {K.~M.}\ \bibnamefont
  {Fijalkowski}}, \bibinfo {author} {\bibfnamefont {{Nan}}\ \bibnamefont
  {Liu}}, \bibinfo {author} {\bibfnamefont {P.}~\bibnamefont {Mandal}},
  \bibinfo {author} {\bibfnamefont {S.}~\bibnamefont {Schreyeck}}, \bibinfo
  {author} {\bibfnamefont {K.}~\bibnamefont {Brunner}}, \bibinfo {author}
  {\bibfnamefont {C.}~\bibnamefont {Gould}}, \ and\ \bibinfo {author}
  {\bibfnamefont {L.~W.}\ \bibnamefont {Molenkamp}},\ }\bibfield  {title}
  {\enquote {\bibinfo {title} {Quantum anomalous {Hall} edge channels survive
  up to the {Curie} temperature},}\ }\href {\doibase
  10.1038/s41467-021-25912-w} {\bibfield  {journal} {\bibinfo  {journal} {Nat.
  Commun.}\ }\textbf {\bibinfo {volume} {12}},\ \bibinfo {pages} {5599}
  (\bibinfo {year} {2021})}\BibitemShut {NoStop}%
\bibitem [{\citenamefont {Yang}\ \emph {et~al.}(1995)\citenamefont {Yang},
  \citenamefont {Nakajima},\ and\ \citenamefont {Sakai}}]{Yang_1995}%
  \BibitemOpen
  \bibfield  {author} {\bibinfo {author} {\bibfnamefont {Tao}\ \bibnamefont
  {Yang}}, \bibinfo {author} {\bibfnamefont {Sadanojo}\ \bibnamefont
  {Nakajima}}, \ and\ \bibinfo {author} {\bibfnamefont {Shiro}\ \bibnamefont
  {Sakai}},\ }\bibfield  {title} {\enquote {\bibinfo {title} {Electronic
  structures of wurtzite {GaN}, {InN} and their alloy
  {$\mathrm{Ga}_{1-x}\mathrm{In}_x\mathrm{N}$} calculated by the tight-binding
  method},}\ }\href {\doibase 10.1143/JJAP.34.5912} {\bibfield  {journal}
  {\bibinfo  {journal} {Jpn. J. Appl. Phys.}\ }\textbf {\bibinfo {volume}
  {34}},\ \bibinfo {pages} {5912} (\bibinfo {year} {1995})}\BibitemShut
  {NoStop}%
\end{thebibliography}%

\end{document}